\documentclass[]{jfm}

\usepackage{graphicx}
\usepackage{psfrag}
\usepackage{tikz}
\usepackage{amssymb}
\usepackage{siunitx}
\usepackage{newtxtext}
\usepackage{newtxmath}
\usepackage{natbib}
\usepackage{hyperref}
\usepackage{comment}
\usepackage{pstool}
\usepackage{cleveref}
\usepackage{amsfonts}
\usepackage{setspace}
\usepackage{color}
\usepackage{floatrow}
\usepackage{subfig}
\usepackage{scalerel}

\hypersetup{
    colorlinks = true,
    urlcolor   = blue,
    citecolor  = black,
}
\definecolor{magenta2}{RGB}{255,0,255}
\crefformat{section}{\S#2#1#3} 
\crefformat{subsection}{\S#2#1#3}
\crefformat{subsubsection}{\S#2#1#3}
\tikzstyle{longdashed}=                  [dash pattern=on 6pt off 2pt]
\tikzstyle{dashdotdot}=              [dash pattern=on 4pt off 2pt on \the\pgflinewidth off 1pt on \the\pgflinewidth off 2pt]
\tikzstyle{dot}=              [dash pattern=on 1pt off 1pt on 1pt off 1pt on 1pt off 1pt]
\DeclareRobustCommand{\redsolid}{\raisebox{2pt}{\tikz{\draw[red,solid,line width=0.9pt](0,0) -- (5mm,0);}}}
\DeclareRobustCommand{\bluedashed}{\raisebox{2pt}{\tikz{\draw[blue,dashed,line width=1.0pt](0,0) -- (5mm,0);}}}
\DeclareRobustCommand{\blackdashed}{\raisebox{2pt}{\tikz{\draw[black,longdashed,line width=1.0pt](0,0) -- (5mm,0);}}}
\DeclareRobustCommand{\greendashdotted}{\raisebox{2pt}{\tikz{\draw[green,dashdotted,line width=1.0pt](0,0) -- (4.7mm,0);}}}

\DeclareRobustCommand{\magentadashed}{\raisebox{2pt}{\tikz{\draw[magenta2,longdashed,line width=1.0pt](0,0) -- (5mm,0);}}}
\DeclareRobustCommand{\blackdashdotdot}{\raisebox{2pt}{\tikz{\draw[black,dashdotdot,line width=1.0pt](0,0) -- (5mm,0);}}}

\DeclareRobustCommand{\blackdot}{\raisebox{2pt}{\tikz{\draw[black,dot,line width=1.0pt](0,0) -- (4.7mm,0);}}}
\newcommand{\overbar}[1]{\mkern 1.5mu\overline{\mkern-1.5mu#1\mkern-1.5mu}\mkern 1.5mu}

\newcommand{\sqdiamond}[1][fill=green, draw=green]{\tikz [x=1.2ex,y=1.5ex,line width=.1ex,line join=round, yshift=-0.285ex] \draw  [#1]  (0,.4) -- (.4,0.8) -- (0.8,.4) -- (.4,0) -- (0,.4) -- cycle;}
\newcommand{\sqtri}[1][fill=blue, draw=blue]{\tikz [x=1.2ex,y=1.65ex,line width=.1ex,line join=round, yshift=-0.285ex] \draw  [#1]  (0,0) -- (.45,0.7) -- (0.9,0) -- (0,0) -- cycle;}

\newcommand{\sqcirgreen}[1][fill=green, draw=green]{\tikz [x=1.2ex,y=1.2ex,line width=.1ex,line join=round, yshift=-0.285ex] \filldraw  [#1]  circle (0.5);}
\newcommand{\sqcirmag}[1][fill=magenta2, draw=magenta2]{\tikz [x=1.2ex,y=1.2ex,line width=.1ex,line join=round, yshift=-0.285ex] \draw  [#1]  circle (0.5);}
\newcommand\dif{\mathrm{d}}

\newcommand{\RomanNumeralCaps}[1]

\title{Structure of an axisymmetric turbulent boundary layer under adverse pressure gradient: a large-eddy simulation study}

\author {Di Zhou\aff{1,2},
  Kan Wang\aff{1,3}
 \and Meng Wang\aff{1}\corresp{\email{m.wang@nd.edu}}}

\affiliation{\aff{1}Department of Aerospace and Mechanical Engineering, Institute for Flow Physics and Control, University of Notre Dame, Notre Dame, IN 46556, USA
\aff{2}Present address: Graduate Aerospace Laboratories, California Institute of Technology, Pasadena, CA 91125, USA
\aff{3}Present address: Cadence Design Systems Inc., San Jose, CA 95134, USA}

\begin{document}
\maketitle

\begin{abstract}
The spatial characteristics and structure of an axisymmetric turbulent boundary layer under strong adverse pressure gradient and weak transverse curvature are investigated using incompressible large-eddy simulation. The boundary layer is on a $20^{\circ}$ tail cone of a body of revolution at a length-based Reynolds number of $1.9\times10^6$. The simulation results are in agreement with the experimental measurements of Balantrapu \textit{et al}. (\textit{J. Fluid Mech.}, vol. 929, 2021) and significantly expand the experimental results with new flow-field details and physical insights. The mean streamwise velocity profiles exhibit a shortened logarithmic region and a longer wake region compared with planar boundary layers at zero pressure gradient. With the embedded-shear-layer scaling, self-similarity is observed for the mean velocity and all three components of turbulence intensity. The azimuthal-wavenumber spectra of streamwise velocity fluctuations possess two peaks in the wall-normal direction, an inner peak at the wavelength of approximately 100 wall units and an outer peak in the wake region with growing strength, wavelength and distance to the wall in the downstream direction. Two-point correlations of streamwise velocity fluctuations show significant downstream growth and elongation of turbulence structures with increasing inclination angle. However, relative to the boundary-layer thickness, the correlation structures decrease in size in the downstream direction. The distributions of streamwise and wall-normal integral lengths across the boundary-layer thickness resemble those of zero-pressure-gradient planar boundary layers, whereas the azimuthal integral length deviates from the planar boundary-layer behavior at downstream stations.
\end{abstract}


\section{Introduction}
\label{sec:intro}

Turbulent boundary layers (TBLs) are among the most extensively studied topics in fluid dynamics. In recent decades, a wealth of experimental and numerical simulation data has been amassed over a broad range of Reynolds numbers and flow conditions. Analysis of the data has led to significant advancements in understanding of TBLs, as highlighted in the recent reviews of \citet{marusic2010wall}, \citet{smiths2011high} and \citet{jimenez2018coherent}. Despite these advancements, knowledge of TBLs developing under adverse pressure gradient (APG) conditions remains less mature compared to that of canonical zero-pressure-gradient (ZPG) TBLs. This knowledge gap presents ongoing challenges to the accurate prediction and control of such flows, which are prevalent in engineering applications. The presence of APG can alter the TBL behavior significantly, potentially leading to flow separation and associated detrimental effects. An enhanced understanding of the flow physics and structures in these conditions is therefore vital to the design and optimization of engineering flow systems, and  forms the foundation for developing effective flow-control strategies.

APG TBLs have been studied actively since the 1950s. Earlier investigations, such as those by \citet{clauser1954turbulent} and \citet{coles1956law}, were predominantly experimental and analytical, and limited to narrow ranges of flow conditions due to the constraints of experimental techniques at the time. The advancement in sensors and measurement technology since the late 1960s facilitated more detailed flow field measurements, and by the 1990s, high-fidelity numerical simulations were introduced as a powerful new tool to the study.  It is now generally understood that (e.g., \citet{spalart1993experimental,maruvsic1995wall,aubertine2005turbulence, monty2011parametric}),
under the influence of APG, the mean-velocity profile develops a larger wake region, and the log-region profile shifts below the classical log-law profile for ZPG TBLs.  Additionally, the turbulent kinetic energy decreases near the wall while a dominant secondary peak develops in the outer layer.  Beyond low-order turbulence statistics, APG also has a significant impact on turbulence structures \citep{krogstad1995influence, lee2009structures, harun2013pressure, maciel2017coherent, lee2017large, volino2020non, gungor2024turbulent}.  Near-wall streaks are weakened and become wider, large-scale turbulent motions in the outer region are energized, and streamwise correlations are decreased with increased inclination angles of coherent structures.  Multiple scaling prameters have been proposed for APG TBLs \citep{zagarola1998mean, aubertine2006reynolds, schatzman2017experimental, maciel2018outer, wei2023outer, chen2023universal}.  In particular, \citet{schatzman2017experimental} experimentally investigated the TBL on a two-dimensional (2-D) ramp with steady and unsteady APG imposed by an airfoil from above with plasma actuation, and proposed an embedded-shear-layer scaling in terms of the shear-layer vorticity thickness and the velocity defect at the upper inflection point of the mean streamwise velocity profile. Using this scaling, similarity was observed in both space and time for the measured mean and phase-averaged velocity profiles and, to a lesser degree, the phase-averaged streamwise turbulence intensity and Reynolds shear stress. \citet{schatzman2017experimental} also tested the embedded-shear-layer scaling with some previously published APG TBL measurements, and confirmed its general applicability.

Most of the previous APG TBL studies were focused on plane-wall or 2-D ramp geometries. When flow encounters an axisymmetric body of revolution (BOR) in the axial direction, an axisymmetric TBL forms in the presence of streamwise pressure gradients. This configuration is common in various aerodynamic and hydrodynamic engineering applications, such as aircraft fuselages and submarine hulls, and has thus motivated many investigations as well. \citet{patel1974measurements} performed detailed measurements of a thick axisymmetric TBL in the tail region of a BOR, revealing different behavior compared to thin planar TBLs, such as significant static pressure variations across the boundary layer and decreasing turbulence levels toward downstream. \citet{hammache2002whole} conducted particle-image-velocimetry (PIV) measurements over an axisymmetric Stratford ramp \citep{stratford1959experimental} constituting the tail section of a BOR. Under streamwise APG, the TBL is continuously on the verge of separation and thickens rapidly. The outer-scaled mean streamwise velocity profiles exhibited a pronounced wake region and self-similarity. A widely studied BOR in recent years is the DARPA SUBOFF geometry \citep{groves1989geometric, huang1992measurements} with or without appendages.  \citet{posa2016numerical,posa2020numerical} conducted large-eddy simulation (LES) of flow over a fully appended SUBOFF to examine the effects of appendages and their junction flows on the development of the downstream boundary layer and wake, and their Reynolds number dependence. The surface geometry was treated using an immersed-boundary method, and good agreement with experimental data was demonstrated. \citet{kumar2018large} performed LES of flow over the SUBOFF hull without appendages at a reduced Reynolds number and zero angle of attack, and also obtained agreement with higher-Reynolds-number experimental results.  Their axisymmetric TBL on the cylindrical midsection of the hull exhibited higher skin friction and more pronounced radial decay of turbulence compared with a ZPG planar TBL.  Additionally, the axisymmetric wake showed self-similarity in mean streamwise velocity but not in turbulence intensities.  Note that the aforementioned LES studies are primarily focused on the behavior of turbulence statistics in the boundary layer and the wake, with little attention to the turbulence structures in the boundary layer and the pressure gradient effect.

In a recent experimental study, \citet{balantrapu2021structure, balantrapu2023wall} at Virginia Tech (VT) focused on a thick axisymmetric TBL that develops on the tail cone of a BOR at zero angle of attack. The BOR features a cylindrical centerbody of diameter $D=\SI{0.432}{\meter}$ and equal length, an ellipsoidal nose also of length $D$, and a tail cone of length $1.17D$ with a $20^{\circ}$ half apex angle.  Measurements were conducted with a free-stream velocity of $\SI{22}{\meter/\second}$, corresponding to a Reynolds number of $1.9\times10^6$ based on the BOR length. Hotwire velocity measurements were supplemented by non-time-resolved PIV measurements in a smaller region, and surface pressure fluctuations were measured using a longitudinal array of microphones. Their single hotwire measurements revealed self-similarity for the streamwise mean-velocity and turbulence intensity profiles when analyzed using the embedded-shear-layer scaling of \cite{schatzman2017experimental}.  Through frequency-spectral analysis of the fluctuating streamwise velocity, they observed that, as in planar TBLs, large-scale motions were energized and grew roughly in proportion to the boundary-layer thickness.  By comparing the two-point streamwise velocity correlations estimated from single-point hotwire data and Taylor's hypothesis with those from PIV measurements, it was noted that the convection velocity was significantly larger than the local mean velocity.  All analyses were based on outer scales because of limited near-wall access in the measurements.

The present numerical study was started in parallel with the VT experiment of \citet{balantrapu2021structure, balantrapu2023wall}, with the objectives of assisting the design of the BOR, particularly in the choice of the tail-cone apex angle that produces a thick boundary layer without separation, cross-validating the experiment and computational results, and obtaining more comprehensive spatial and temporal details of the flow and thereby new insight into the evolution of the axisymmetric TBL under strong APG. Accurate prediction of the tail-cone TBL also served as a precursor for a related study of the noise from a rotor ingesting the TBL at the tail-cone end \citep{zhou2024rotor}. Given the high Reynolds number and the expansive surface area of the BOR, conducting LES posed significant computational challenges, and a side goal was therefore to explore means to mitigate the computational cost while providing a highly accurate description of the TBL in the tail-cone region. To this end, a wall model was employed in the LES in the nose and centerbody sections of the BOR, but the crucial tail-cone section was wall-resolved. This approach, discussed in more detail in \cref{sec:numerical_method}, was validated by comparing its results with the results of fully wall-resolved LES \citep{zhou2020large}. Using the simulation data, a detailed analysis of the tail-cone TBL has been carried out to gain a better understanding of how APG and transverse curvature influence the evolution of the TBL statistical properties and structure. The results are presented in this paper. In addition to new understanding and insights, the simulation results complement the experimental data of \citet{balantrapu2021structure, balantrapu2023wall} for this fundamentally interesting flow with significantly expanded flow regions and details.

The remainder of the paper is structured as follows: Section \ref{sec:Computation_method} describes the numerical approach, flow configuration, and simulation set-up including grids and boundary conditions. In \cref{sec:validation}, the simulation results are validated through grid refinement and comparison with the experimental data of \citet{balantrapu2021structure,balantrapu2023wall} in terms of velocity and surface-pressure statistics, spectra and correlations. Section \ref{sec:analysis} is dedicated to a detailed analysis of the turbulence statistics and structures of the tail-cone TBL and their evolution under the APG. Quantities examined include the velocity statistics profiles and their scaling, pre-multiplied energy spectra, two-point correlation structures and their inclination angles, and integral length scales. Finally, the key findings of the study are summarized in \cref{sec:conclusion}.

\section{Computational methodology}
\label{sec:Computation_method}
\subsection{Numerical approach}
\label{sec:numerical_method}
Flow simulations are conducted using a finite-volume, unstructured-mesh LES code developed at Stanford University \citep{you2008discrete}. The spatially-filtered, incompressible Navier-Stokes equations are solved with second-order accuracy using cell-based, low-dissipative and energy-conservative spatial discretization and a fully implicit, fractional-step time-advancement method with the Crank–Nicolson scheme. The Poisson equation for pressure is solved using the method of Generalized Product Bi-Conjugate Gradient with safety convergence (GPBiCGSafe) proposed by \citet{fujino2012performance}. The subgrid-scale stress is modeled using the dynamic Smagorinsky model \citep{germano1991dynamic,lilly1992proposed}.

Because of the large surface area of the BOR and high Reynolds number, LES of the flow is computationally very expensive if it resolves all the energetic eddies down to the wall. In order to reduce the computational cost, a zonal wall-modeled LES (WMLES) approach is used, where a wall model is applied to the nose and centerbody sections, whereas in the downstream tail-cone section, which is of primary interest, the LES is wall-resolved. In the sections with WMLES, the equilibrium stress-balance wall model \citep{cabot2000approximate,wang2002dynamic} is employed to account for the effect of the near-wall dynamic eddies in terms of approximate wall shear-stress boundary conditions provided to the LES. The accuracy of this approach has been validated previously \citep{zhou2020large} by comparing the results of the zonal WMLES with those from a wall-resolved LES (WRLES) for the entire BOR. In particular, the comparison demonstrates that the tail-cone TBL is relatively insensitive to the detailed near-wall turbulence structures in the upstream boundary layer that are present in the WRLES but unresolved in the WMLES, possibly due to the strong perturbation to the flow produced by the sharp corner of the junction between the centerbody and the tail cone. A similar zonal WMLES approach was used by \citet{posa2020numerical} in their LES of flow over the SUBOFF geometry using an immersed boundary method, also with satisfactory results for the tail-cone TBL in comparison with experimental data at equivalent Reynolds numbers.

\subsection {Configuration and simulation set-up}
\label{sec:setup}

\begin{figure}
\centering
\includegraphics[width=.86\textwidth,trim={0cm 0.1cm 0.0cm 0.1cm},clip]{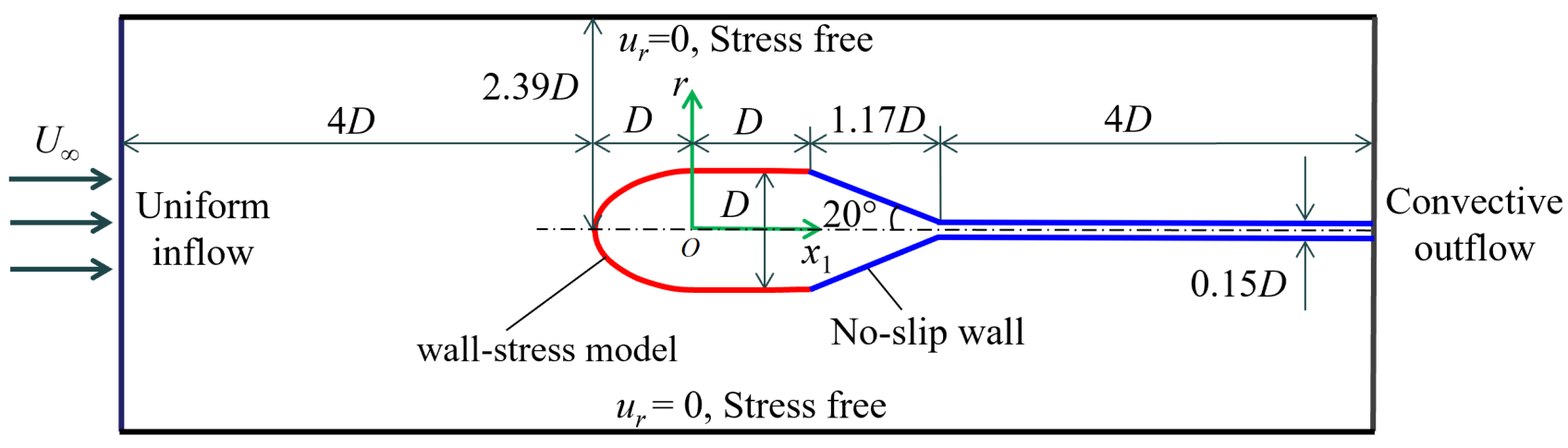}
\caption{Simulation set-up for flow over a body of revolution.}
\label{set_up}
\end{figure}

The physical conditions for the present simulations are consistent with those in the experiment of \citet{balantrapu2021structure}. The flow configuration and boundary conditions are shown schematically in figure~\ref{set_up}. As mentioned eariler, the BOR consists of a cylindrical section in the middle, a 2:1 ellipsoidal nose, and a tail cone that connects to the centerbody at one end and a cylindrical support pole at the other end. For convenience, the diameter $D$ of the centerbody is used as the length scale for normalization. The nose and the centerbody both have a length equal to $D$, the tail cone has a length of $1.17D$ with a $20^{\circ}$ half apex angle, and the support pole has a diameter of $0.15D$. A circumferential trip ring with a rectangular cross-section of $0.002D\times0.001D$ is placed at the downstream end of the nose to induce transition in the boundary layer. The height of the trip ring is one half of the experimental value, which is found to produce a closer match with the experimental boundary-layer thickness in the downstream. Note that the flow-details around the trip is not well-resolved. The BOR is at zero angle of attack, and the Reynolds number is $Re_L=U_\infty L/\nu=1.9\times10^6$ based on the free-stream velocity $U_\infty$ and the total length of the BOR, $L=3.17D$. 

Simulations are conducted in a cylindrical domain of length $11.2D$ and radius $2.39D$. The radius of the domain is chosen to provide the same blockage ratio as in the VT wind tunnel. The center of the cross-section at the nose end, chosen as the origin of the coordinates, is $5D$ downstream from the inlet. For convenience, both a Cartesian coordinate system ($x_1$-$x_2$-$x_3$) with velocity components ($u_1$, $u_2$, $u_3$) and a cylindrical coordinate system ($x_1$-$r$-$\theta$) with velocities ($u_1$, $u_r$, $u_\theta$) are used simultaneously in this paper, where the $x_3$- and $\theta$-coordinates obeying the right-hand rule. The boundary conditions consist of a uniform inflow at the inlet, stress-free conditions with radial velocity $u_r=0$ on the outer boundary, approximate wall-shear stress for WMLES and no-slip condition for WRLES on the respective solid surfaces, and convective outflow conditions at the exit. 

The computational mesh consists of structured-mesh blocks around the surfaces of the BOR and the support pole and unstructured-mesh blocks in the outer region. Over the wall-modeled centerbody, the meshes are relatively coarse with 380 streamwise cells and 1472 azimuthal cells. In the middle of the centerbody, the grid spacings in wall units are $\Delta x^+\approx130$, $\Delta r_{\text{min}}^+\approx25$ and $(R\Delta\theta)^+\approx68$, where $R$ is the centerbody radius, and there are 25 mesh cells across the boundary-layer thickness. Since the tail-cone section is the focus of the investigation, it is wall-resolved, and the mesh is significantly finer in wall units. The transition to the wall-resolved mesh is gradual and starts at $0.2D$ upstream of the centerbody-tail cone junction. To check grid convergence, simulations are performed on two meshes with different resolutions in the tail-cone section. On the coarse mesh, the tail-cone section has 800 streamwise cells, 1472 uniformly spaced azimuthal cells (same as on the centerbody), and 90 cells across the thickness of the TBL at the end of the cone. Except in the immediate neighborhood of the sharp centerbody-tail cone corner, the streamwise and azimuthal grid spacings are less than 30 and 60 wall units, respectively, and decrease to 12 and 2 at the end of the tail cone. The wall-normal spacing for the first off-wall cell is less than 2 wall units and decreases to 0.8 at the tail-cone end. On the fine mesh, the tail-cone section has 1200 and 2944 uniformly spaced cells in the streamwise and azimuthal directions, respectively, and 170 cells across the thickness of the TBL at the tail-cone end. The streamwise and azimuthal grid spacings in wall units are both less than 30 slightly (0.05D) downstream of the upper tail-cone corner, and decrease to 6 and 1, respectively, at the tail-cone end. The wall-normal spacing for the first off-wall cell on the tail cone decreases from 1.2 to 0.3 along the streamwise direction. In total, the coarse mesh contains $3.43\times10^8$ cells, and the fine mesh has $9.16\times10^8$ cells.

A maximum Courant–Friedrichs–Lewy number of 1.8 is employed for time advancement in all simulations. This corresponds to time step sizes of $\Delta t U_\infty/D \approx 1.32\times 10^-4$ for the coarse mesh and $8.96\times 10^-5$ for the fine mesh. The simulations are first run for over one and half flow-through times ($16.8D/U_\infty$) to wash out initial transients, and then another two flow-through times ($22.3D/U_\infty$) to obtain converged statistics.
 
\section{Validation}\label{sec:validation}


\floatsetup[figure]{style=plain,subcapbesideposition=top}
\begin{figure}
\centering

\sidesubfloat[]{
{\includegraphics[width=.75\textwidth,trim={0.1cm 0.1cm 0.1cm 0cm},clip]{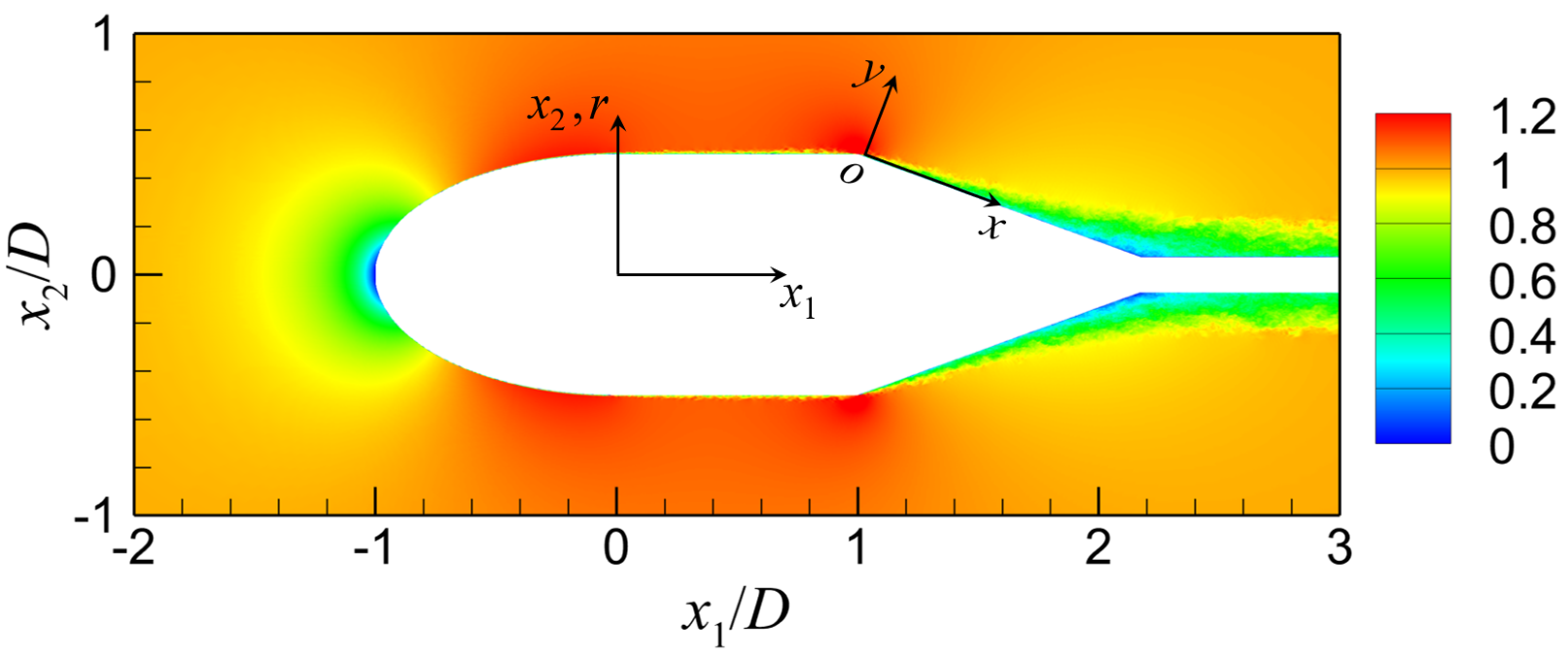}\label{mean}}}

\sidesubfloat[]{
{\includegraphics[width=.47\textwidth,trim={0.0cm 0.1cm 0.0cm -1.5cm},clip]{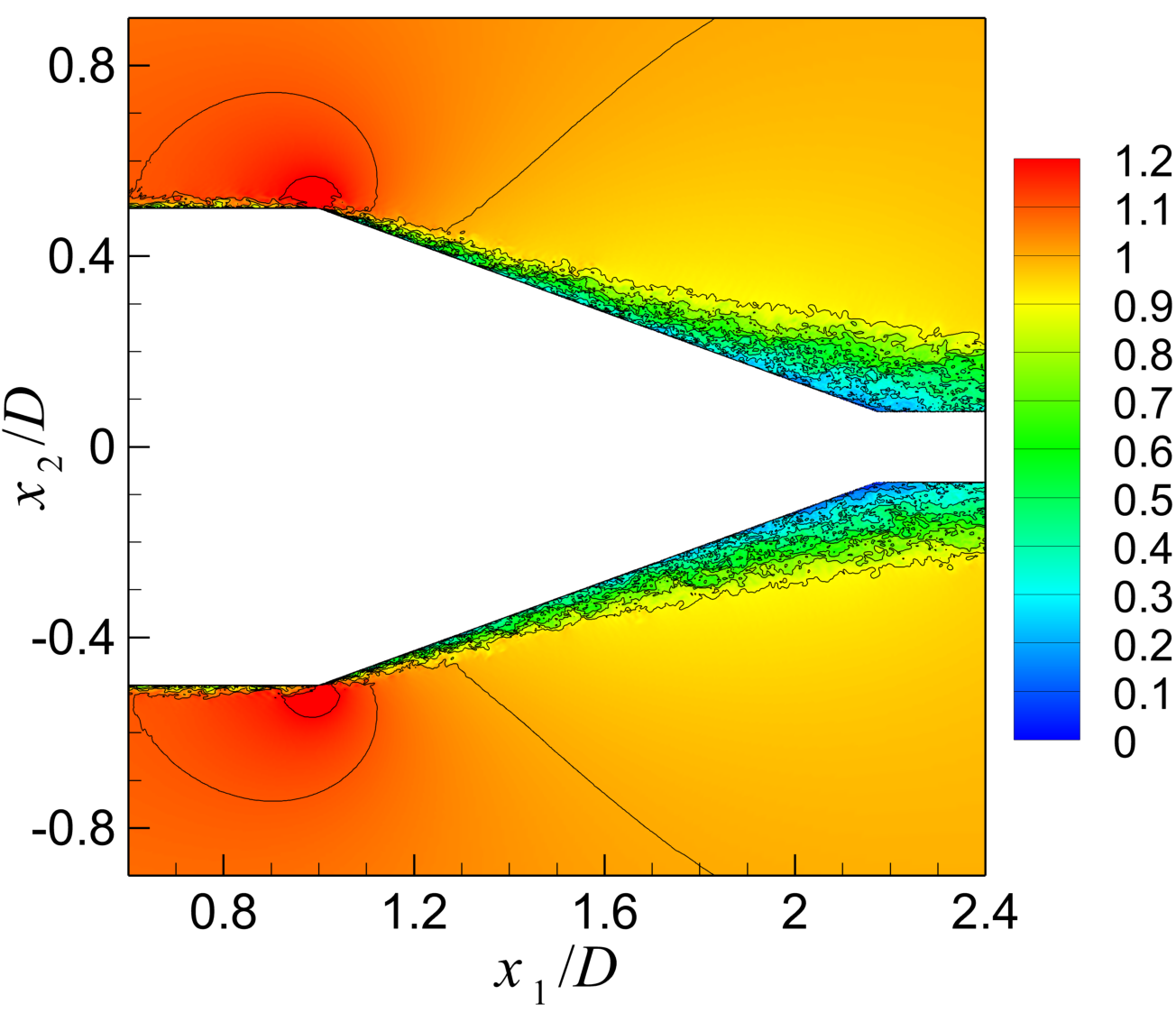}}\label{inst_cone}}
\sidesubfloat[]{
{\includegraphics[width=.49\textwidth,trim={0.4cm 0.0cm 0.6cm 0.1cm},clip]{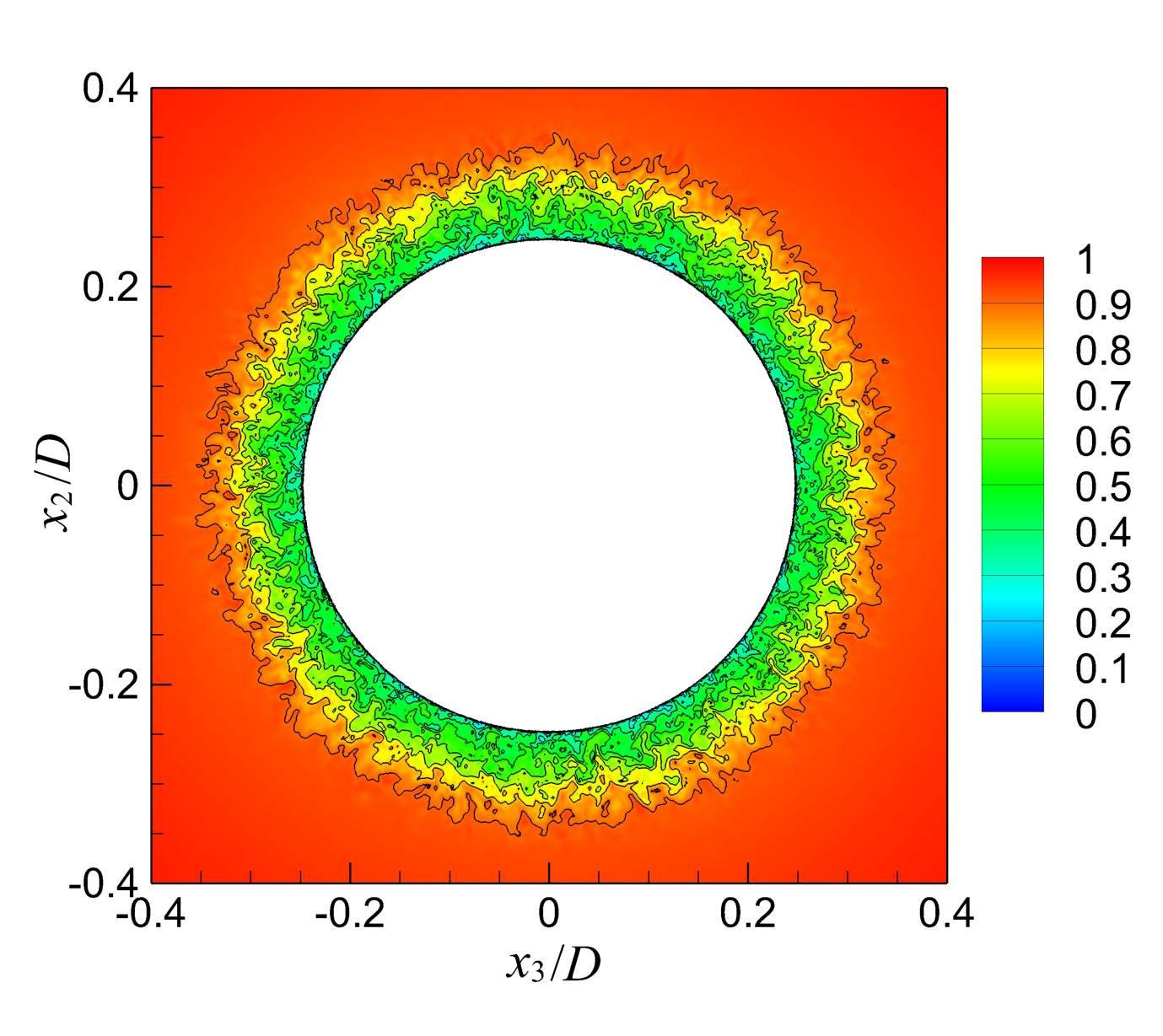}}\label{inst_cone_cross}}
\vspace{-0.5cm}
\caption{Isocontours of instantaneous axial velocity $u_1/U_\infty$ in (a) the $x_3=0$ plane; (b) the $x_3=0$ plane around the tail cone; (c) the $x_1/D=1.69$ plane. Part (a) also illustrates the coordinate systems employed in this study.}
\label{u1_inst&mean}
\end{figure}

The velocity field around the BOR is illustrated in figure~\ref{u1_inst&mean} in terms of the instantaneous axial-velocity contours in two perpendicular planes through the BOR axis. The results are obtained from the fine-mesh simulation. The flow accelerates around the nose, transitions quickly to turbulence after the trip ring at the downstream end of the nose, and accelerates again as it approaches the sharp corner between the centerbody and the tail cone. Downstream of this corner, under the influence of APG, the flow decelerates over the tail cone, causing a rapid thickening of the boundary layer. Notably, the tail-cone TBL remains attached except for a very small separation (too small to be seen in the figure) immediately after the corner, which is consistent with the observations in the VT experiment \citep{balantrapu2021structure}.

\begin{figure}
\centering
{\psfrag{b}[][]{\large{$x_{1}/D$}}
\psfrag{d}[][]{\large{$C_p$}}
\includegraphics[width=.55\textwidth,trim={0.7cm 6cm 0 1.5cm},clip]{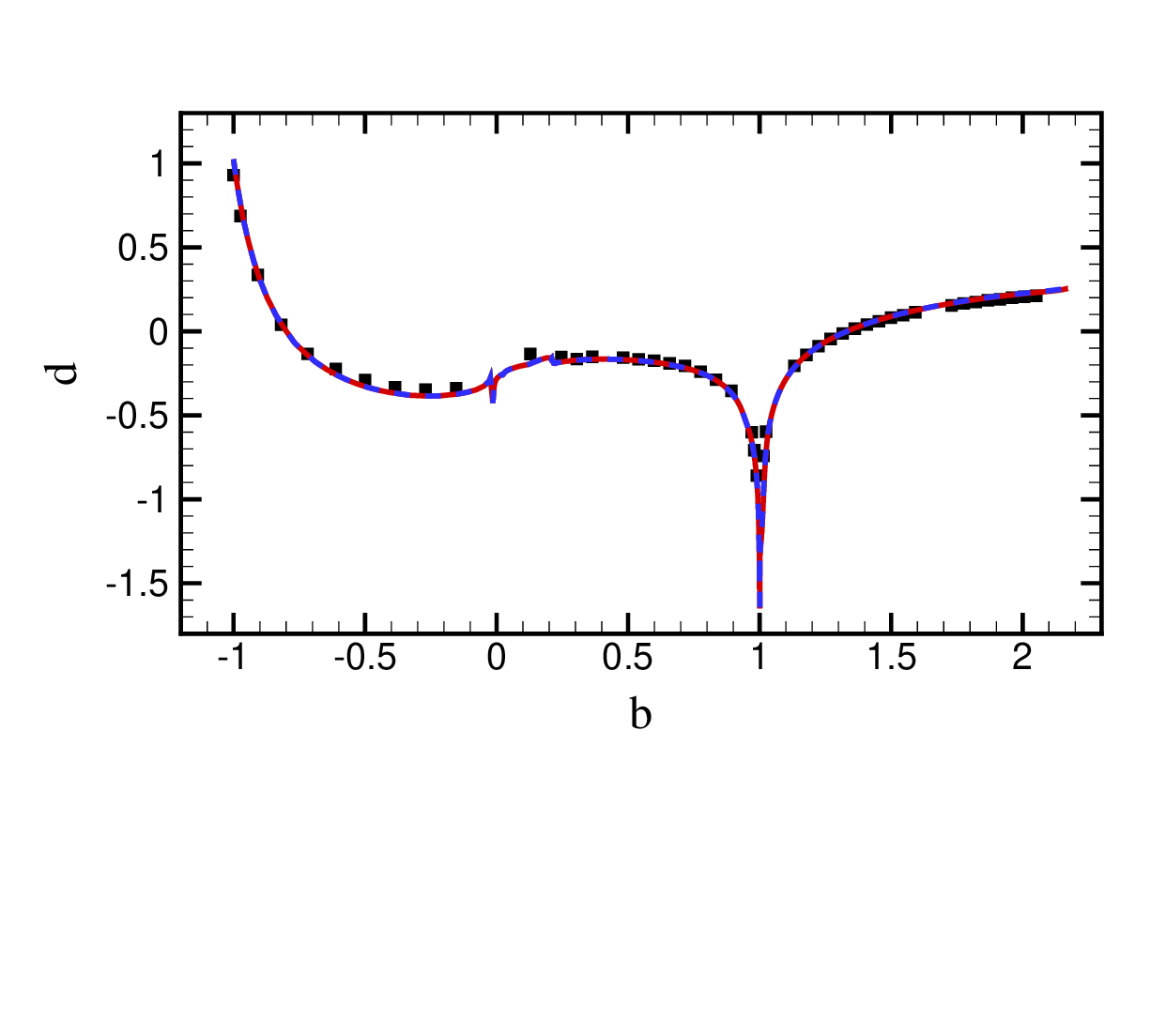}
}
\caption{Mean pressure coefficient on the BOR: {\redsolid}, fine mesh; {\bluedashed}, coarse mesh; $\blacksquare$, experiment \citep{balantrapu2021structure}}
\label{Cp}
\end{figure}

\begin{figure}
\centering
{\psfrag{x}[][]{\large{$x_{1}/D$}}
\psfrag{y}[][]{\large{$C_f$}}
\includegraphics[width=.54\textwidth,trim={1.1cm 6cm 0 1.5cm},clip]{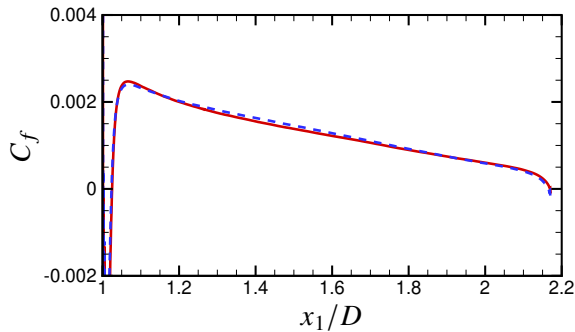}
}
\caption{Mean skin-friction coefficient on the tail cone of the BOR: {\redsolid}, fine mesh; {\bluedashed}, coarse mesh.}
\label{Cf}
\end{figure}

In figure~\ref{Cp}, the axial distribution of the mean pressure coefficient, $C_p=(P_w-P_\infty)/(\frac{1}{2}\rho U_\infty^2)$, on the BOR is compared with the experimental data of \citet{balantrapu2021structure}. The reference pressure $P_\infty$ is taken at the inlet near the radial boundary. The distributions obtained from the two simulations are indistinguishable and agree well with the experimental data. The $C_p$ distribution illustrates a strong favorable pressure gradient in the nose region and immediately upstream of the centerbody-tail cone corner. Downstream of the corner, the flow is first subjected to a very strong APG, followed by a more mild APG over the majority of the tail cone. The small kink seen at $x_1/D\approx0$ is caused by the transition trip. Figure~\ref{Cf} shows the axial distribution of the mean skin-friction coefficient, $C_f=\tau_{w,x}/(\frac{1}{2}\rho U_\infty^2)$, on the tail-cone section of the BOR, where $\tau_{w,x}$ is the mean streamwise shear stress on the wall. A tiny separation bubble with negative $C_f$ is observed at the beginning of the tail cone, and differences between the results from the two meshes are very small, indicating that grid convergence has been achieved.

\begin{figure}
\centering
{\psfrag{x}[][]{{$x_1/D$}}
\psfrag{y}[][]{{$p'_{w, \text{rms}}/(\rho U_{\infty}^2)$}}
\includegraphics[width=.55\textwidth,trim={0.1cm 6.5cm 1cm 1.5cm},clip]{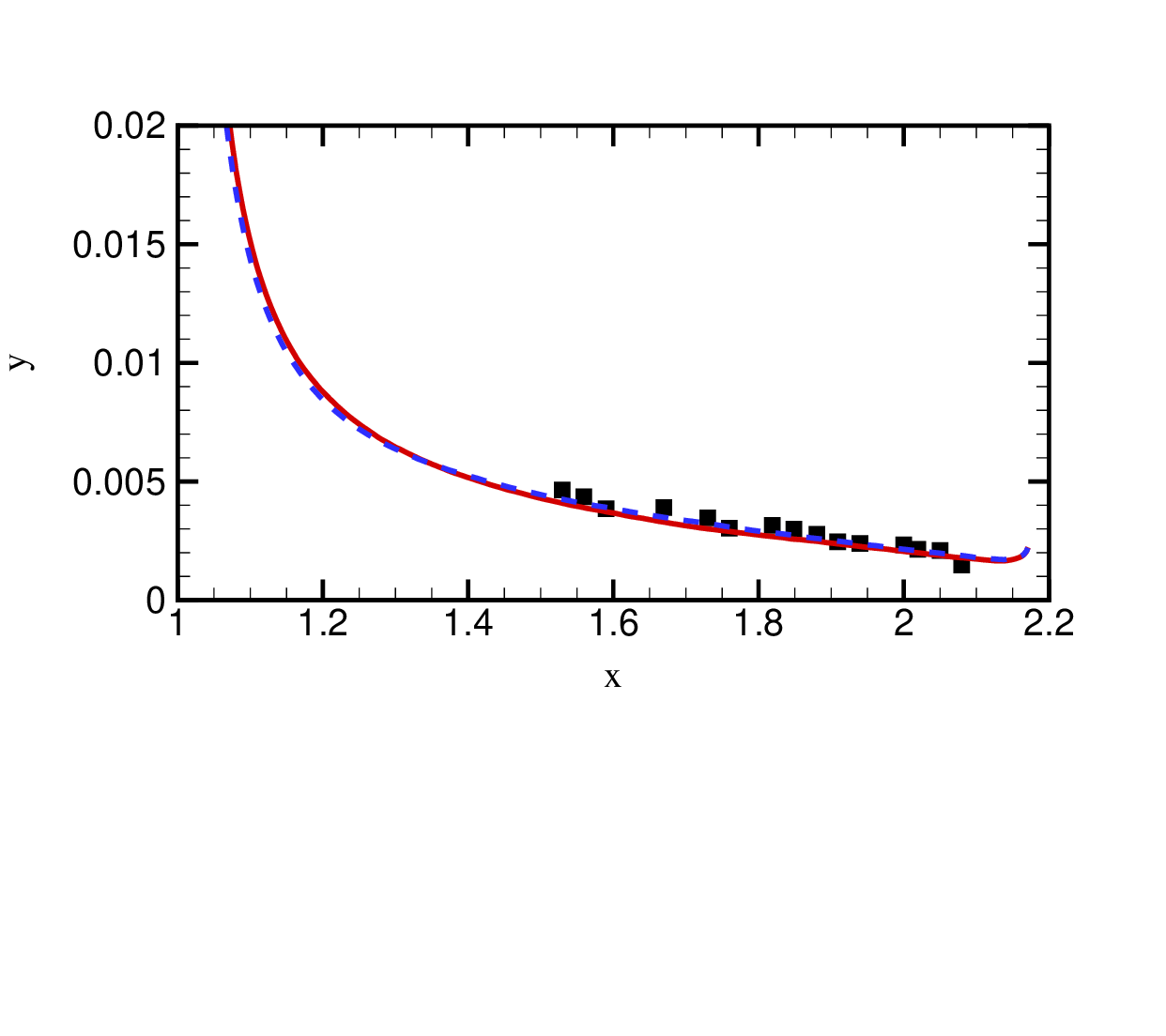}}
\caption{R.M.S. values of pressure fluctuations on the tail-cone surface: {\redsolid}, fine mesh; {\bluedashed}, coarse mesh; $\blacksquare$, experiment \citep{balantrapu2023wall}.}
\label{prms}
\end{figure}

\begin{figure}
\centering
\sidesubfloat[]{
{\psfrag{b}[][]{\small{$fD/U_\infty$}}
\psfrag{d}[][]{\small{$\Phi_{pp}U_\infty/[(\rho U_\infty^2)^2 D]$}}
\includegraphics[width=0.46\textwidth,trim={0.0 6cm 2.0cm 1.5cm},clip]{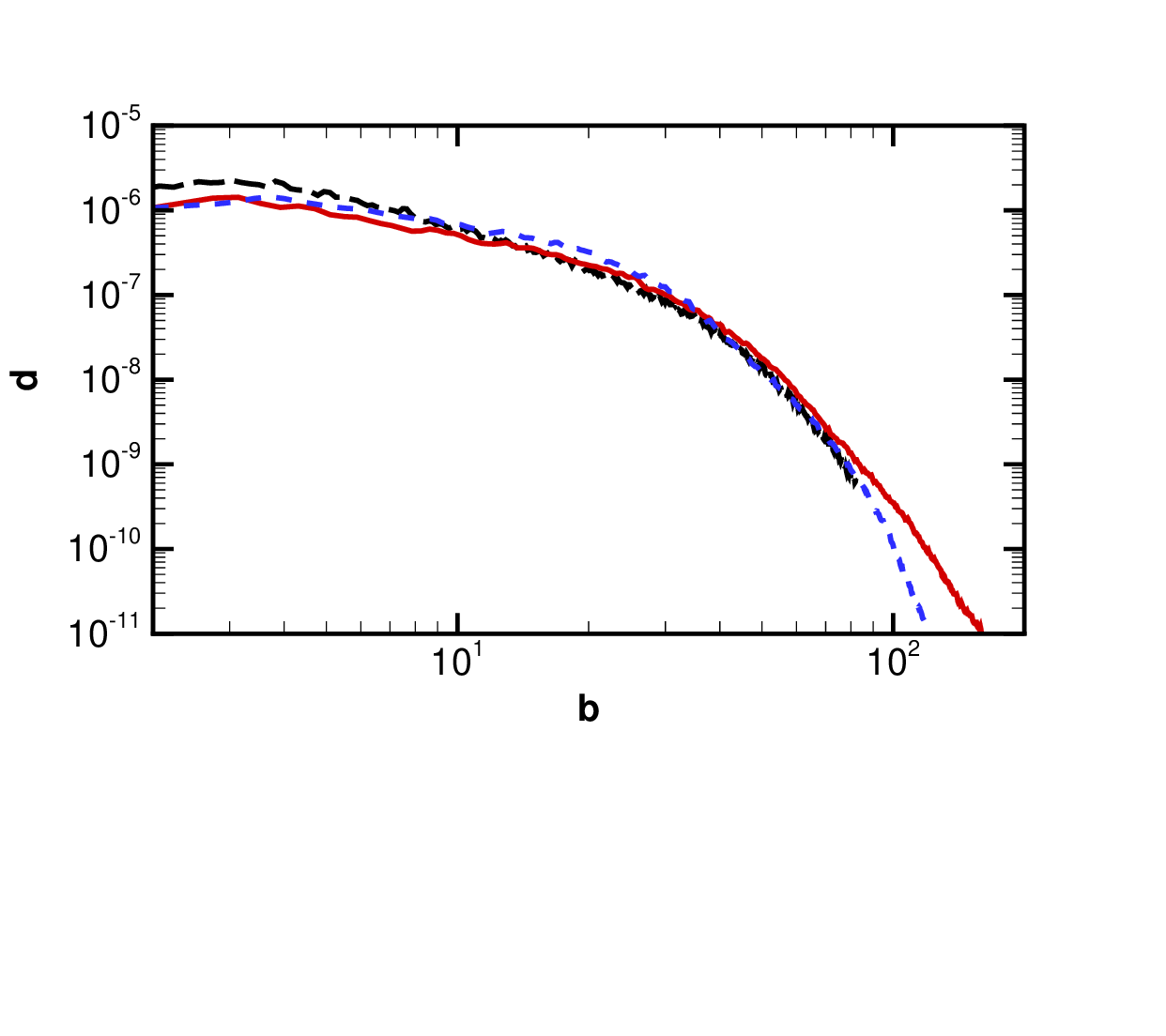}}}
\sidesubfloat[]{
{\psfrag{b}[][]{\small{$fD/U_\infty$}}
\psfrag{d}[][]{\small{$\Phi_{pp}U_\infty/[(\rho U_\infty^2)^2 D]$}}
\includegraphics[width=0.46\textwidth,trim={0.0 6cm 2.0cm 1.5cm},clip]{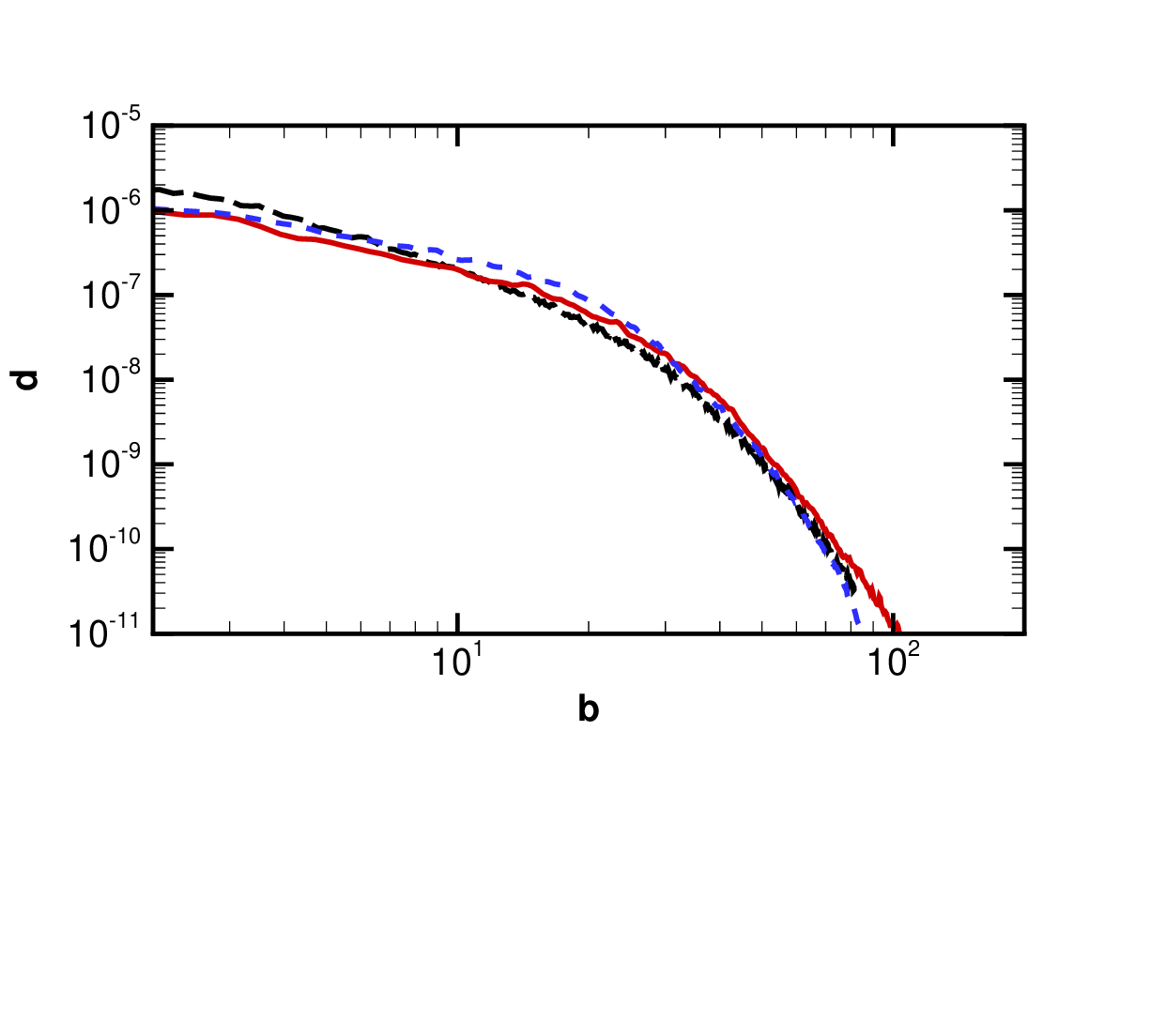}}}

\caption{Frequency spectra of surface-pressure fluctuations on the tail cone at: (a) $x_1/D = 1.53$; (b) $x_1/D = 1.76$. {\redsolid}, fine mesh; {\bluedashed}, coarse mesh; {\blackdashed}, experiment \citep{balantrapu2023wall}.}
\label{wallpressure_fluct}
\end{figure}

Figure~\ref{prms} shows the root-mean-square (r.m.s.) values of pressure fluctuations $p'_w$ on the tail-cone surface. The results from coarse- and fine-mesh simulations again agree well with each other and with the experimental data of \citet{balantrapu2023wall}. The frequency spectra of pressure fluctuations at two axial locations on the tail-cone surface, $x/D=1.53$ and $1.76$, are shown in figure~\ref{wallpressure_fluct}. The simulations predict the wall-pressure fluctuations well over nearly the entire frequency range compared with the experimental data \citep{balantrapu2023wall}. The grid resolution effect is small and mainly affects the high-frequency content.

\begin{figure}
\centering
\sidesubfloat[]{
{\psfrag{b}[][]{{$U_1/U_\infty$}}
\psfrag{d}[][]{{$r/D$}}\includegraphics[width=0.31\textwidth,trim={0.3cm 0.5cm 7.6cm 1.8cm},clip]{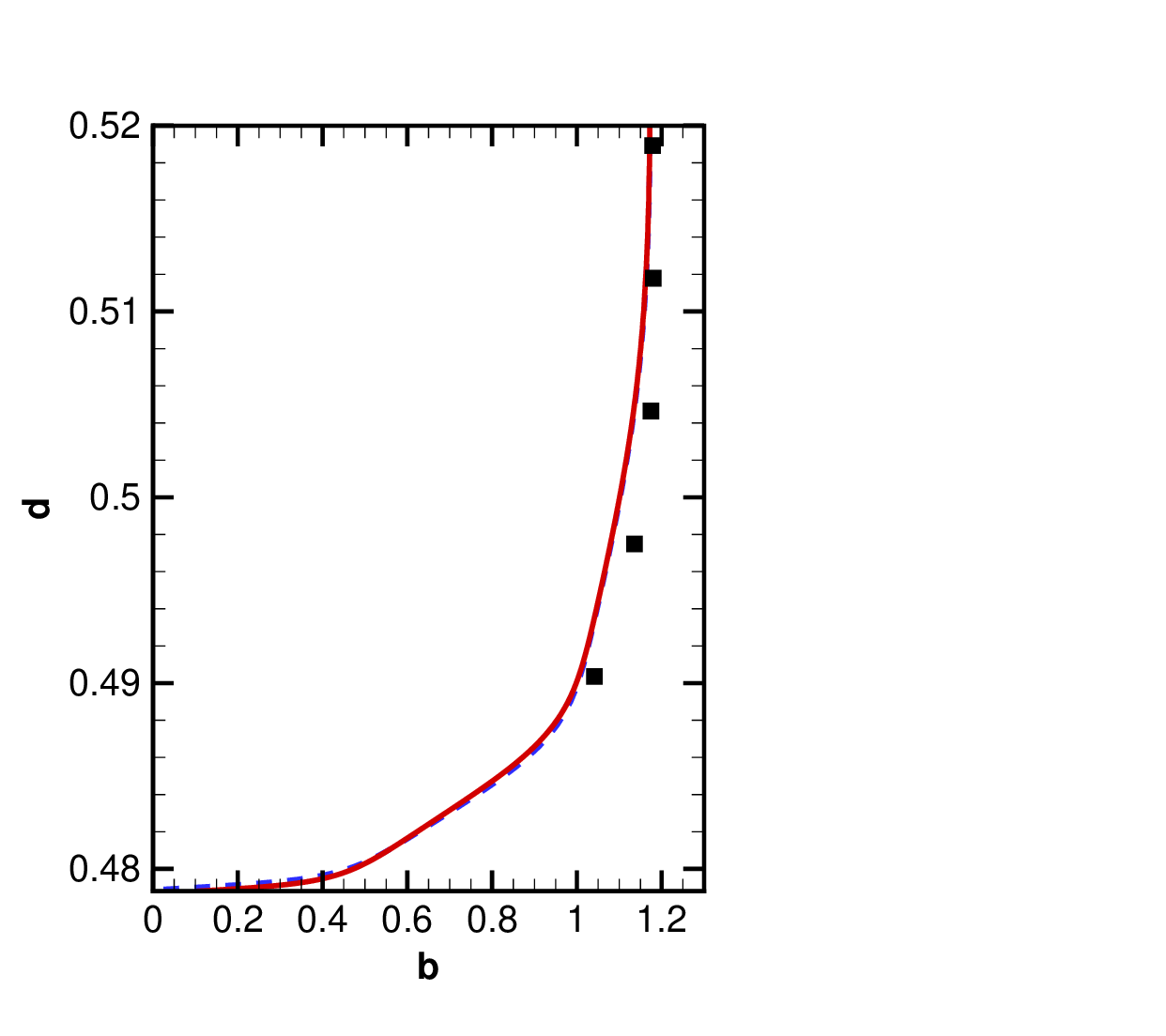}} 
{\psfrag{b}[][]{{$u'_{1, \text{rms}}/U_\infty$}}
\psfrag{d}[][]{{$r/D$}}\includegraphics[width=0.31\textwidth,trim={0.3cm 0.5cm 7.6cm 1.8cm},clip]{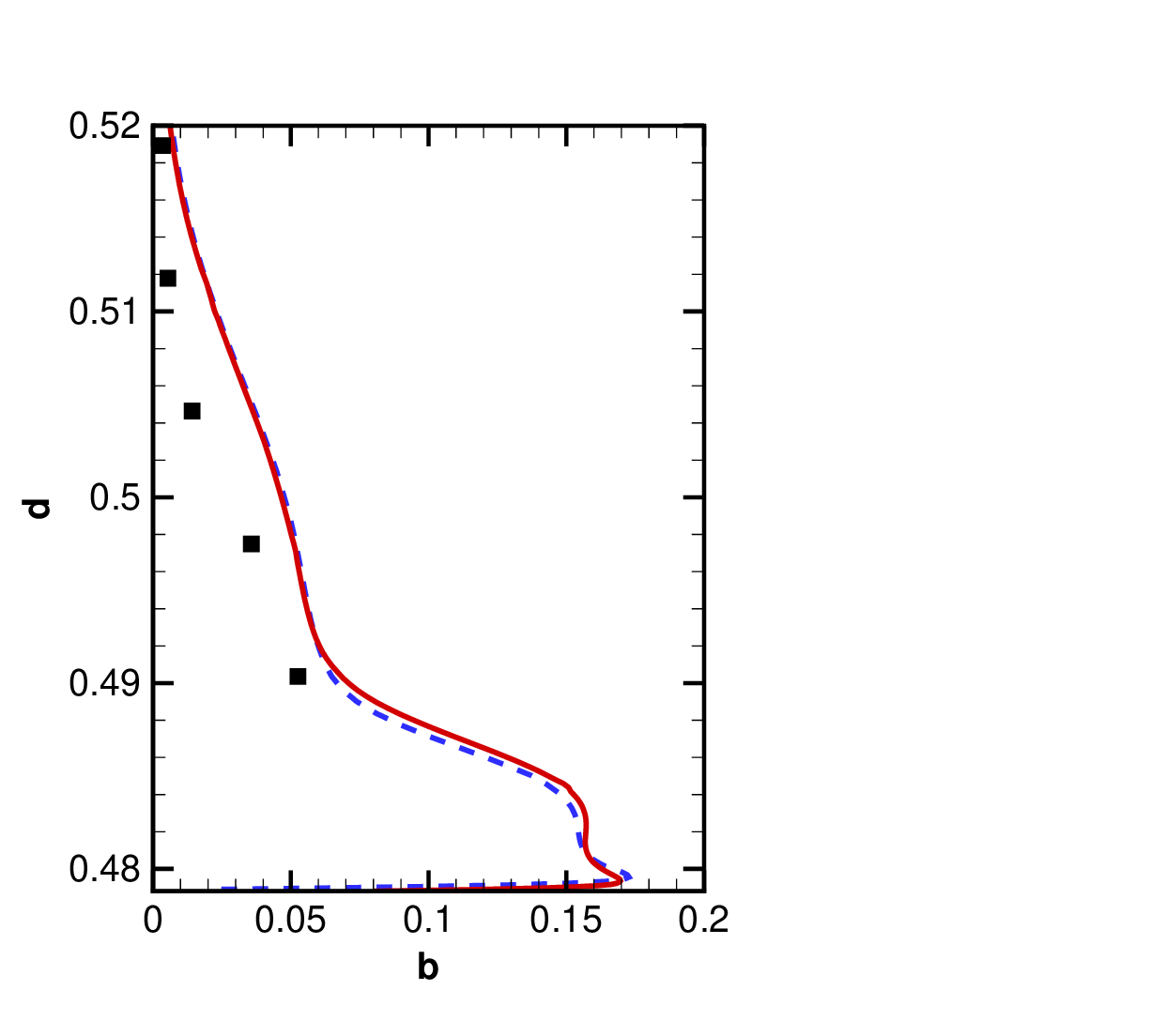}}
{\psfrag{b}[][]{{$u'_{\theta, \text{rms}}/U_\infty$}}
\psfrag{d}[][]{{$r/D$}}\includegraphics[width=0.31\textwidth,trim={0.3cm 0.5cm 7.6cm 1.8cm},clip]{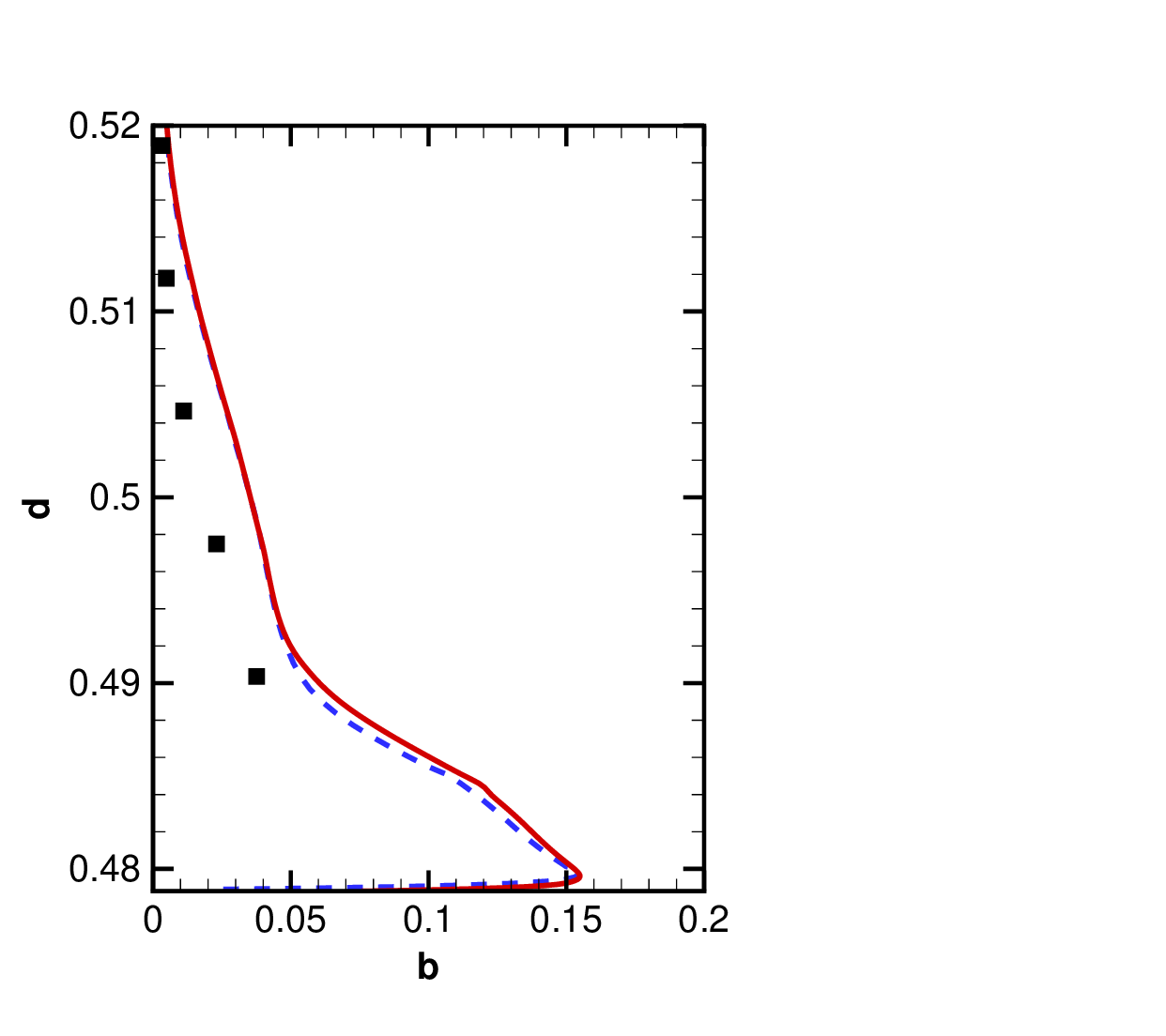}}
}

\sidesubfloat[]{
{\psfrag{b}[][]{{$U_1/U_\infty$}}
\psfrag{d}[][]{{$r/D$}}\includegraphics[width=0.31\textwidth,trim={0.3cm 0.5cm 7.6cm 1.8cm},clip]{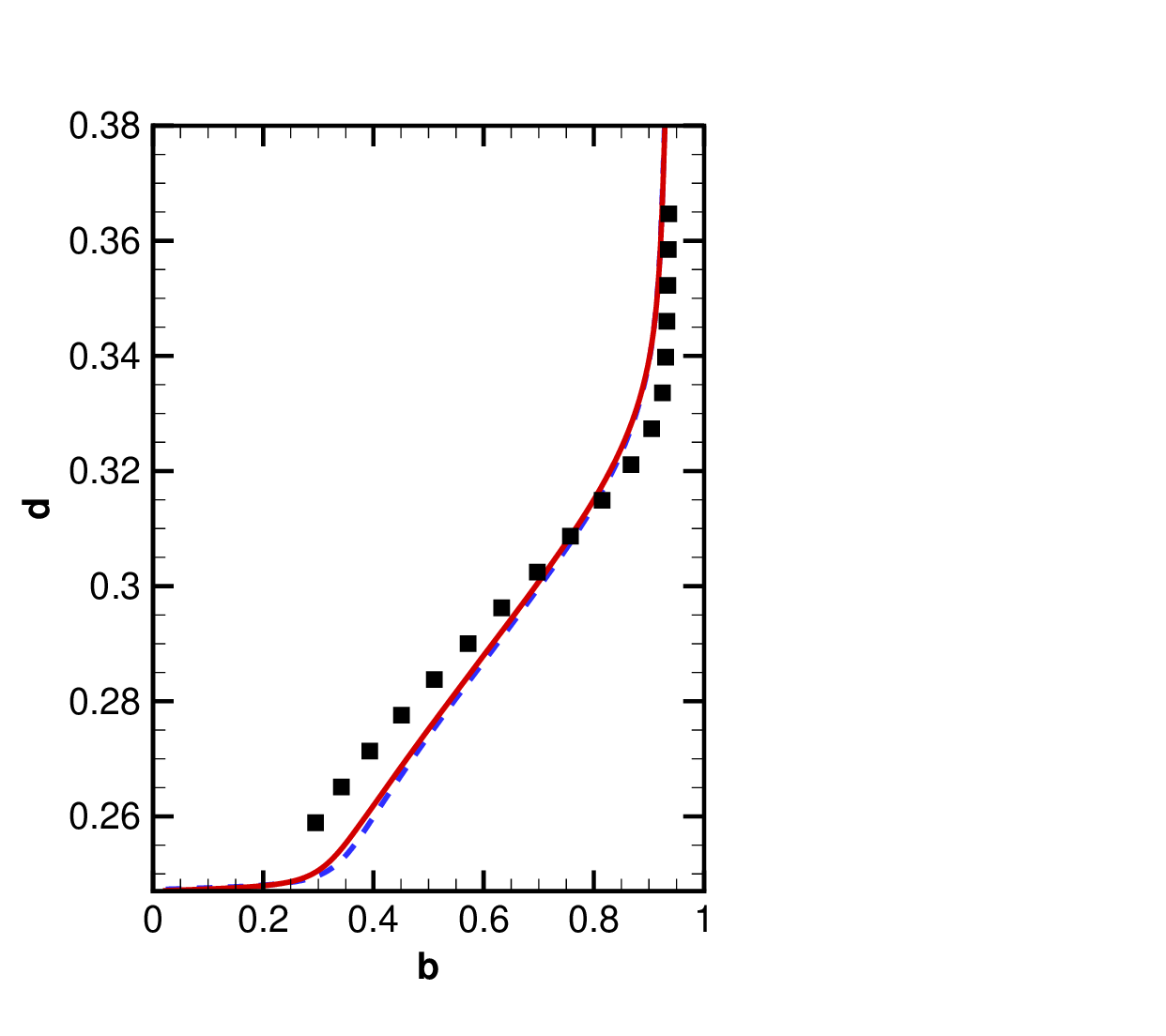}} 
{\psfrag{b}[][]{{$u'_{1, \text{rms}}/U_\infty$}}
\psfrag{d}[][]{{$r/D$}}\includegraphics[width=0.31\textwidth,trim={0.3cm 0.5cm 7.6cm 1.8cm},clip]{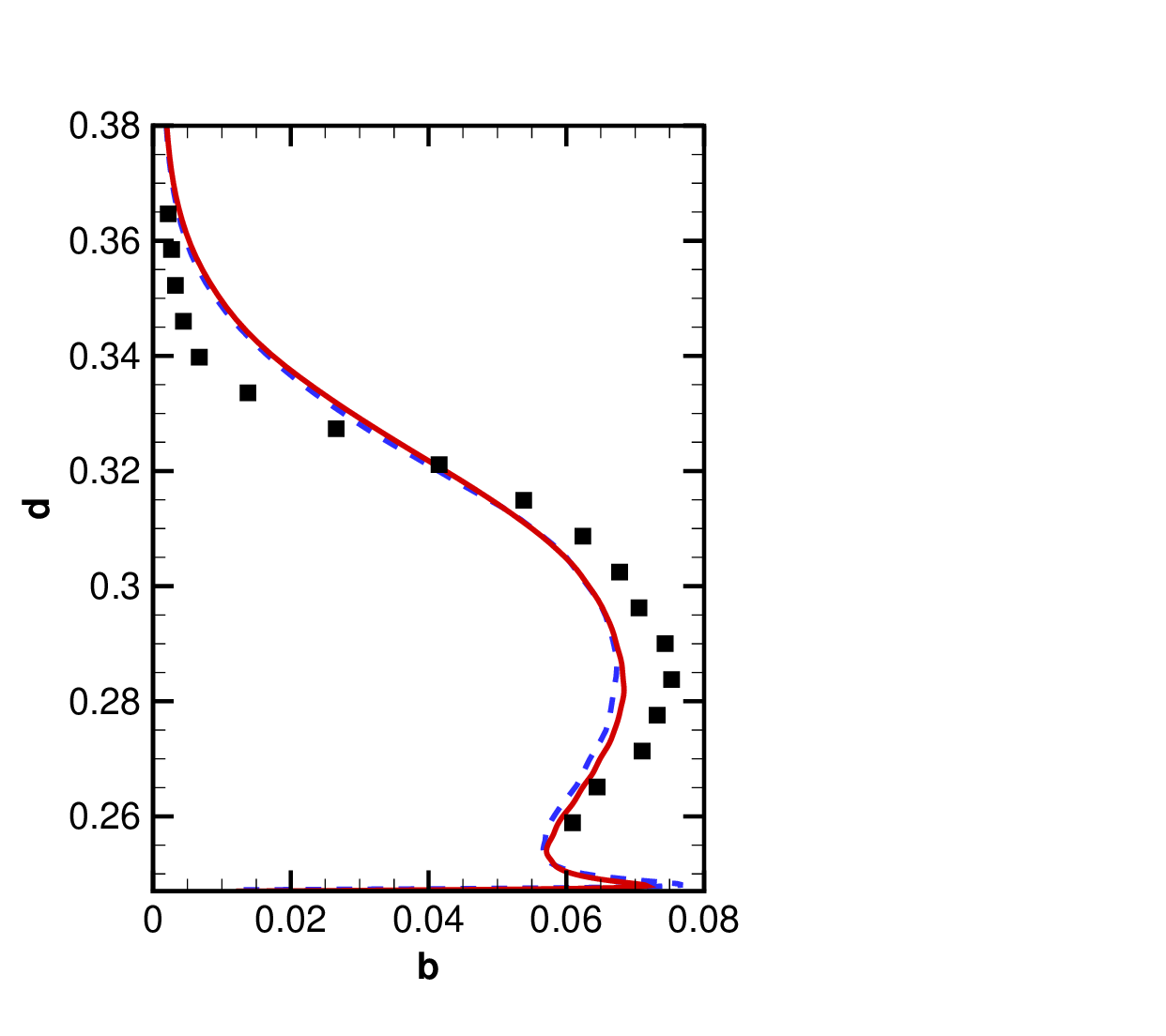}}
{\psfrag{b}[][]{{$u'_{\theta, \text{rms}}/U_\infty$}}
\psfrag{d}[][]{{$r/D$}}\includegraphics[width=0.31\textwidth,trim={0.3cm 0.5cm 7.6cm 1.8cm},clip]{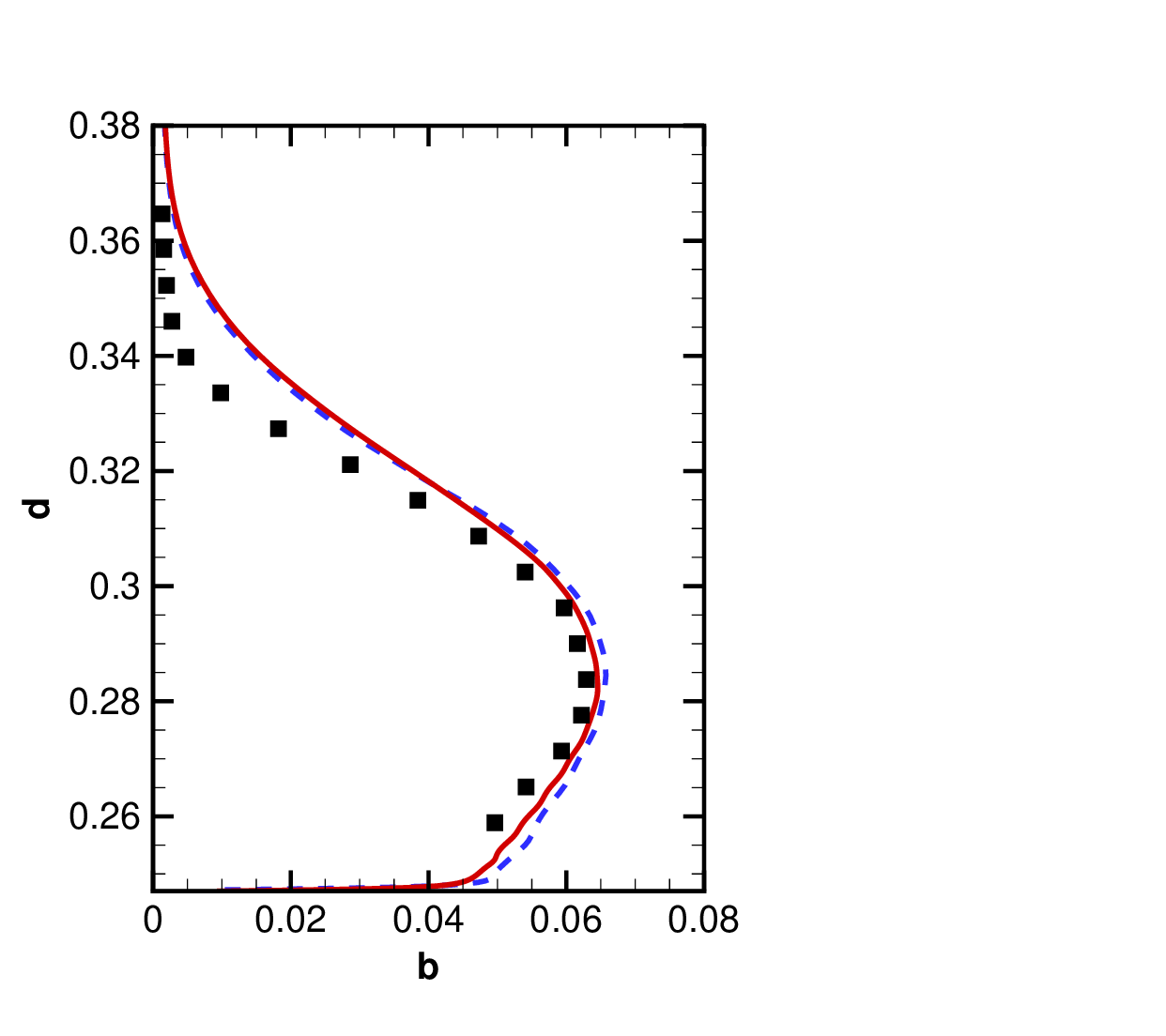}}
}

\sidesubfloat[]{
{\psfrag{b}[][]{{$U_1/U_\infty$}}
\psfrag{d}[][]{{$r/D$}}\includegraphics[width=0.31\textwidth,trim={0.3cm 0.5cm 7.6cm 1.8cm},clip]{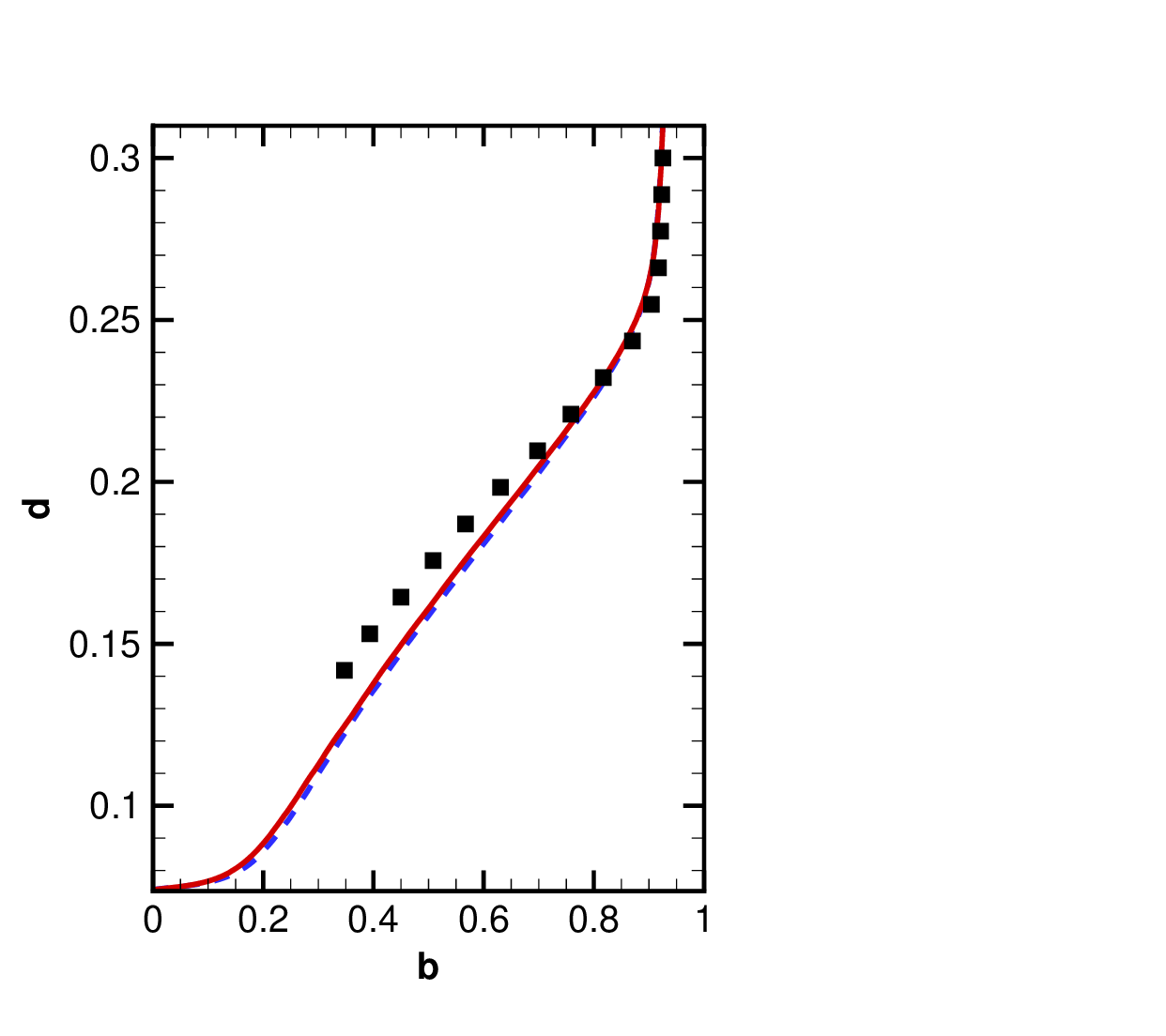}} 
{\psfrag{b}[][]{{$u'_{1, \text{rms}}/U_\infty$}}
\psfrag{d}[][]{{$r/D$}}\includegraphics[width=0.31\textwidth,trim={0.3cm 0.5cm 7.6cm 1.8cm},clip]{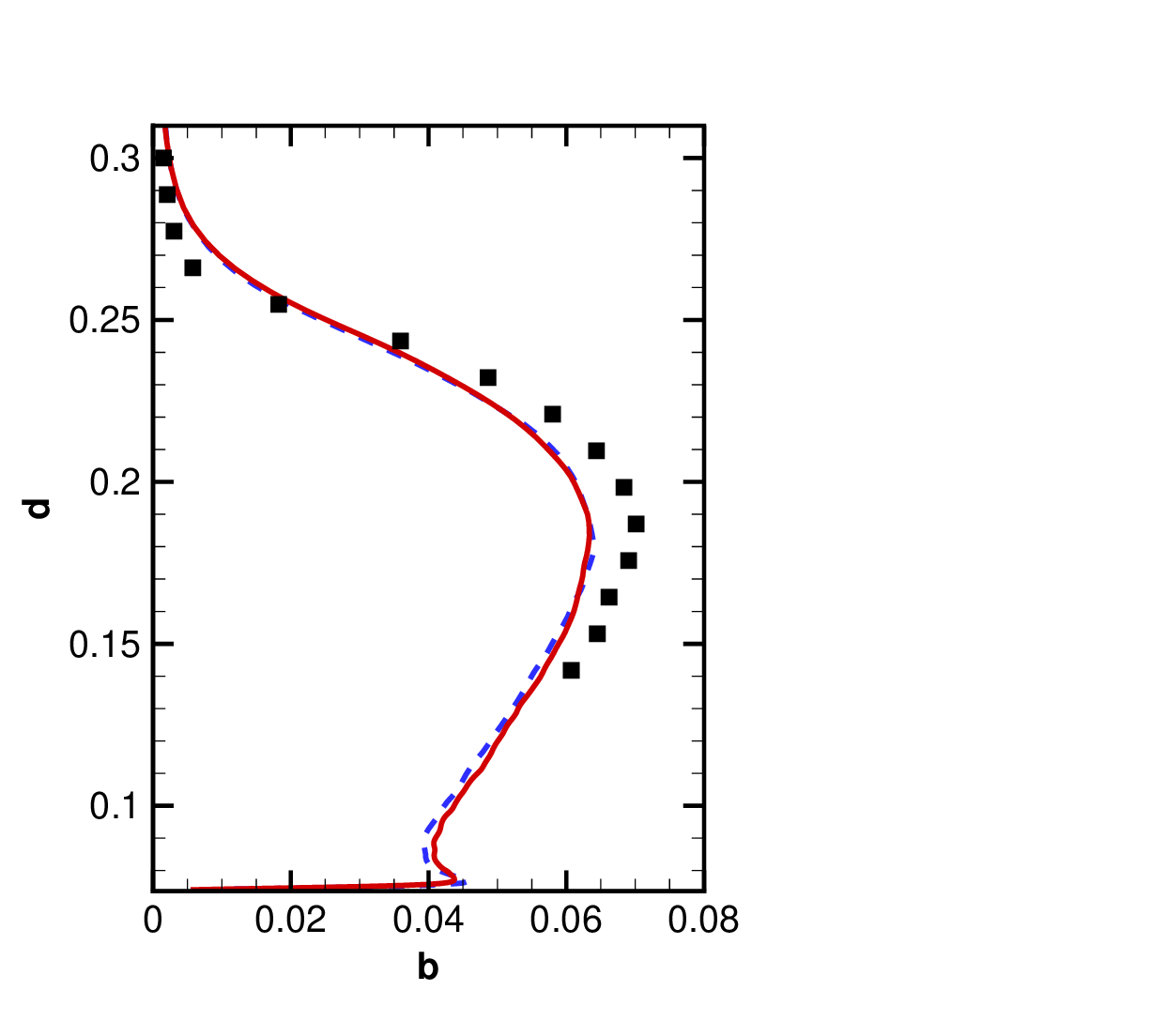}}
{\psfrag{b}[][]{{$u'_{\theta, \text{rms}}/U_\infty$}}
\psfrag{d}[][]{{$r/D$}}\includegraphics[width=0.31\textwidth,trim={0.3cm 0.5cm 7.6cm 1.8cm},clip]{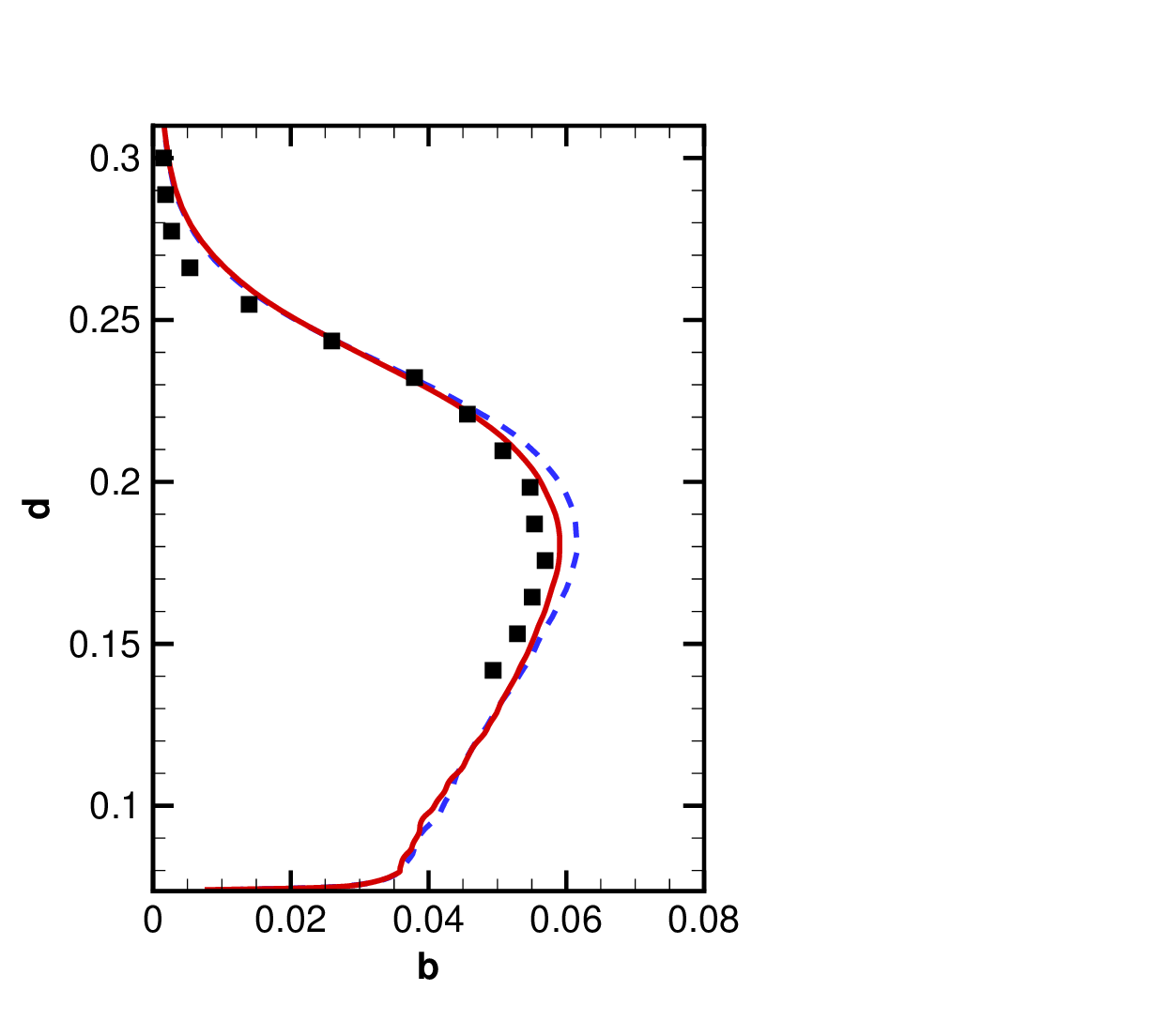}}
}

\caption{Profiles of mean axial velocity and the r.m.s. values of axial and azimuthal velocity fluctuations at: (a) $x_1/D = 1.06$; (b) $x_1/D = 1.69$; (c) $x_1/D = 2.17$. {\redsolid}, fine mesh; {\bluedashed}, coarse mesh; $\blacksquare$, experiment \citep{hickling2019turbulence}.}
\label{vel_cone}
\end{figure}

A comparison of boundary layer profiles with the measurements of \citet{balantrapu2021structure} is made in figure~\ref{vel_cone}, where the profiles of mean axial velocity $U_1$ and r.m.s.~values of both axial $(u'_{1,\text{rms}})$ and azimuthal $(u'_{\theta,\text{rms}})$ velocity fluctuations are depicted at three axial stations on the tail cone. The numerical profiles demonstrate grid convergence and are in reasonable agreement with the measurement data obtained using a four-sensor hot-wire \citep{balantrapu2021structure}. These results show quantitatively the flow deceleration and boundary-layer thickening along the tail cone. At the downstream end of the tail cone, $x_1/D=2.17$, the boundary-layer thickness reaches nearly one half of the BOR radius $R$, or approximately ten times the boundary-layer thickness immediately upstream of the tail cone. As the TBL develops along the tail cone with decreasing radius, the peak turbulence intensities shift towards the outer region of the boundary layer and the axisymmetric TBL behaves increasingly like an axisymmetric wake.

\begin{figure}
\centering

\sidesubfloat[]{
{\psfrag{b}[][]{{$fD/U_\infty$}}
\psfrag{d}[][]{{$E_{u_{1}u_{1}}/(U_\infty D)$}}
\includegraphics[width=0.46\textwidth,trim={1cm 4cm 0cm 1.8cm},clip]{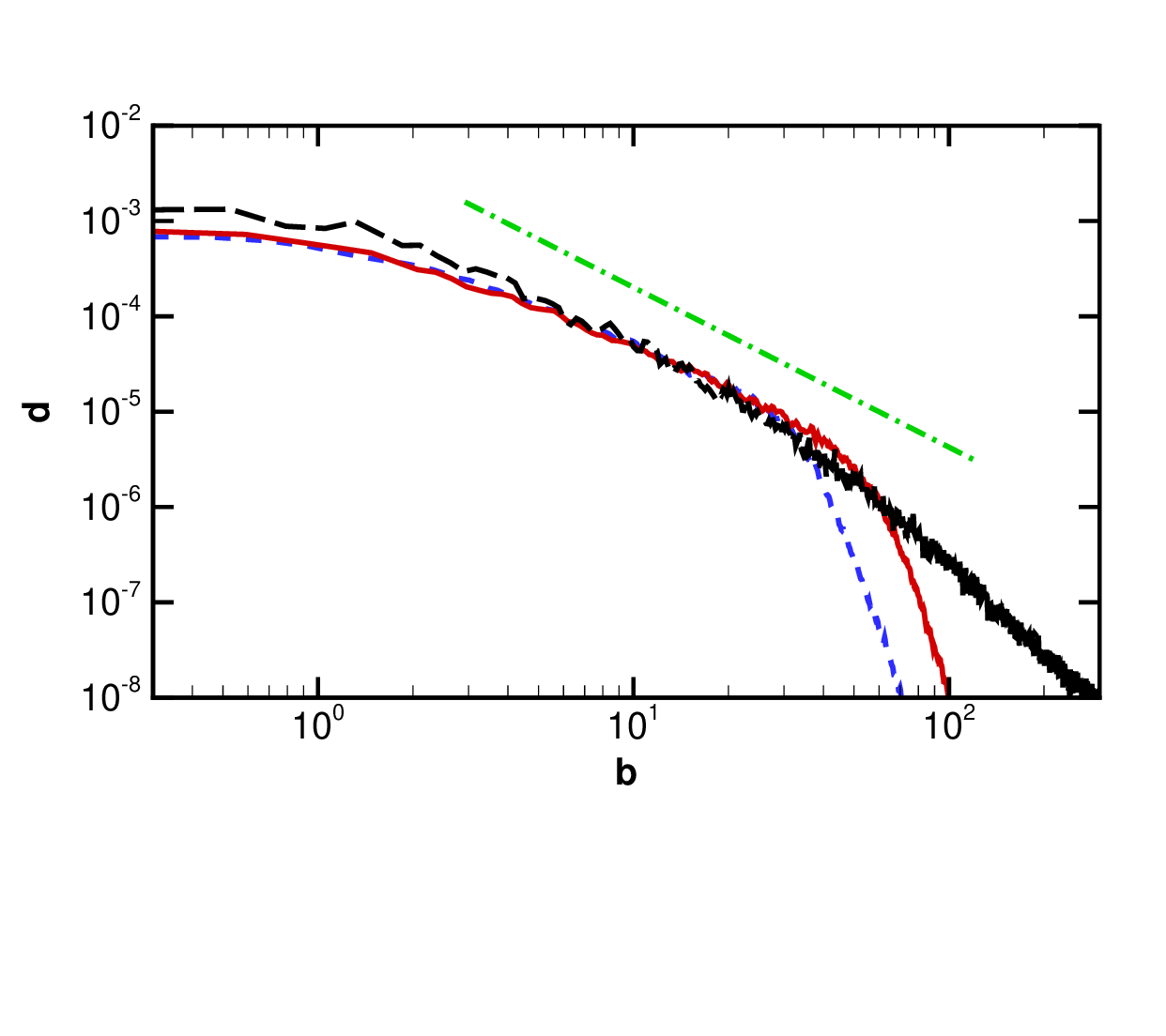}}}
\sidesubfloat[]{
{\psfrag{b}[][]{{$fD/U_\infty$}}
\psfrag{d}[][]{{$E_{u_{\theta}u_{\theta}}/(U_\infty D)$}}
\includegraphics[width=0.46\textwidth,trim={1cm 4cm 0cm 1.8cm},clip]{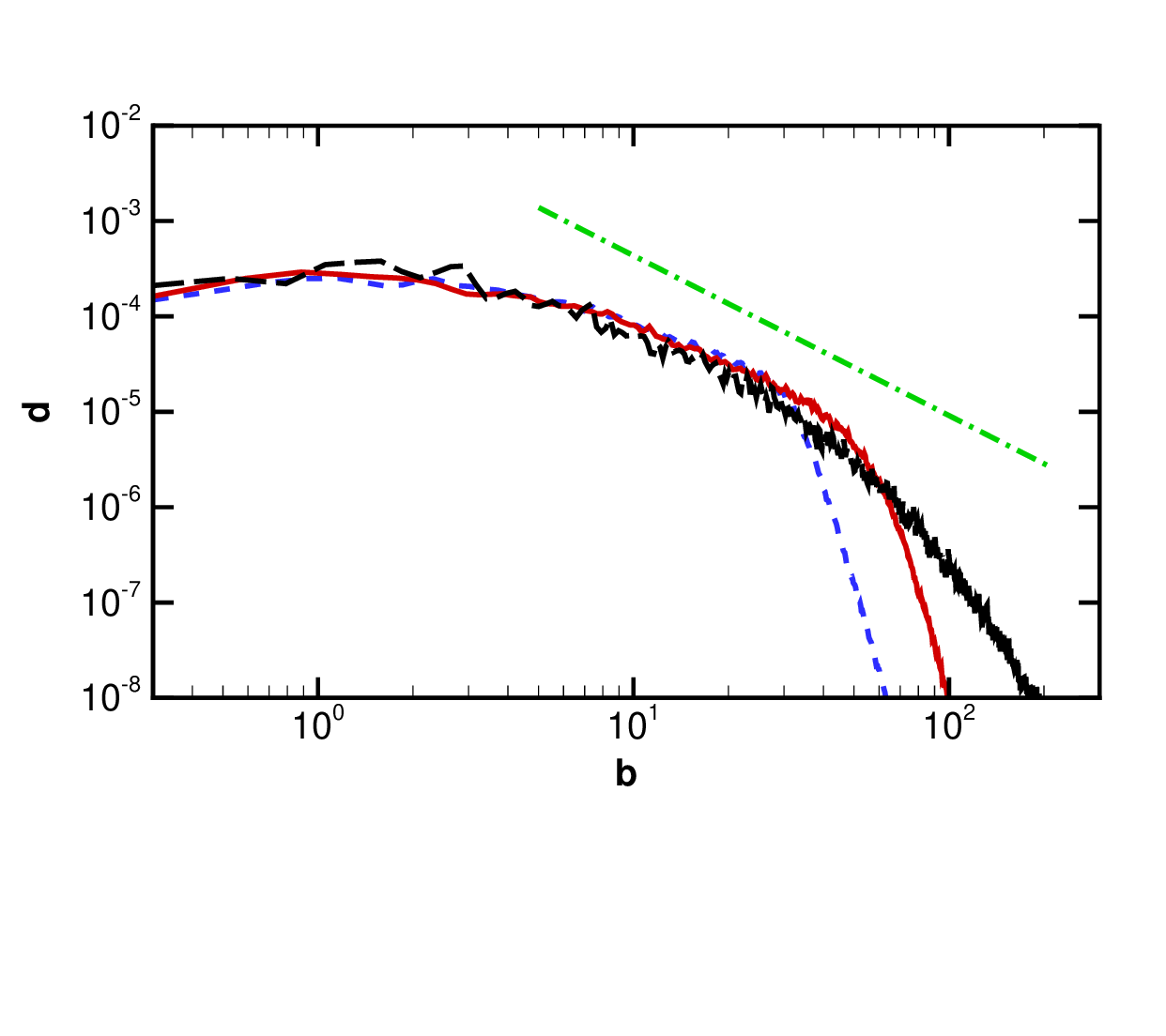}}}

\sidesubfloat[]{
{\psfrag{b}[][]{{$fD/U_\infty$}}
\psfrag{d}[][]{{$E_{u_{1}u_{1}}/(U_\infty D)$}}
\includegraphics[width=0.46\textwidth,trim={1cm 4cm 0cm 1.8cm},clip]{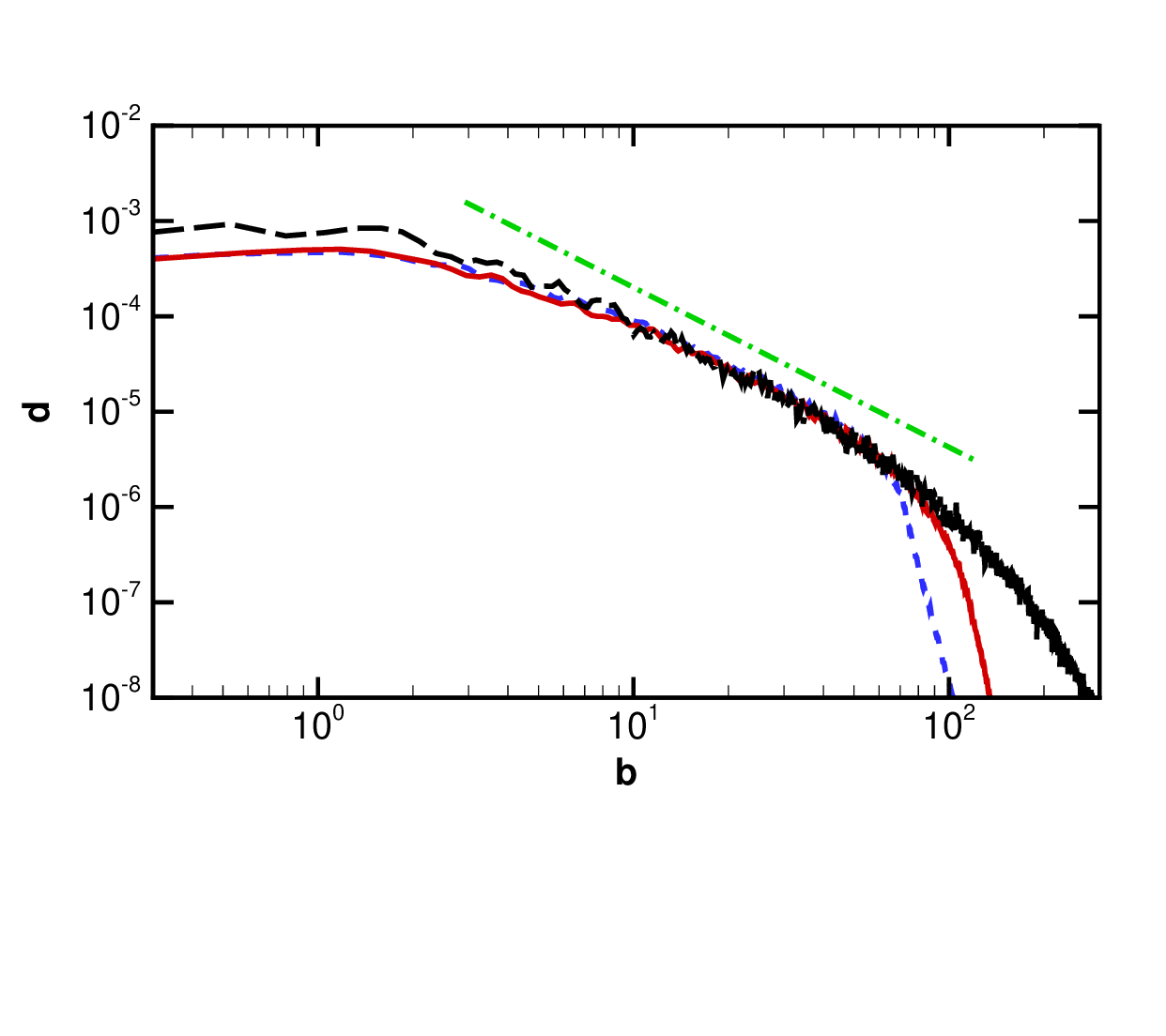}}}
\sidesubfloat[]{
{\psfrag{b}[][]{{$fD/U_\infty$}}
\psfrag{d}[][]{{$E_{u_{\theta}u_{\theta}}/(U_\infty D)$}}
\includegraphics[width=0.46\textwidth,trim={1cm 4cm 0cm 1.8cm},clip]{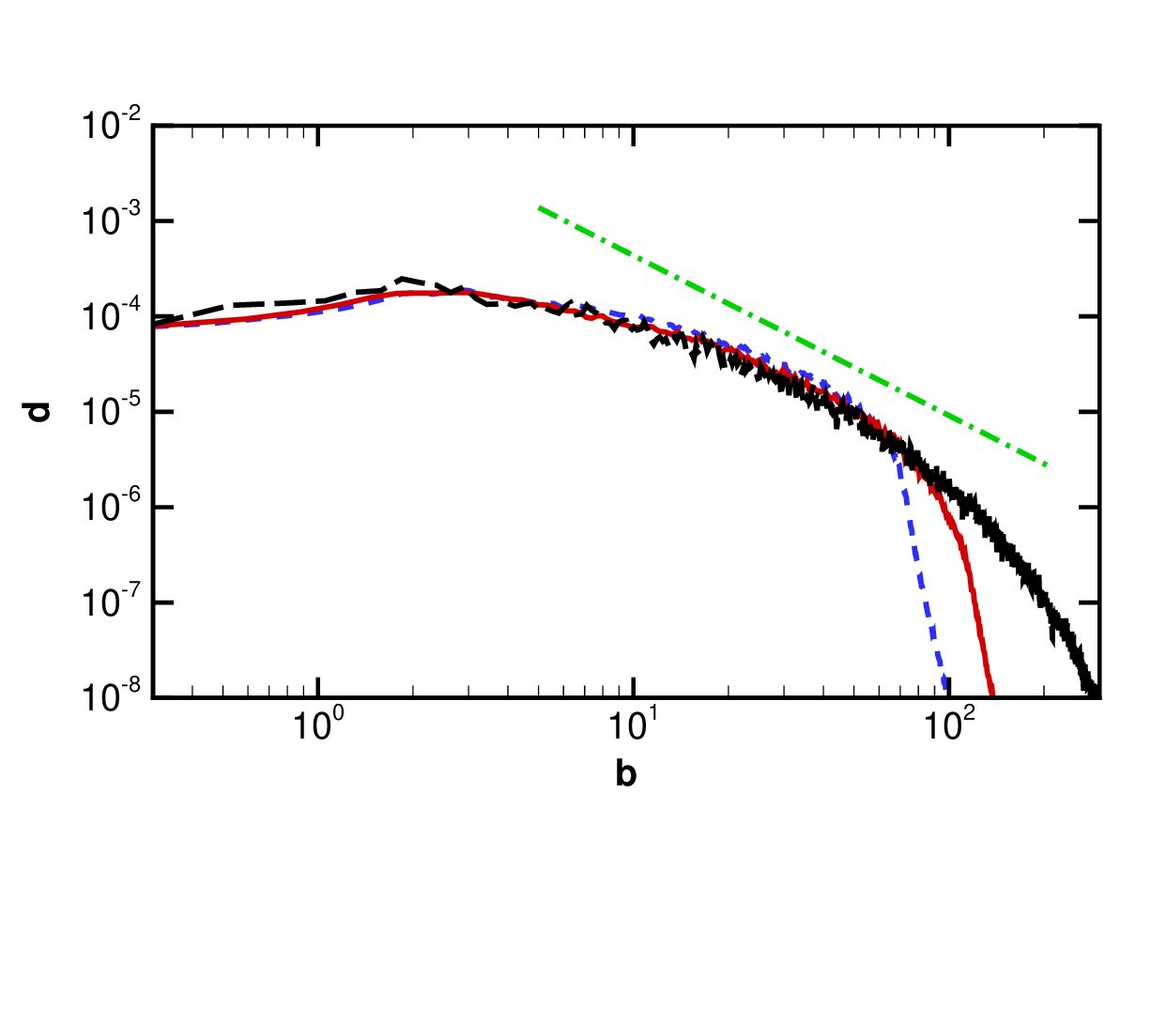}}}
\caption{Frequency spectra of fluctuating axial (a,c) and azimuthal (b,d) velocities at two radial locations at the tail-cone end, $x_1/D = 2.17$: (a,b) $r/D=0.14$; (c,d) $r/D=0.21$. {\redsolid}, fine mesh; {\bluedashed}, coarse mesh; {\blackdashed}, experiment \citep{hickling2019turbulence}; {\greendashdotted}, -5/3 slope.}
\label{uspectra_cone}
\end{figure}

\begin{figure}
\centering

\sidesubfloat[]{
{\psfrag{x}[][]{{$\Delta r/D$}}
\psfrag{y}[][]{{$C_{u u}$}}\includegraphics[width=0.46\textwidth,trim={1.6cm 3.8cm 1.2cm 1.86cm},clip]{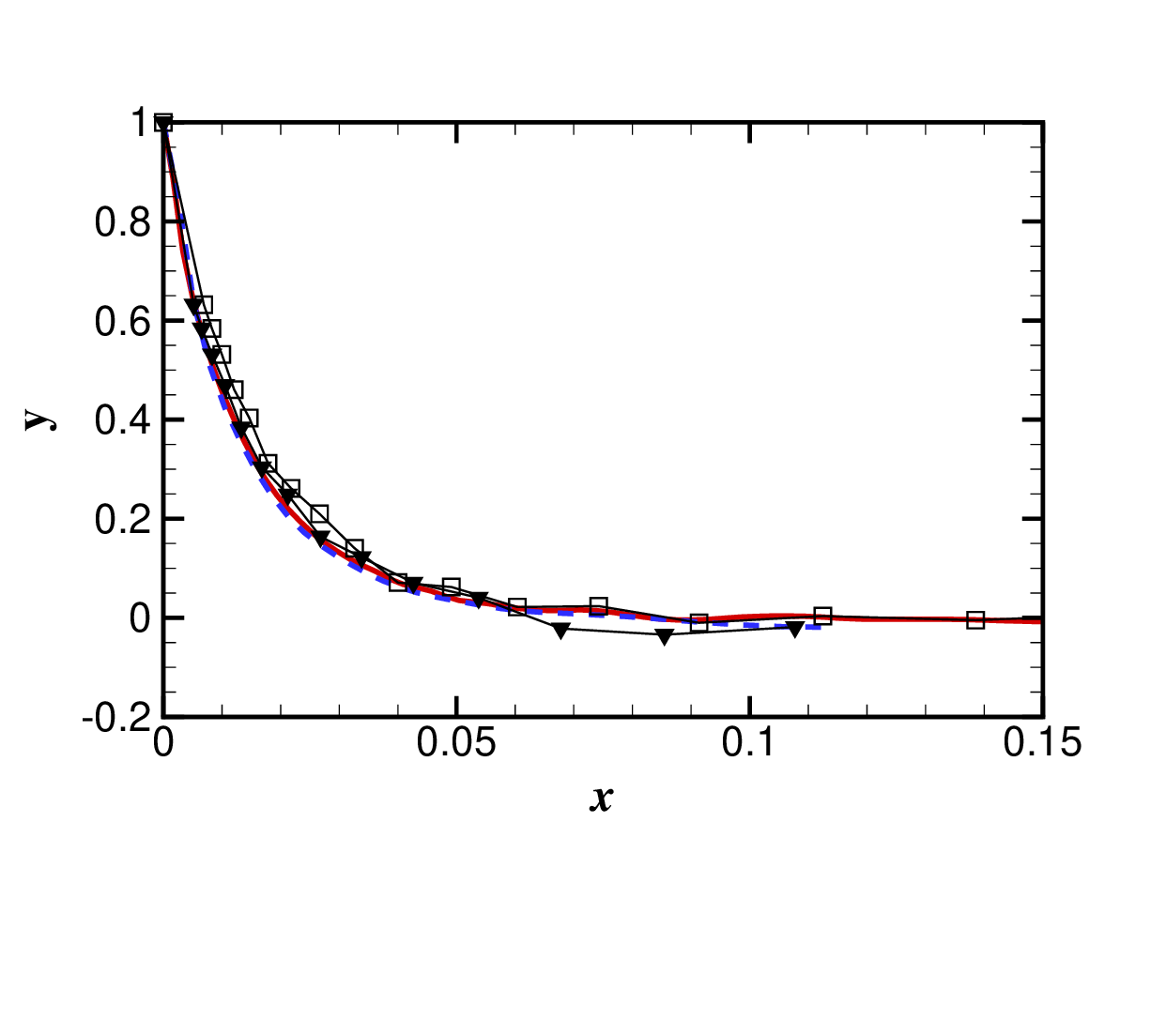}}}
\sidesubfloat[]{
{\psfrag{x}[][]{{$(r\Delta\theta)/D$}}
\psfrag{y}[][]{{$C_{uu}$}}\includegraphics[width=0.46\textwidth,trim={1.6cm 3.8cm 1.2cm 1.8cm},clip]{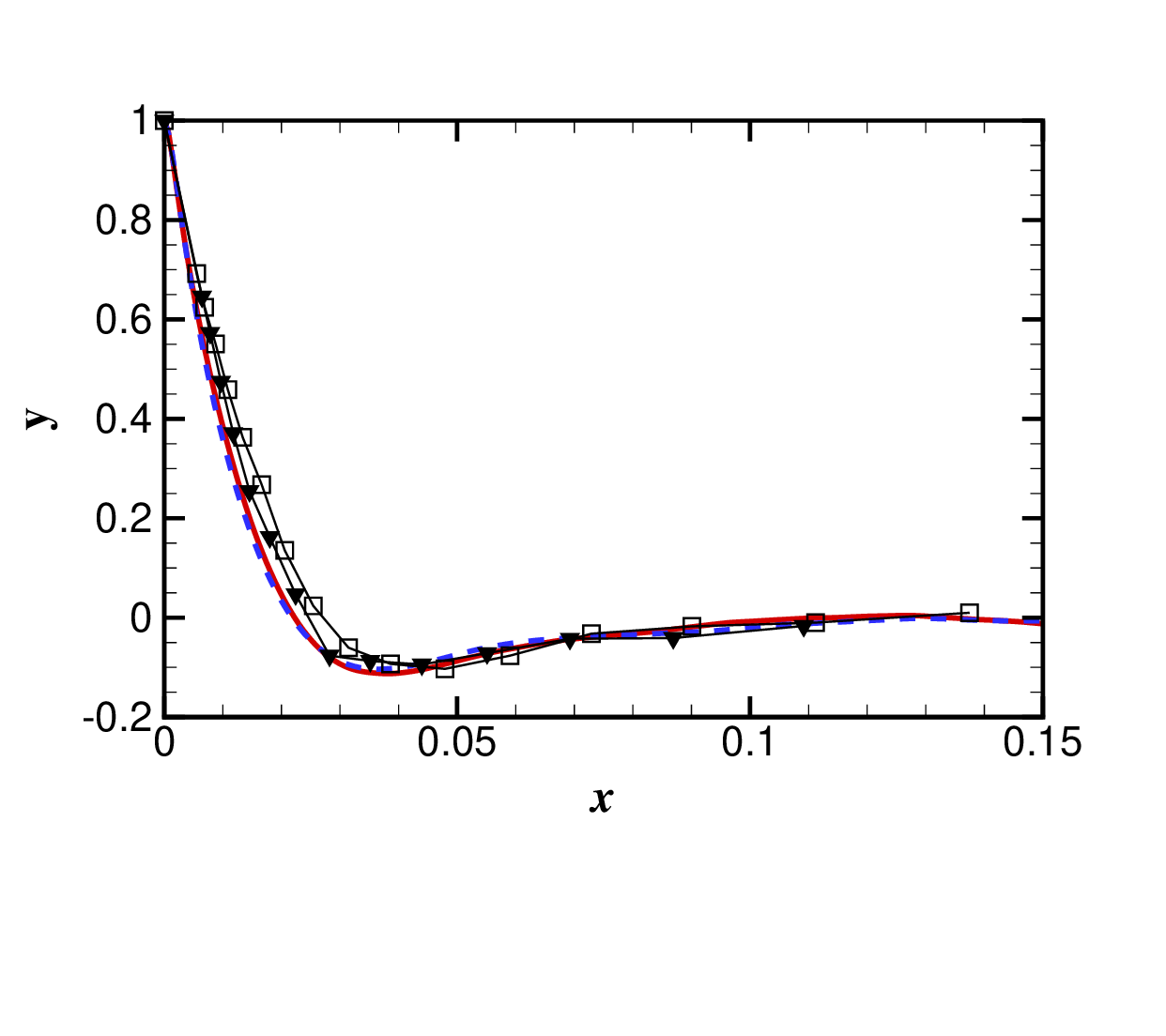}}}

\caption{Two-point correlation coefficients of fluctuating streamwise velocity anchored at two radial positions at the tail-cone end ($x_1/D = 2.17$): (a) as a function of radial separation $\Delta r/D$; (b) as a function of azimuthal separation $(r\Delta\theta)/D$. {\redsolid}, LES for $r/D=0.14$; {\bluedashed}, LES for $r/D=0.21$; $\square$, experiment for $r/D=0.14$ \citep{balantrapu2021structure}; $\blacktriangledown$ experiment for $r/D=0.21$ \citep{balantrapu2021structure}.}
\label{u_twocor}
\end{figure}

The frequency spectra of axial and azimuthal velocity fluctuations at two radial locations at the tail-cone end are shown in figure~\ref{uspectra_cone}. The simulation results exhibit good agreement with the experimental data of Balantrapu \textit{et al.} (2021; private communication 2019) except at high frequencies, where the spectral content is limited by grid resolution and the fine-mesh simulation produces a better comparison. In figure~\ref{u_twocor}, the two-point correlation coefficients of the streamwise velocity fluctuations $u'$ anchored at two radial positions at the end of the tail cone are compared with the single hot-wire data of \citet{balantrapu2021structure}. The two-point correlation coefficients are calculated from 
\begin{equation}
    C_{uu}(x_1, r,\Delta r, \Delta \theta) = \frac{\overbar{\langle u'(x_1,r,\theta,t)u'(x_1,r+\Delta r, \theta+\Delta \theta,t)\rangle}}{\sqrt{\overbar{\langle u'^2(x_1,r,\theta,t)\rangle}}\sqrt{\overbar{\langle u'^2(x_1,r+\Delta r, \theta+\Delta \theta,t)\rangle}}},
\end{equation}
where the angle brackets denote spatial averaging over the homogeneous azimuthal direction, and the overbar denotes temporal averaging. The computed results, obtained from the fine-mesh simulation, agree well with the experimental data \citep{balantrapu2021structure}. They illustrate that variations in correlation length scales along the radial direction is small. More comprehensive discussions of the spatial correlations and turbulence structures are presented in \cref{sec_corr}. 

In summary, the simulation results presented hitherto demonstrate good agreement with experimental data in terms of the first and second order statistics as well as the space-time characteristics of the tail-cone TBL. Furthermore, they demonstrate grid convergence for all the quantities examined. The comprehensive validation establishes the validity and accuracy of the simulation data, allowing a detailed analysis of the statistics and structure of the axisymmetric tail-cone TBL and their evolution under the APG. 

\section{Analysis of the tail-cone turbulent boundary layer} \label{sec:analysis}

Following the convention, results are presented in the boundary-layer coordinate system $x$-$y$ shown in figure~\ref{mean}, where $x$ is along the cone surface in the flow direction, $y$ is normal to the surface, and the origin is located at the start of the tail cone. This eliminates any ambiguity that may arise from use of the axial-radial coordinates as in \citet{balantrapu2021structure}. The $x$-component of velocity is denoted by $u$ and the $y$-component is $v$. They are related to the axial velocity $u_1$ and radial velocity $u_r$ through 
\begin{equation}
    u=u_{1} \cos{\alpha} - u_r \sin{\alpha},\quad v=u_1 \sin{\alpha} + u_r \cos{\alpha}
\end{equation}
where $\alpha=20^{\circ}$ is the half apex angle of the tail cone. The third velocity component, $u_{\theta}$, is the same as in the original coordinate systems. All the analyses henceforth are based on data form the fine-mesh simulation.

\subsection{Boundary-layer properties and velocity statistics}

\begin{figure}
\centering
\sidesubfloat[]{
{\includegraphics[width=0.7\textwidth,trim={0.4cm 0.0cm 0.0cm 0.0cm},clip]{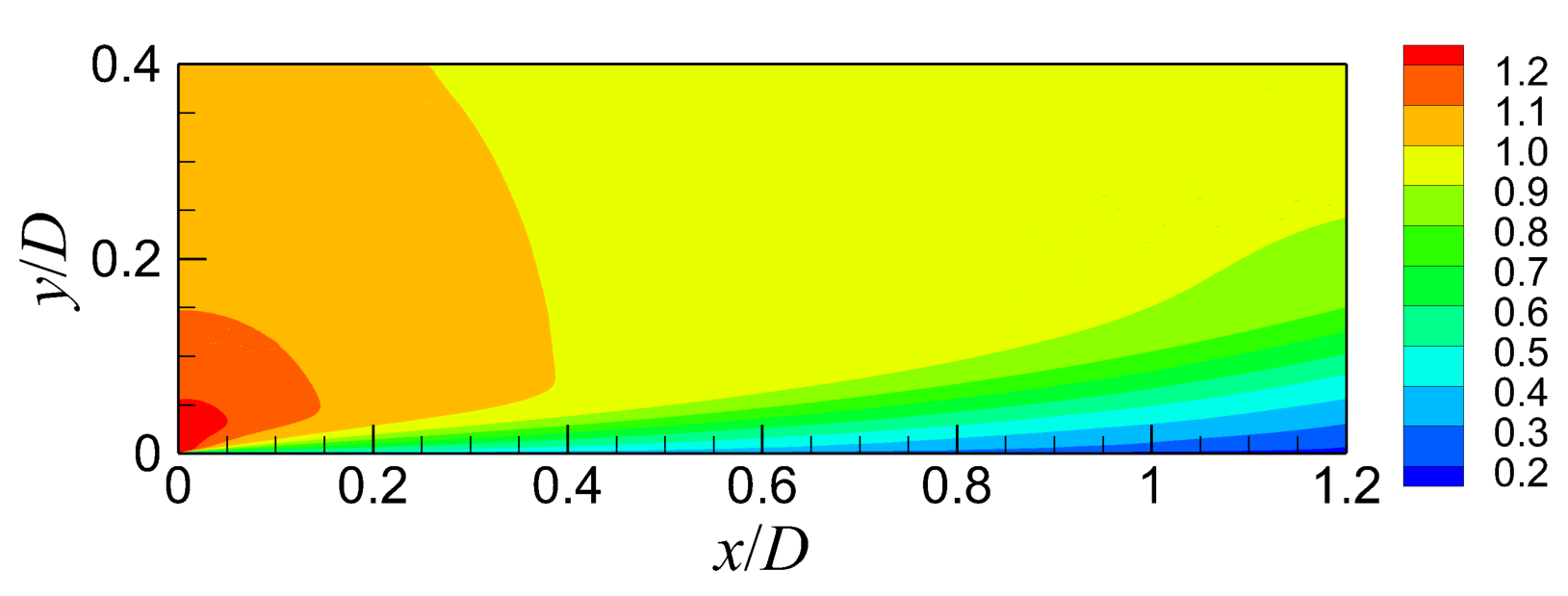}\label{cone_mean_up}}}

\sidesubfloat[]{
{\includegraphics[width=0.7\textwidth,trim={0.37cm 0.2cm 0.7cm 0.0cm},clip]{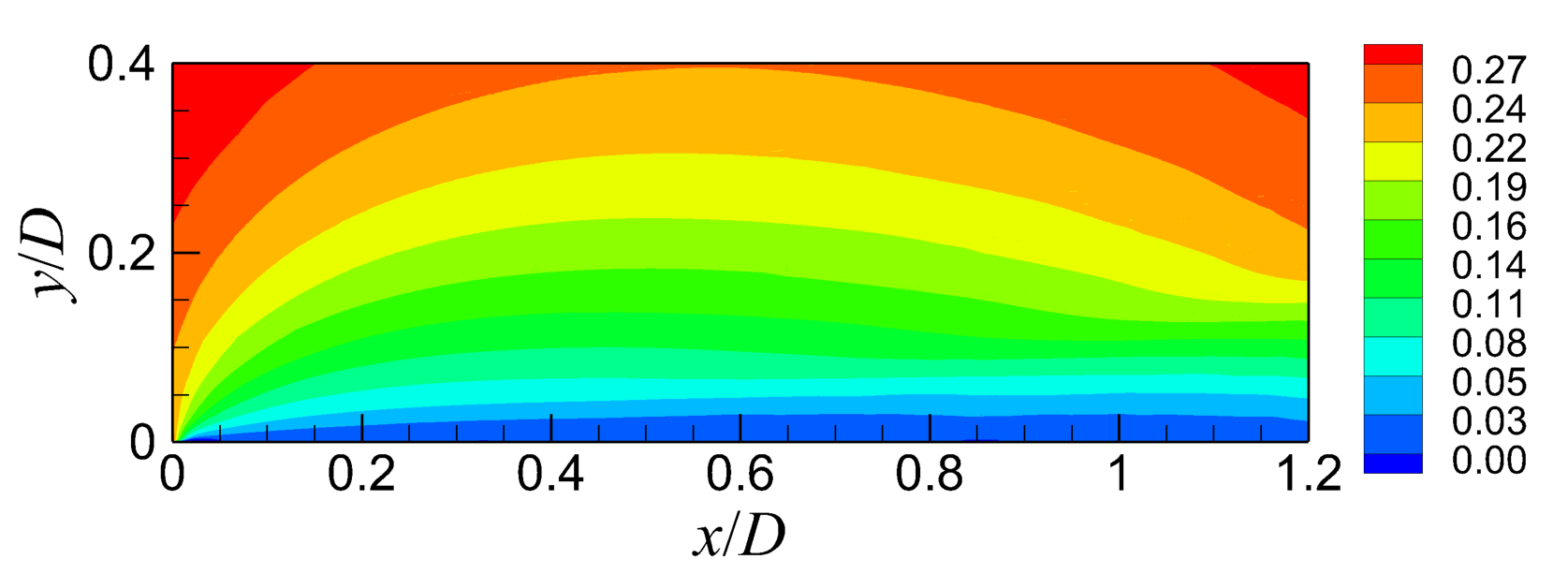}\label{cone_mean_un}}}
\caption{Mean velocity contours in the tail-cone TBL: (a) streamwise velocity $U/U_\infty$; (b) wall-normal velocity $V/U_\infty$.}
\label{cone_mean}
\end{figure}

Figure~\ref{cone_mean} shows isocontours of the streamwise and wall-normal velocity components in the tail-cone TBL. The flow decelerates along the cone as the boundary-layer thickness grows rapidly. Compared to ZPG flat-plate boundary layers, the wall-normal velocity is larger and varies more significantly within the boundary layer, particularly near the beginning and end of the tail cone due to flow turning following the BOR surface. The subsequent flow analyses is focusing on the region between $0.2\leq x/D \leq 1$, which excludes the flow deflection regions near both ends of the tail cone.

\begin{figure}
\centering

\sidesubfloat[b]{
{\psfrag{x}[][]{{$x/D$}}
\psfrag{y}[][]{{$\delta_2/D, \delta_1/D, \delta/D$}}\includegraphics[width=0.3\textwidth,trim={1.3cm 5.3cm 7.3cm 1.6cm},clip]{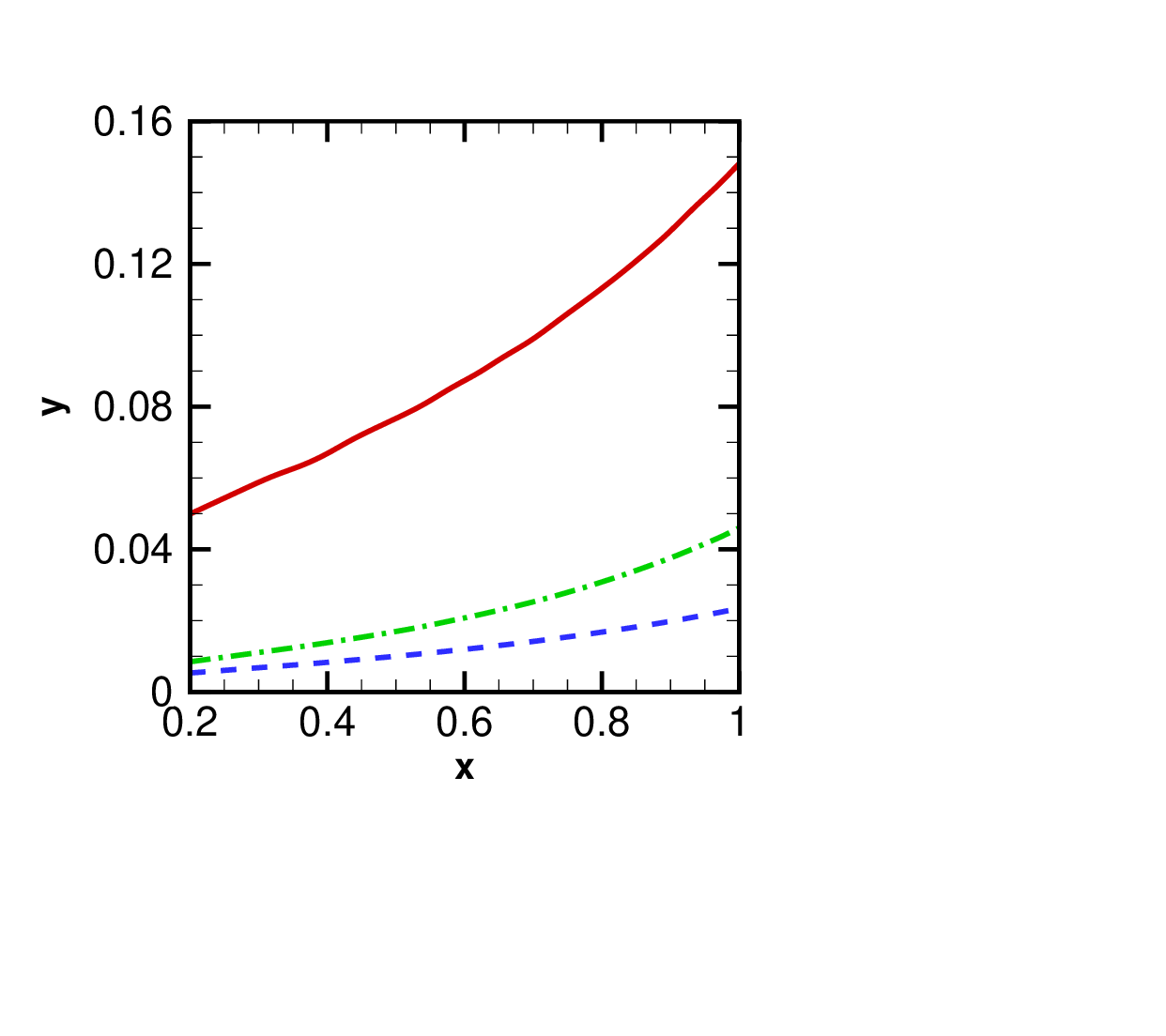}\label{character_cone_a}}}
\sidesubfloat[b]{
{\psfrag{x}[][]{{$x/D$}}
\psfrag{y}[][]{{$Re_{\tau}, Re_{\delta_2}$}}\includegraphics[width=0.30\textwidth,trim={2.2cm 3.5cm 6.7cm 2.0cm},clip]{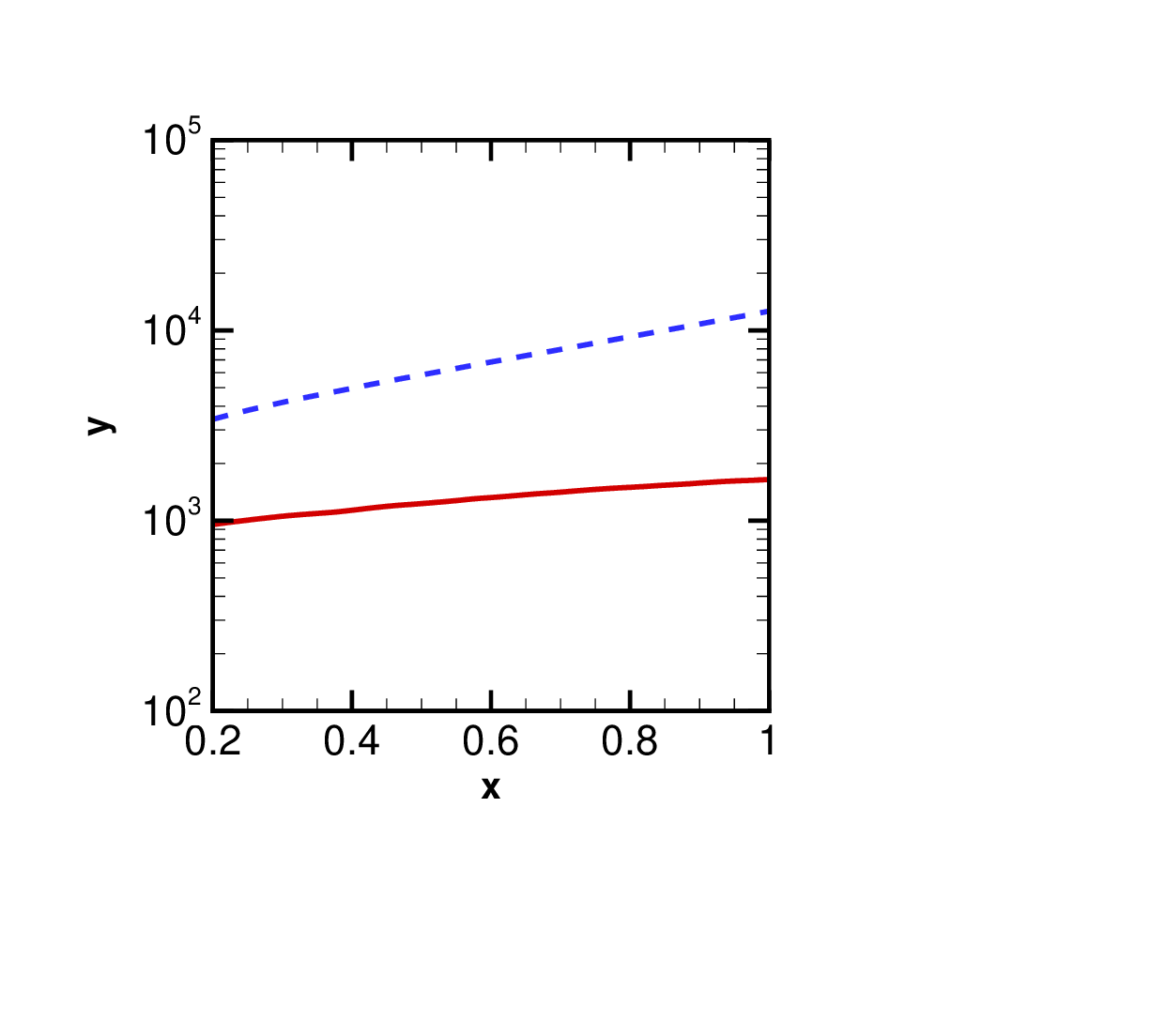}\label{character_cone_b}}}
\sidesubfloat[b]{
{\psfrag{x}[][]{{$x/D$}}
\psfrag{y}[][]{{$H$}}\includegraphics[width=0.30\textwidth,trim={1.3cm 5.3cm 7.7cm 1.8cm},clip]{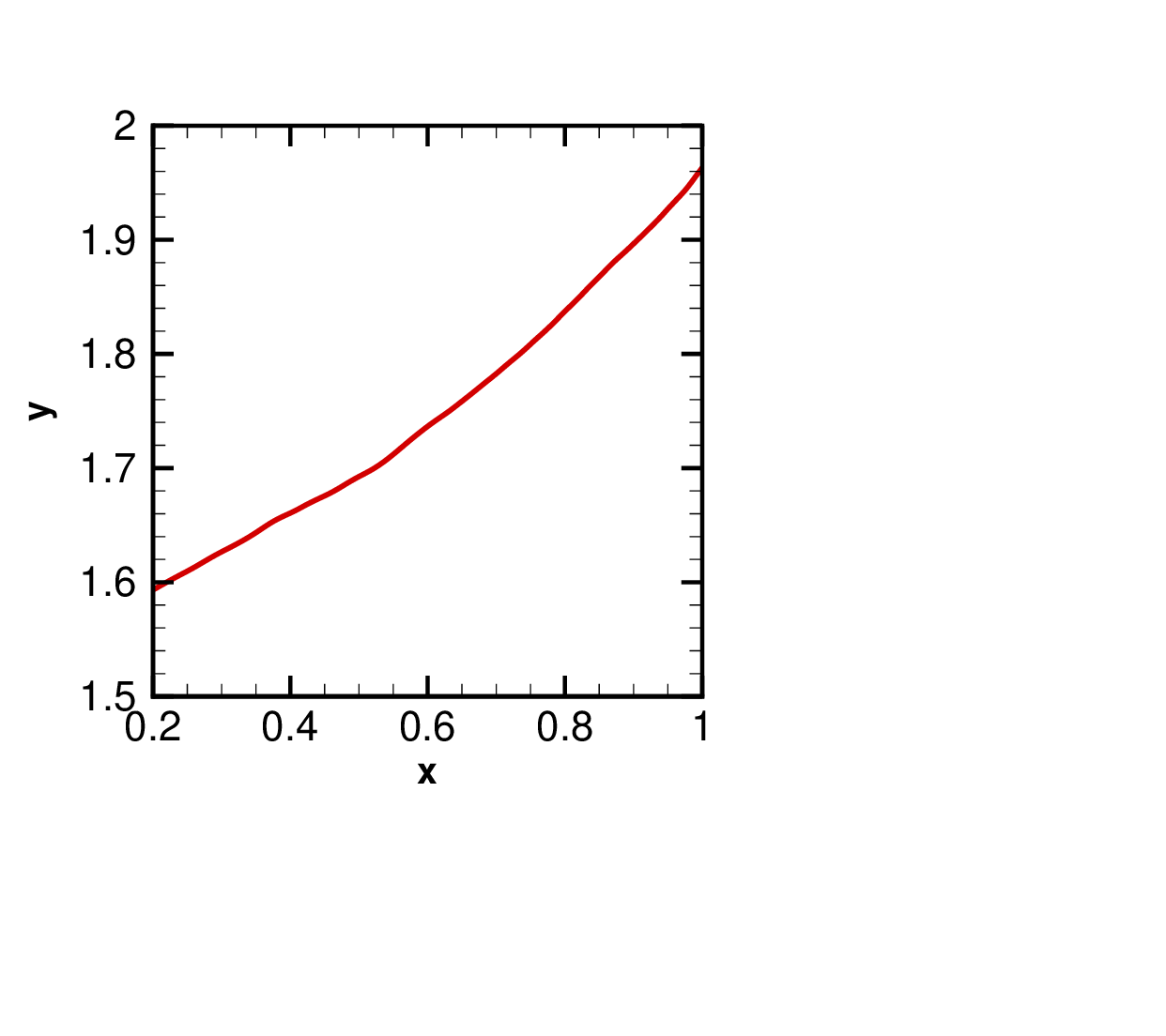}\label{character_cone_c}}}

\caption{Distributions of boundary-layer parameters: (a) boundary-layer thickness $\delta$ ({\redsolid}), displacement thickness $\delta_1$ ({\greendashdotted}) and momentum thickness $\delta_2$ ({\bluedashed}); (b) Reynolds numbers $Re_{\tau}$ ({\redsolid}) and $Re_{\delta_2}$ ({\bluedashed}); (c) shape factor $H$.}
\label{character_cone}
\end{figure}

The streamwise variations of the boundary-layer thickness $\delta$, displacement thickness $\delta_1$ and momentum thickness $\delta_2$ are depicted in figure~\ref{character_cone_a}. It can be noticed that their growth is not linear with the streamwise distance. At $x/D=1$, the values of $\delta$, $\delta_1$ and $\delta_2$ are approximately 4 times of those at $x/D=0.2$. The distributions of the friction Reynolds number $Re_{\tau}$ and the momentum thickness Reynolds number $Re_{\delta_2}$ are shown in figure~\ref{character_cone_b}, and the boundary-layer shape factor $H=\delta_1/\delta_2$ is displayed in figure~\ref{character_cone_c}. The value of $H$ is larger than that in a ZPG flat-plate TBL, and it grows rapidly along the streamwise direction as a result of increasing APG \citep{harun2011boudnary}.  

\begin{figure}
\centering

\sidesubfloat[b]{
{\psfrag{x}[][]{{$(P-P_w)/(\rho U_\infty^2)$}}
\psfrag{y}[][]{{$y/\delta$}}\includegraphics[width=0.46\textwidth,trim={1.1cm 5.1cm 2cm 2.1cm},clip]{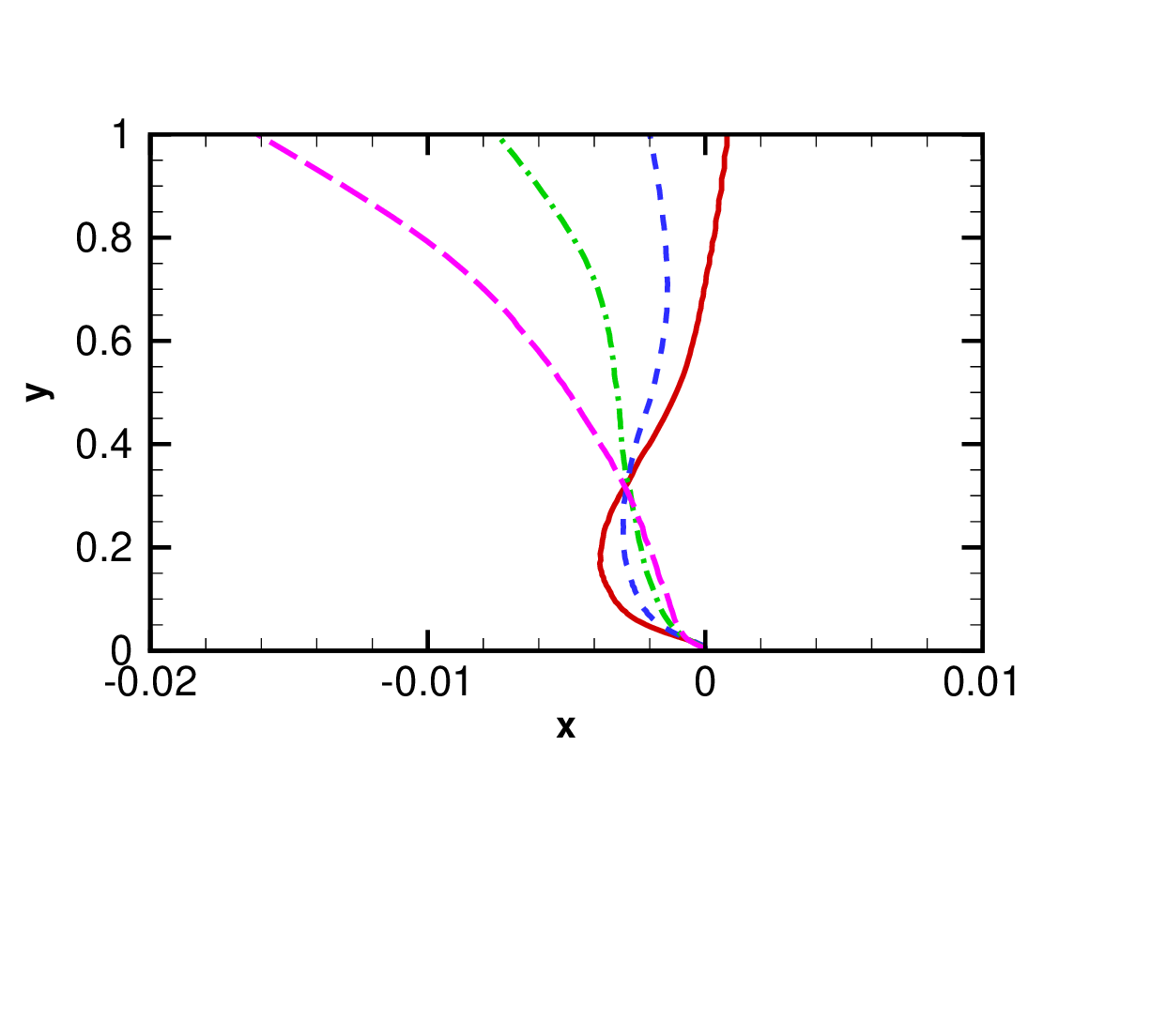}\label{p_cone_a}}}
\sidesubfloat[b]{
{\psfrag{x}[][]{{$x/D$}}
\psfrag{y}[][]{{$\beta$}}\includegraphics[width=0.46\textwidth,trim={1.6cm 5.1cm 1.2cm 1.9cm},clip]{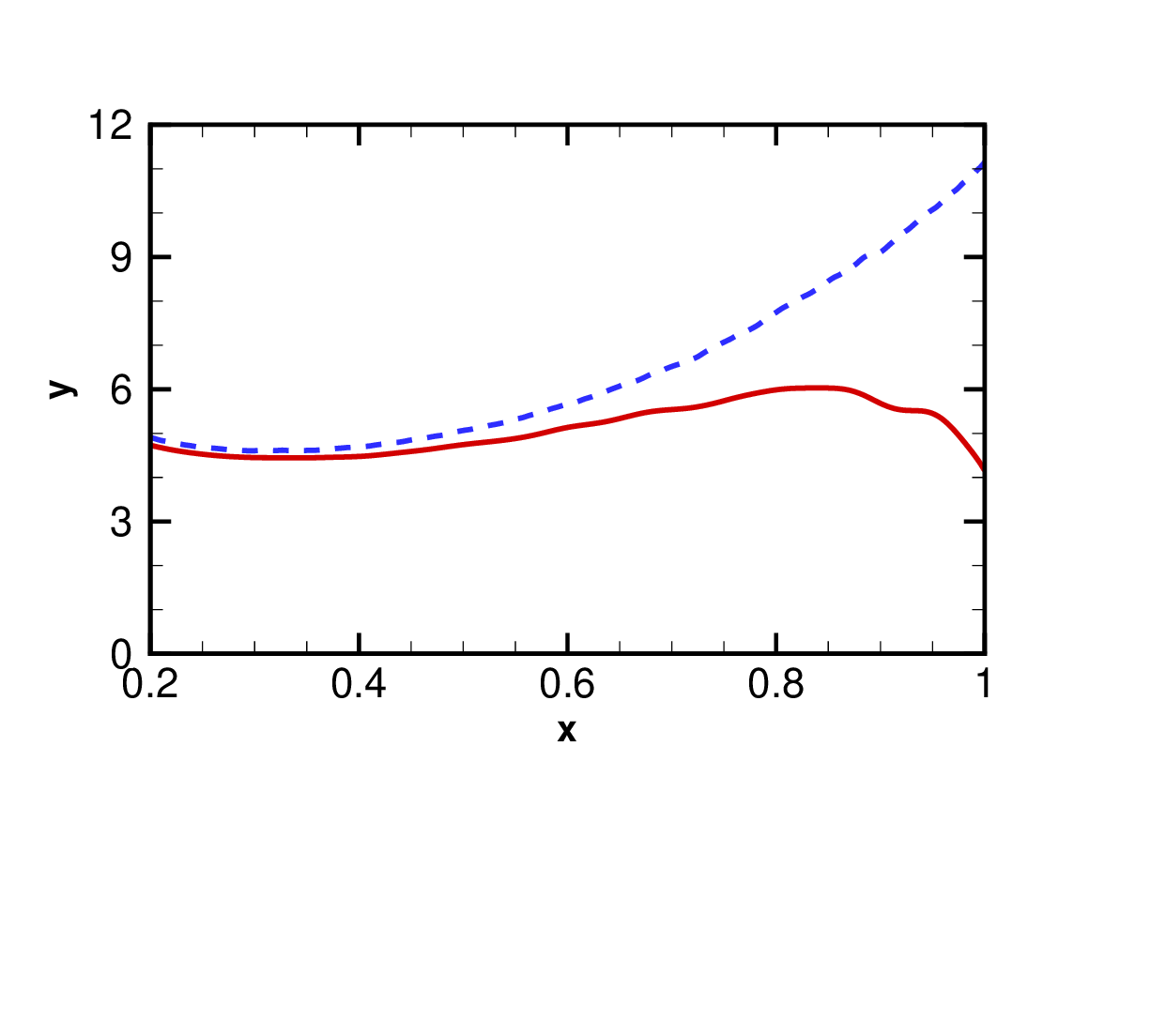}\label{p_cone_b}}}

\caption{(a) Profiles of mean pressure relative to the local wall pressure across the boundary layer at four streamwise locations: {\redsolid}, $x/D=0.21$; {\bluedashed}, $x/D=0.43$; {\greendashdotted}, $x/D=0.71$; {\magentadashed}, $x/D=0.99$. (b) Streamwise variations of the Clauser pressure gradient parameter $\beta$: {\redsolid}, based on the mean pressure at the edge of the boundary layer; {\bluedashed}, based on the mean wall pressure.}
\label{p_cone}
\end{figure}

Figure~\ref{p_cone_a} shows profiles of the mean static pressure relative to the wall pressure, $(P-P_w)/(\rho U_\infty^2)$, at four streamwise locations. The pressure variations across the TBL are significant, and the pressure at the boundary-layer edge decreases relative to the local wall pressure in the downstream direction, indicating that the edge pressure rises more slowly than the wall pressure under the APG. The pressure gradient in a boundary layer is commonly characterized by the Clauser pressure-gradient parameter, $\beta=(\dif{P_e}/\dif{x})\delta_1/\tau_{w,x}$, where $P_e$ is the mean pressure at the edge of the boundary layer and $\tau_{w,x}$ is the mean wall-shear stress along the surface. For a thin boundary layer, the Clauser parameter can be equivalently evaluated based on the mean wall pressure $P_w$ since the pressure change across the boundary layer is small. However, this is not the case for the rapidly thickening boundary layer considered here. The streamwise evolutions of $\beta$ calculated based on $P_e$ and $P_w$ are shown in figure~\ref{p_cone_b}. The values of $\beta$ are high overall, which verifies that the flow is under strong APG. The two definitions give similar $\beta$ values in the upstream portion of the boundary layer where it is relatively thin, but their difference increases towards downstream as the boundary layer thickens. The $\beta$ based on $P_e$ increases from 4.7 at $x/D=0.2$ to a peak value of 6.1 at $x/D=0.85$, and then declines to 4.2 at $x/D=1$. In contrast, the $\beta$ based on $P_w$ increases nearly monotonically from 4.9 to 11.1. The smaller pressure gradient outside the thick TBL is consistent with the earlier findings of \citet{patel1974measurements}.

\begin{figure}
\centering

\sidesubfloat[]{
{\psfrag{x}[][]{{$x/D$}}
\psfrag{y}[][]{{$\delta /r_s$}}\includegraphics[width=0.46\textwidth,trim={2.5cm 5cm 0.0cm 2cm},clip]{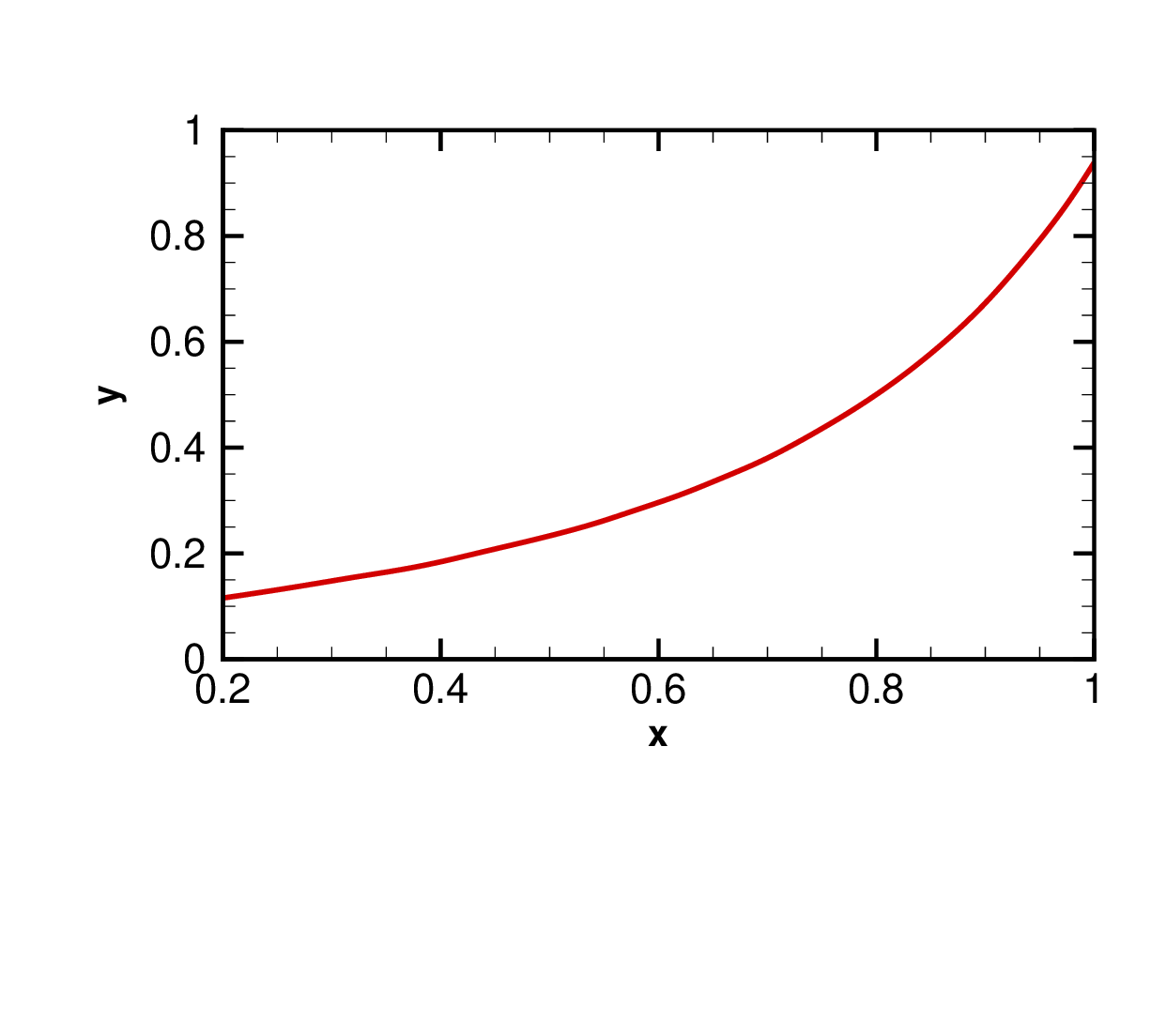}\label{curvature_cone_a}}}
\sidesubfloat[]{
{\psfrag{x}[][]{{$x/D$}}
\psfrag{y}[][]{{$r_s^+$}}\includegraphics[width=0.47\textwidth,trim={1.7cm 6cm 0.7cm 2cm},clip]{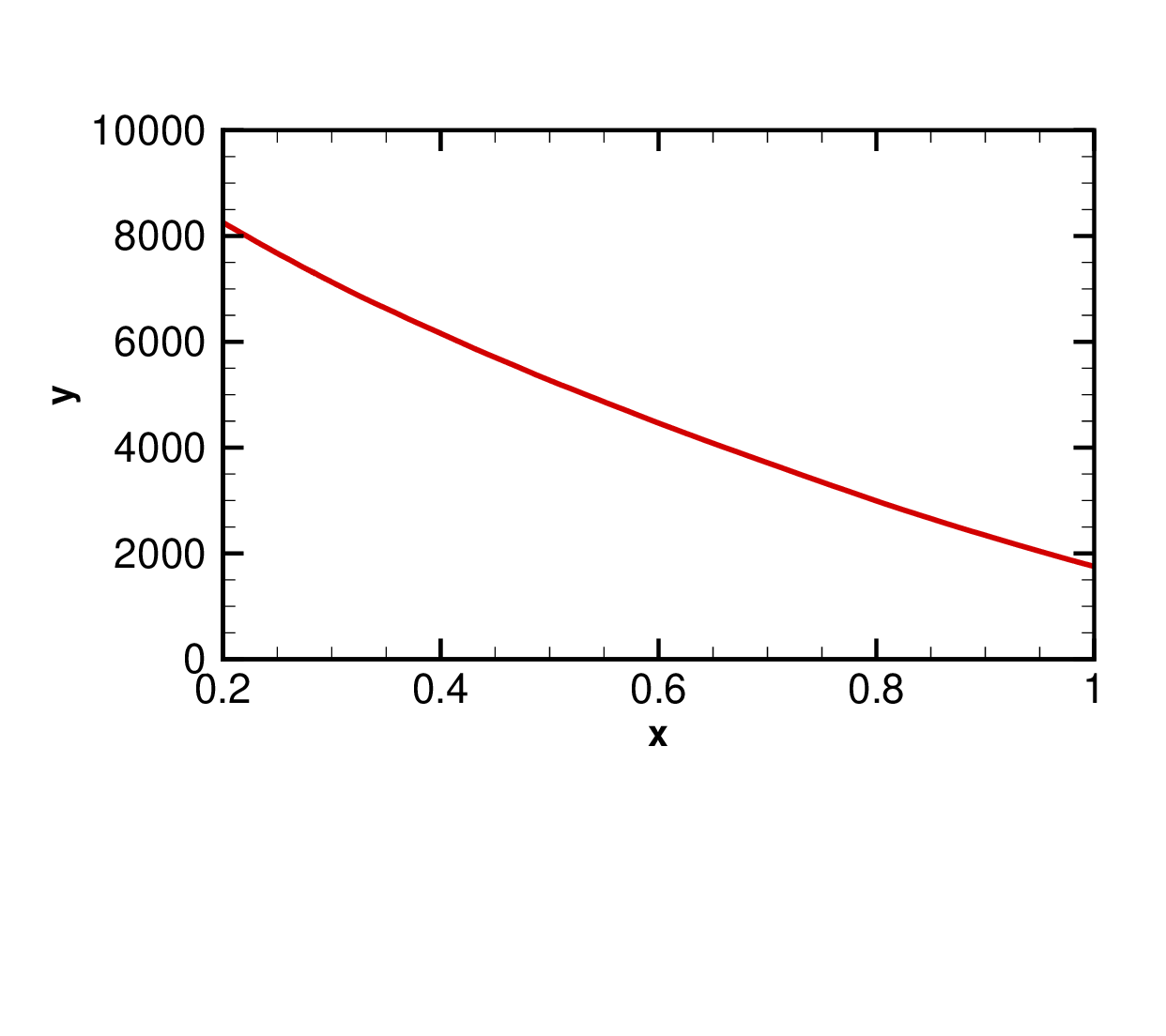}\label{curvature_cone_b}}}

\caption{Distributions of transverse curvature parameters along the tail cone: (a) $\delta/r_s$; (b) $r_s^+$.}
\label{curvature_cone}
\end{figure}

The decreasing radius of the tail cone in the downstream direction can also potentially influence the evolution of the TBL, and thus the transverse curvature effect on the TBL is evaluated. As pointed out by \citet{piquet1999transverse}, the strength of the transverse curvature effect can be estimated based on two parameters, the radius of the surface curvature in inner scale, $r_s^+=u_\tau r_s/\nu$, and the ratio of the boundary-layer thickness to the surface radius, $\delta/r_s$. The distributions of $r_s^+$ and $\delta/r_s$ along the tail-cone are shown in figure~\ref{curvature_cone}. It can be noticed that $\delta/r_s$ increases quickly in the second half of the region, but its value remains less than 1 within the region of analysis. Conversely, the value of $r_s^+$ declines continuously and its smallest value is approximately 1700. According to the review by \citet{piquet1999transverse}, in flows characterized by large $r_s^+$ and small (or order one) $\delta/r_s$ values, the impact of transverse curvature on the boundary layer is small and primarily felt in the outer layer. It can therefore be reasonably assumed that the effect of transverse curvature is small in the present TBL relative to the influence of the APG.

\begin{figure}
\centering

\sidesubfloat[]{
{\psfrag{x}[][]{{$U/U_{e}$}}
\psfrag{y}[][]{{$y/\delta$}}\includegraphics[width=0.46\textwidth,trim={1.2cm 4.6cm 1.8cm 1.8cm},clip]{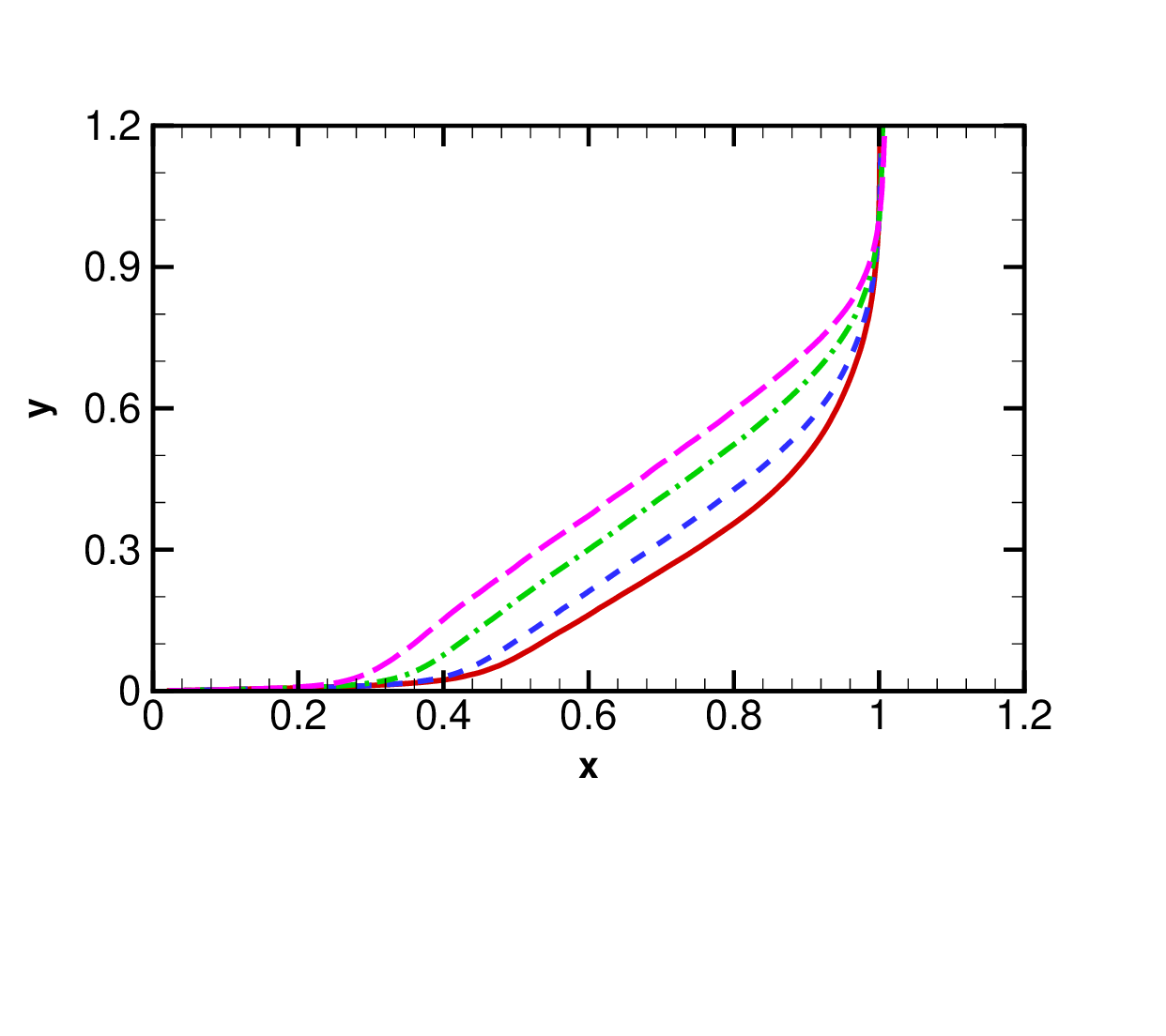}\label{mean_u_profile_a}}}
\sidesubfloat[]{
{\psfrag{x}[][]{{$y^+$}}
\psfrag{y}[][]{{$U^+$}}\includegraphics[width=0.46\textwidth,trim={1.5cm 4.6cm 1.6cm 1.8cm},clip]{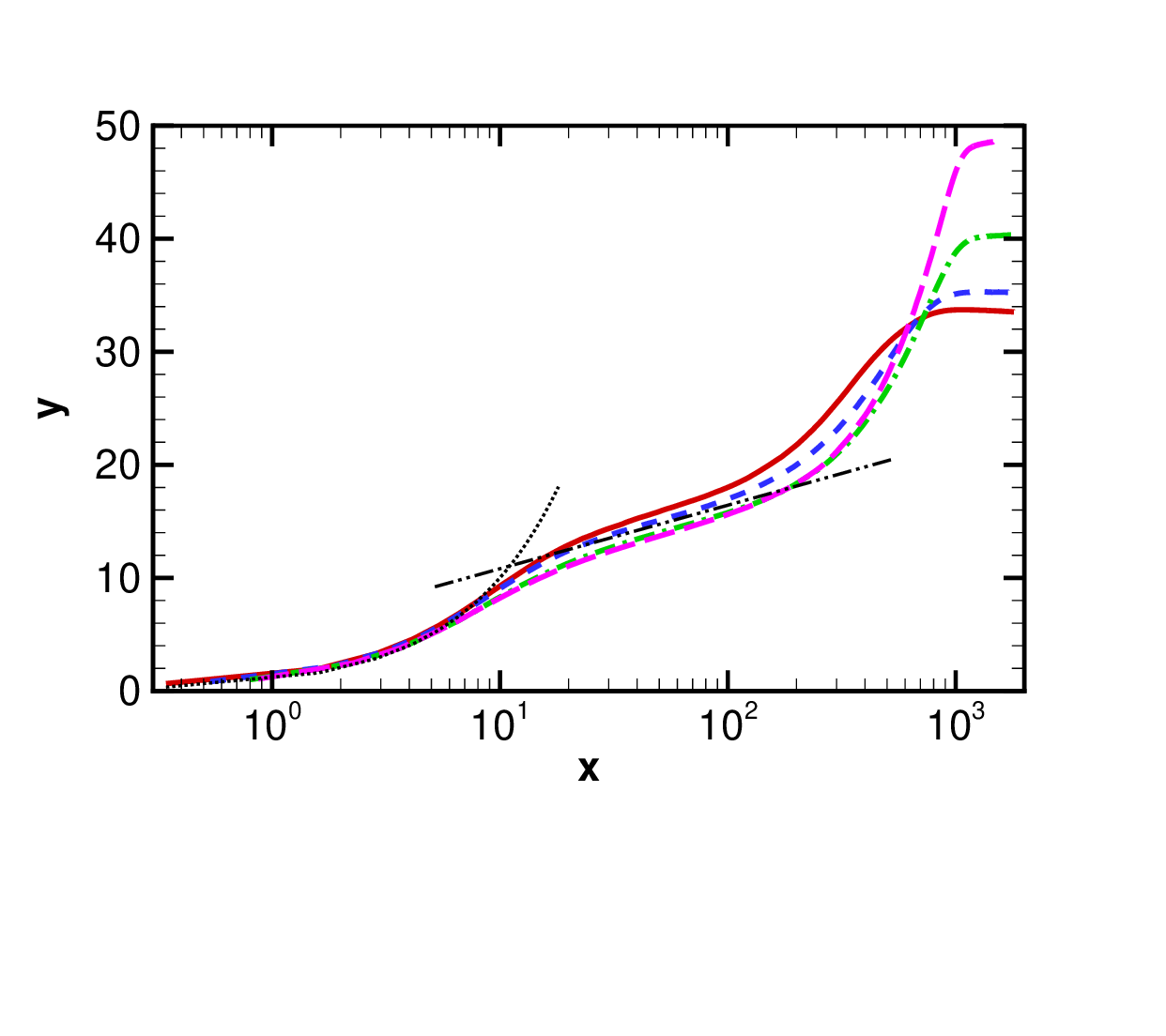}\label{mean_u_profile_b}}}

\caption{Mean streamwise velocity profiles with (a) outer scaling and (b) inner scaling at four streamwise locations: {\redsolid}, $x/D=0.21$; {\bluedashed}, $x/D=0.43$; {\greendashdotted}, $x/D=0.71$; {\magentadashed}, $x/D=0.99$. In (b): {\blackdot}, $U^+=y^+$; {\blackdashdotdot}, $U^+=(1/\kappa)\ln{(y^+)}+B$ with $\kappa=0.41$ and $B=5.2$.}
\label{mean_u_profile}
\end{figure}

The outer- and inner-scaled mean streamwise velocity profiles at four streamwise stations are shown in figures~\ref{mean_u_profile_a} and \ref{mean_u_profile_b}, respectively. In figure~\ref{mean_u_profile_a}, the velocity is scaled by the local mean streamwise velocity $U_e$ at the edge of the boundary layer, and the $y$-coordinate is scaled by the local boundary-layer thickness $\delta$. The profiles become less full in the downstream direction, which is consistent with the monotonic increase of shape factor in figure~\ref{character_cone_c}. Despite the presence of strong APG, the boundary layer remains attached at all four locations and through the end of the tail cone. In Figure~\ref{mean_u_profile_b}, the velocity is scaled by the local friction velocity $u_\tau$, and the $y$-coordinate is scaled using the viscous length scale $\nu/u_\tau$. The inner-scaled profiles collapse within the viscous sublayer onto the linear law. Moreover, each profile features a noticeable logarithmic region extending from $y^+\approx30$ to $130$, which is narrower than that of a canonical ZPG planar TBL. In comparison with the classical log law, $U^+=(1/\kappa)\ln{(y^+)}+B$ with $\kappa=0.41$ and $B=5.2$, these profiles exhibit a slightly steeper slope. At the two downstream stations, the velocity profiles fall below the standard log law due to the increasing $\beta$ value. These changes in the log region are consistent with the results of some previous studies of APG TBLs \citep{nagano1998structure, monty2011parametric}. In the wake region $U^+$ rises faster and attains higher values at downstream locations. 

\begin{figure}
\centering

\sidesubfloat[]{
{\psfrag{x}[][]{{$u'_{\text{rms}}/U_{e}$}}
\psfrag{y}[][]{{$y/\delta$}}\includegraphics[width=0.46\textwidth,trim={1.3cm 0.2cm 2.1cm 0.8cm},clip]{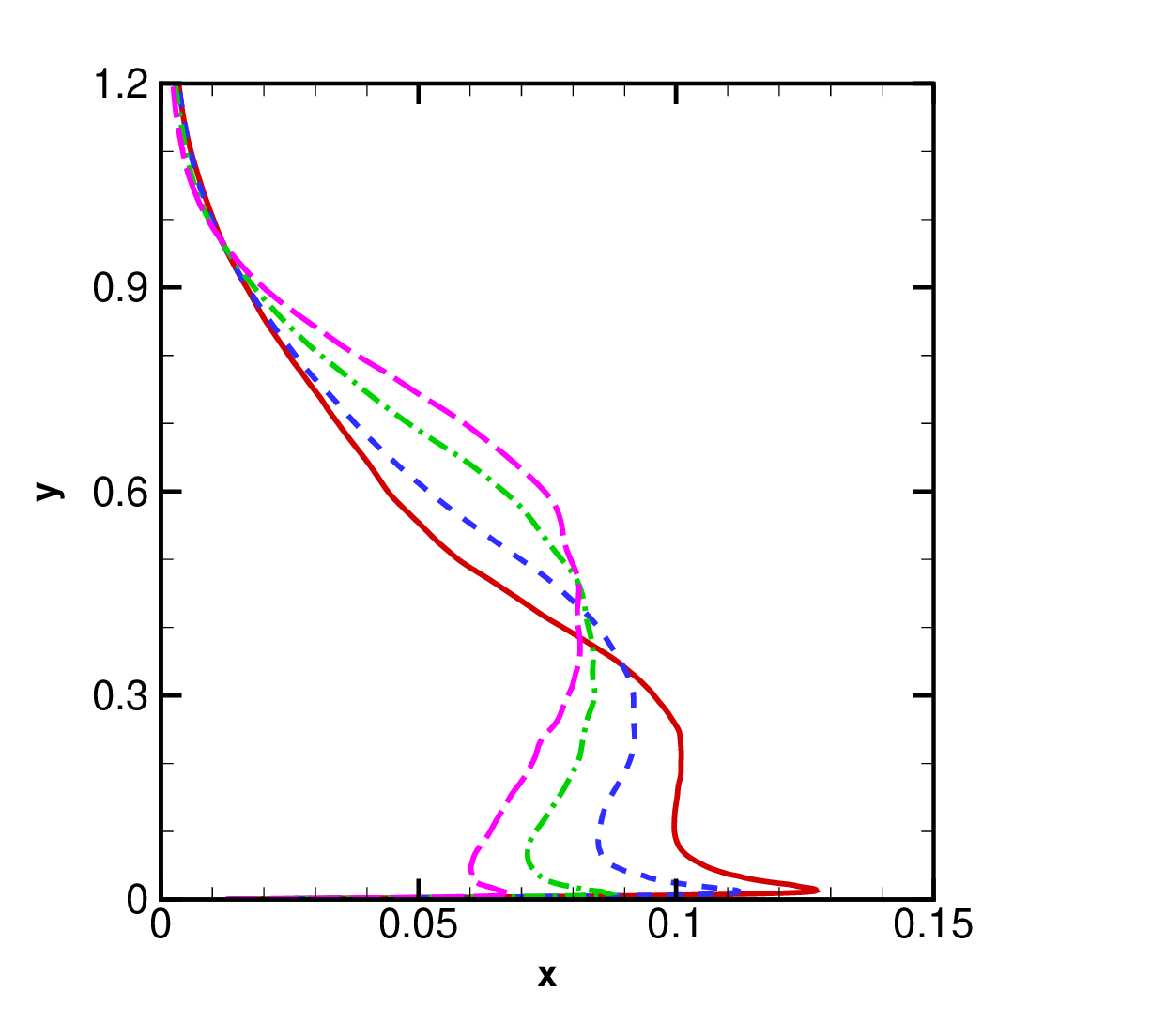}\label{stats_outer_a}}}
\sidesubfloat[]{
{\psfrag{x}[][]{{$v'_{\text{rms}}/U_{e}$}}
\psfrag{y}[][]{{$y/\delta$}}\includegraphics[width=0.46\textwidth,trim={1.2cm 0.0cm 2.2cm 0.9cm},clip]{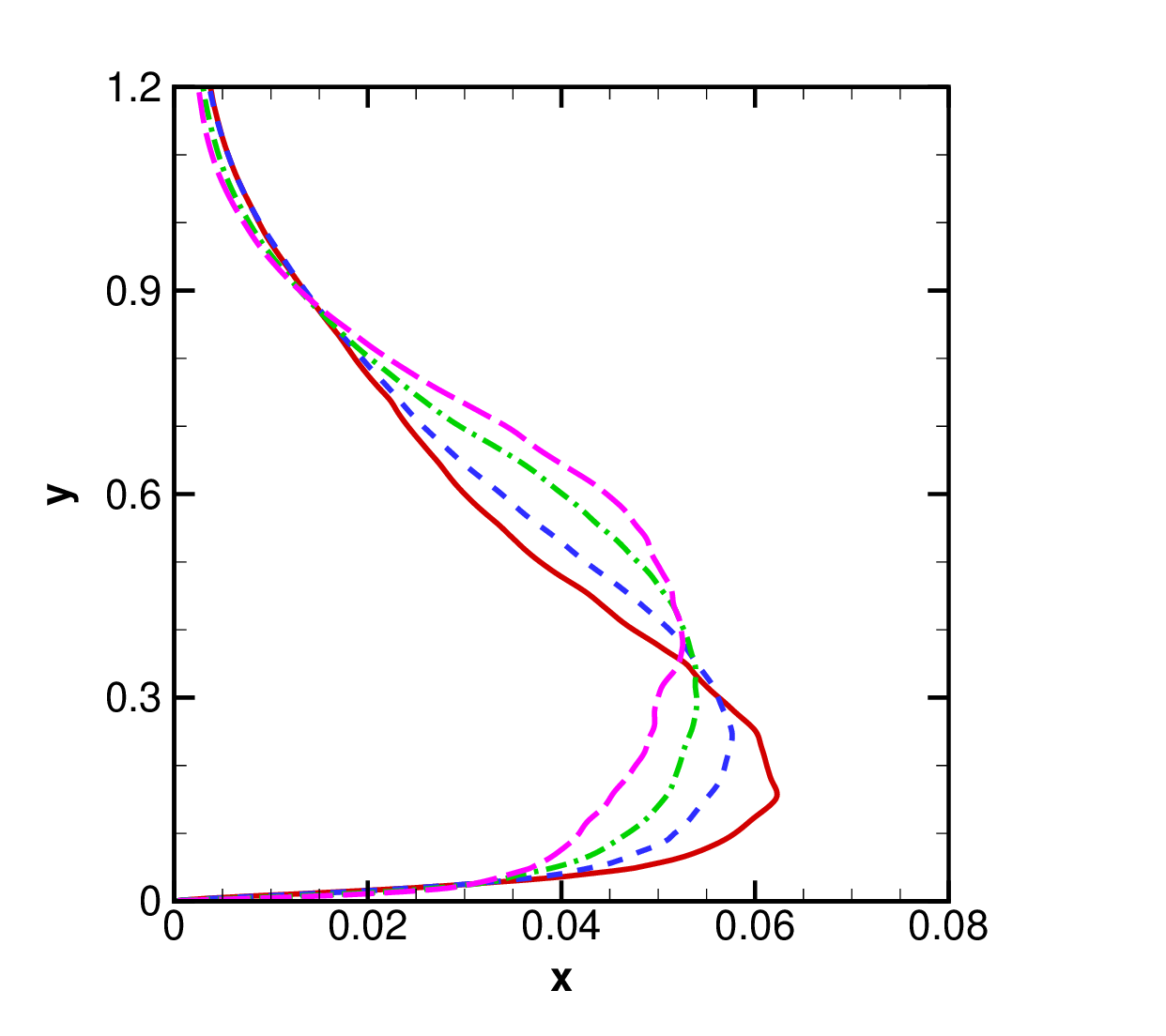}\label{stats_outer_b}}}

\sidesubfloat[]{
{\psfrag{x}[][]{{$u'_{\theta, \text{rms}}/U_{e}$}}
\psfrag{y}[][]{{$y/\delta$}}\includegraphics[width=0.46\textwidth,trim={1.8cm 0.6cm 1.7cm 0.6cm},clip]{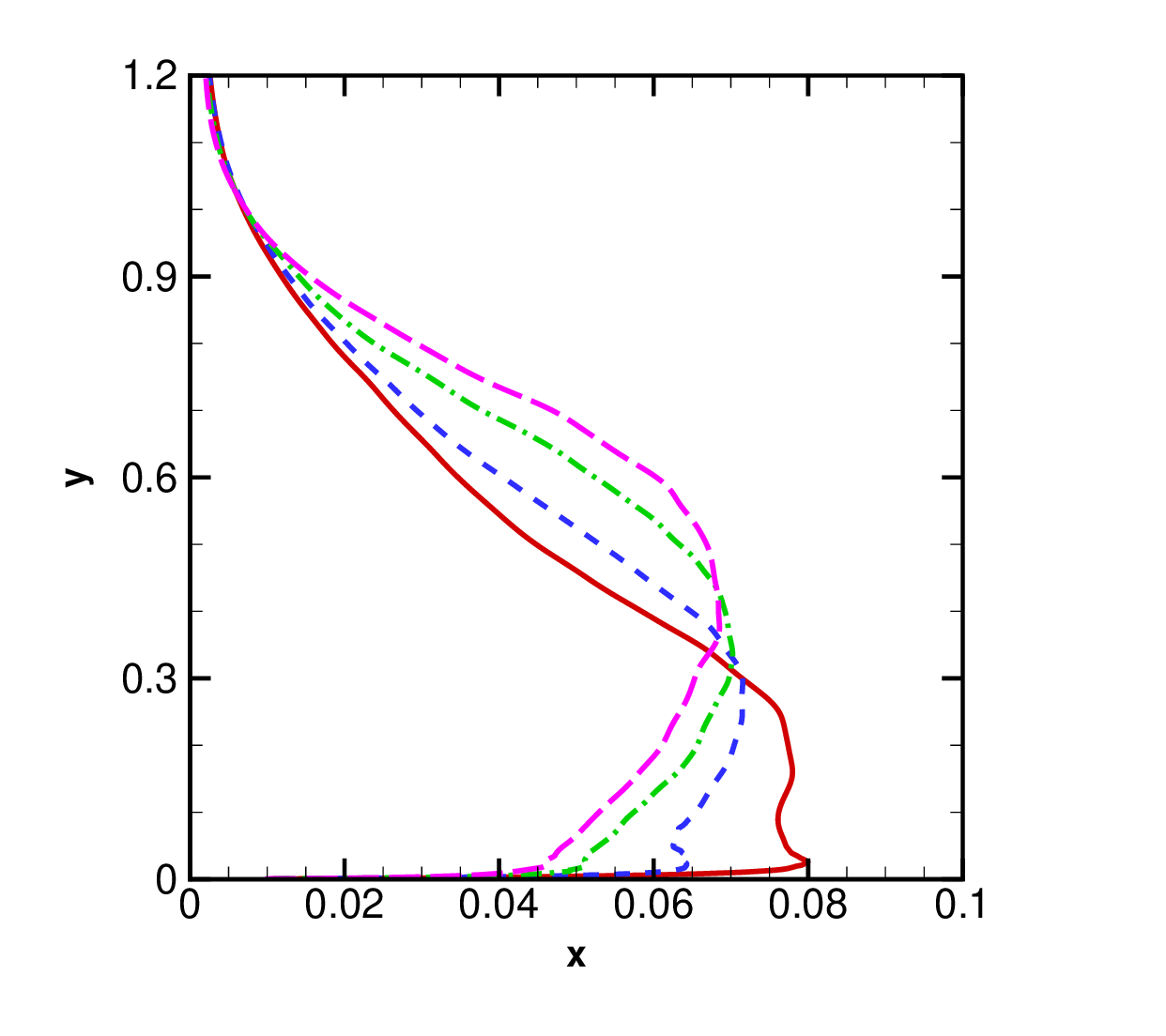}\label{stats_outer_c}}}
\sidesubfloat[]{
{\psfrag{x}[][]{{$\overline{u'v'}/U_{e}^2$}}
\psfrag{y}[][]{{$y/\delta$}}\includegraphics[width=0.46\textwidth,trim={0.9cm 0.4cm 2.5cm 0.9cm},clip]{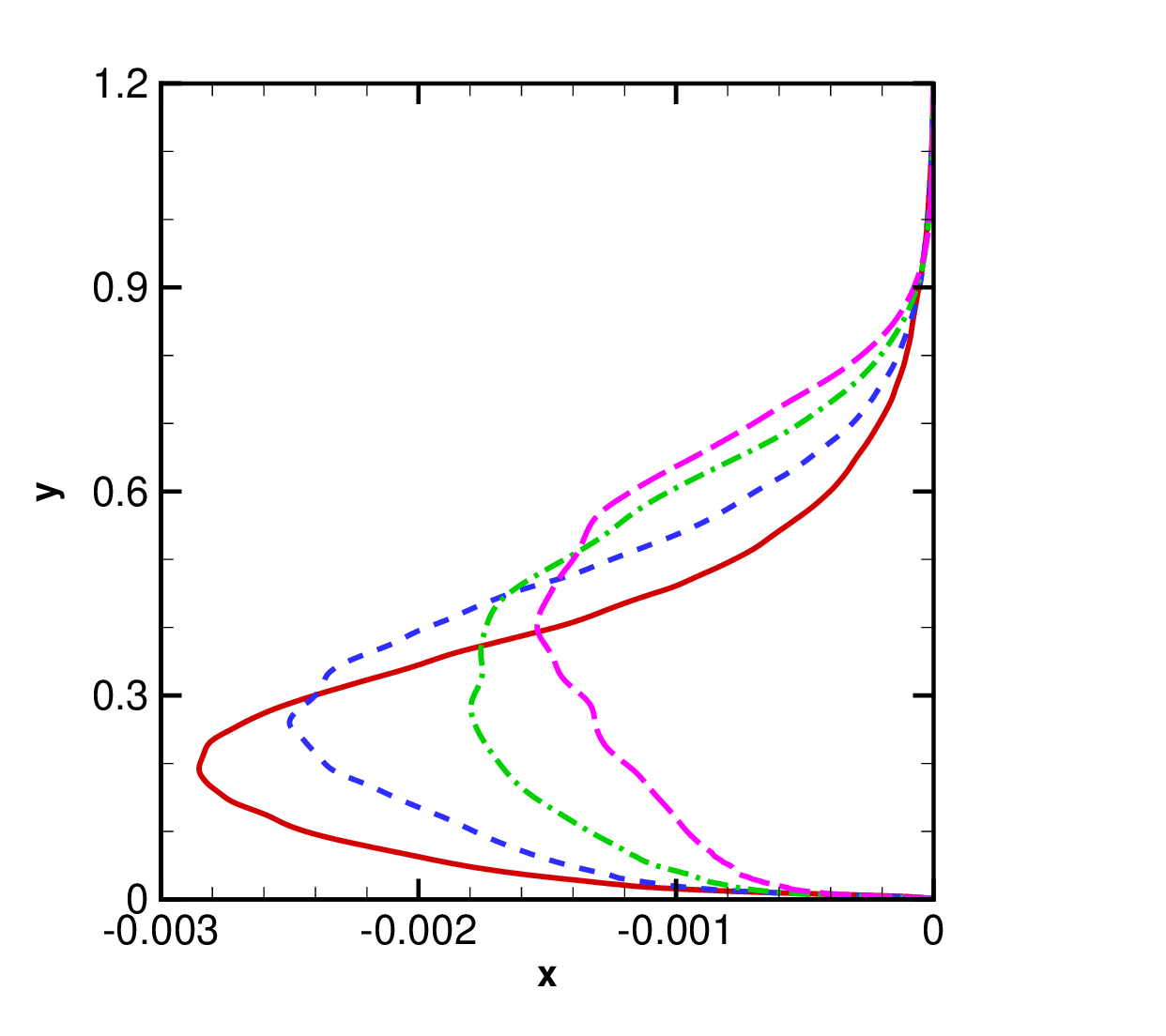}\label{stats_outer_d}}}

\caption{Profiles of (a-c) r.m.s. values of velocity fluctuations and (d) Reynolds shear stress with outer scaling at: {\redsolid}, $x/D=0.21$; {\bluedashed}, $x/D=0.43$; {\greendashdotted}, $x/D=0.71$; {\magentadashed}, $x/D=0.99$.}
\label{stats_outer}
\end{figure}

\begin{figure}
\centering

\sidesubfloat[]{
{\psfrag{x}[][]{{$y^+$}}
\psfrag{y}[][]{{$u'^+_{\text{rms}}$}}\includegraphics[width=0.46\textwidth,trim={1cm 6.8cm 3cm 0.8cm},clip]{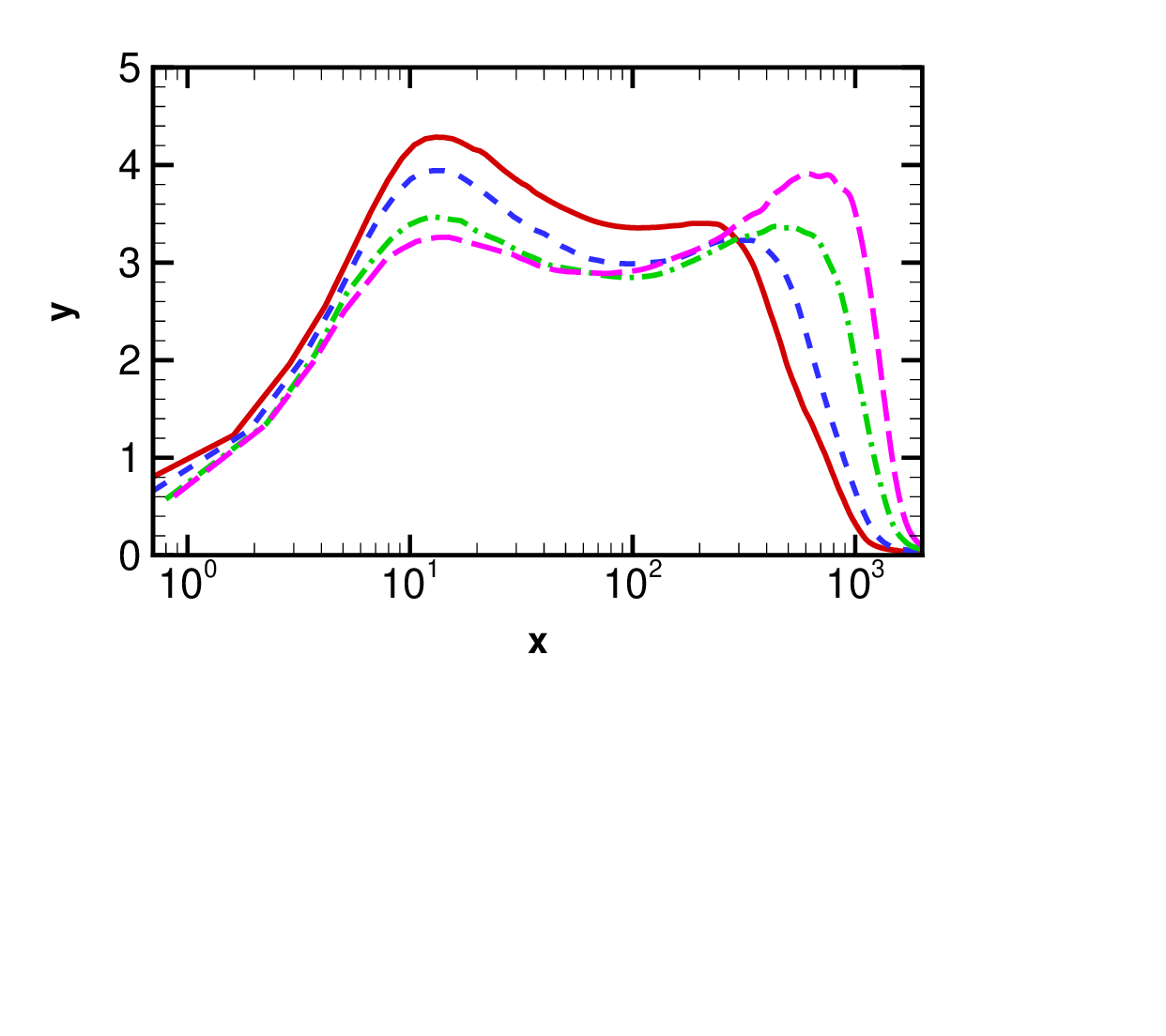}\label{stats_inner_a}}}
\sidesubfloat[]{
{\psfrag{x}[][]{{$y^+$}}
\psfrag{y}[][]{{$v'^+_{\text{rms}}$}}\includegraphics[width=0.47\textwidth,trim={1cm 6.6cm 3cm 0.8cm},clip]{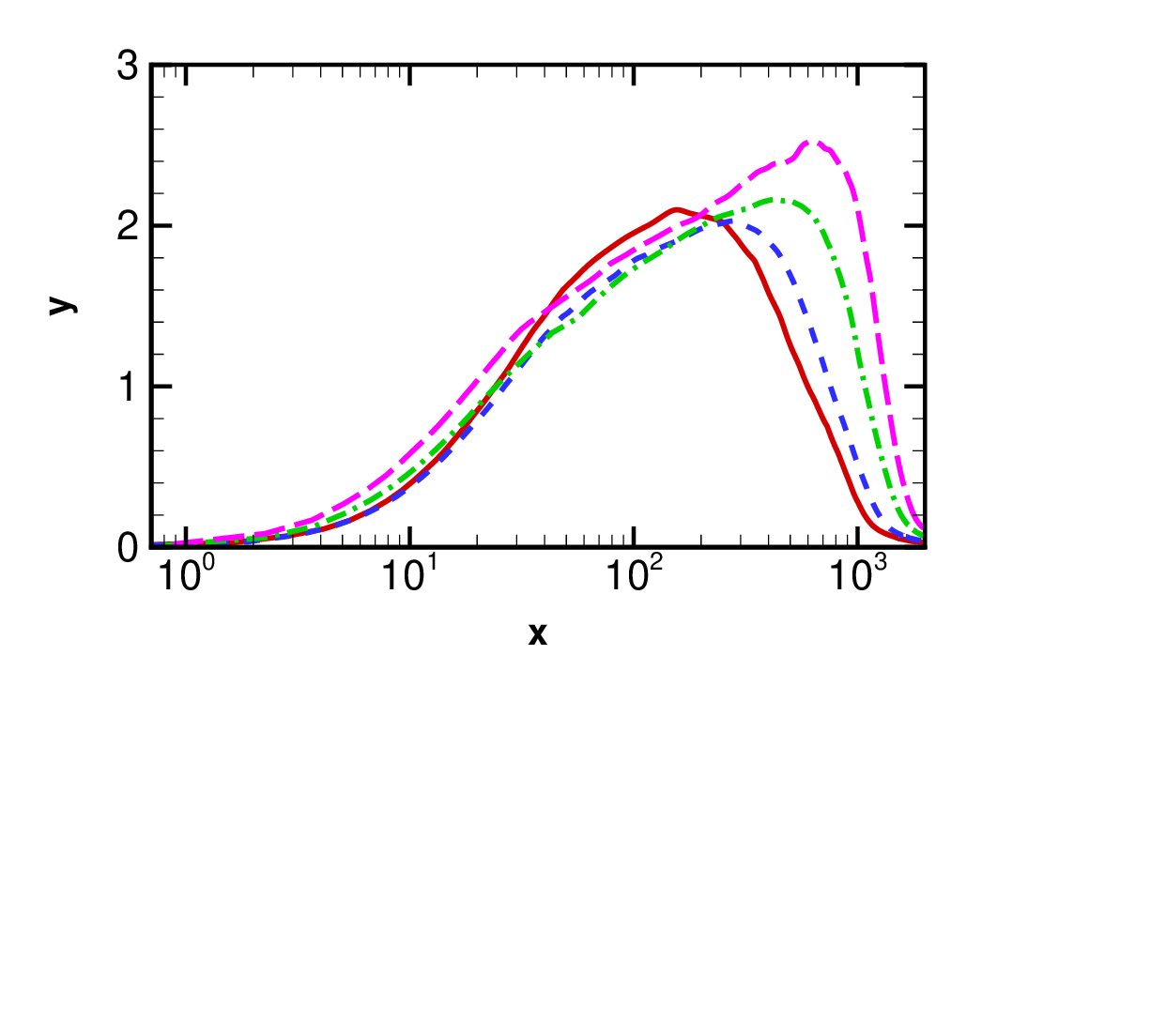}\label{stats_inner_b}}}

\sidesubfloat[]{
{\psfrag{x}[][]{{$y^+$}}
\psfrag{y}[][]{{$u'^+_{\theta, \text{rms}}$}}\includegraphics[width=0.46\textwidth,trim={1cm 6.9cm 3cm 0.7cm},clip]{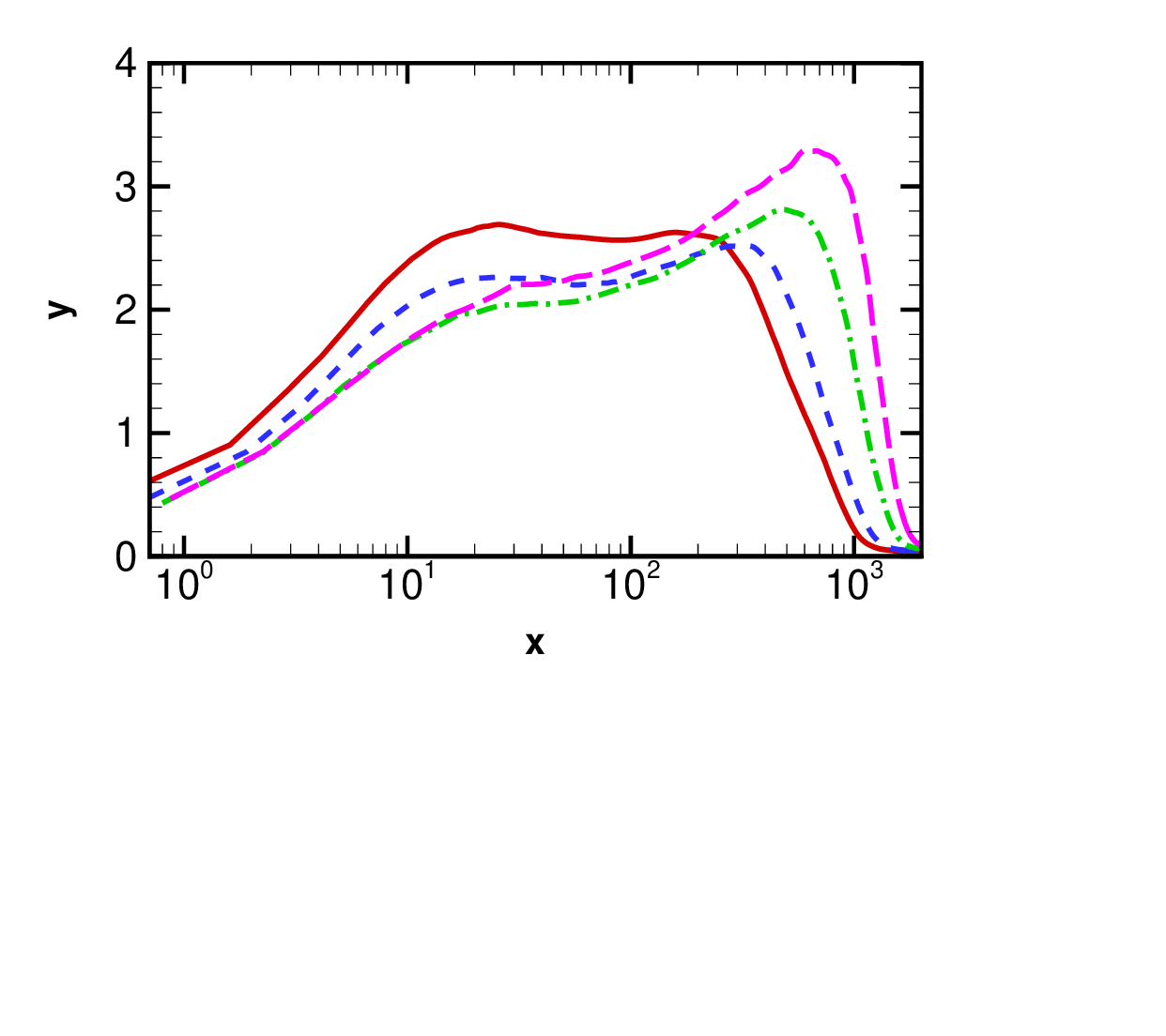}\label{stats_inner_c}}}
\sidesubfloat[]{
{\psfrag{x}[][]{{$y^+$}}
\psfrag{y}[][]{{$\overbar{u'v'}^+$}}\includegraphics[width=0.47\textwidth,trim={1cm 6.6cm 3cm 0.8cm},clip]{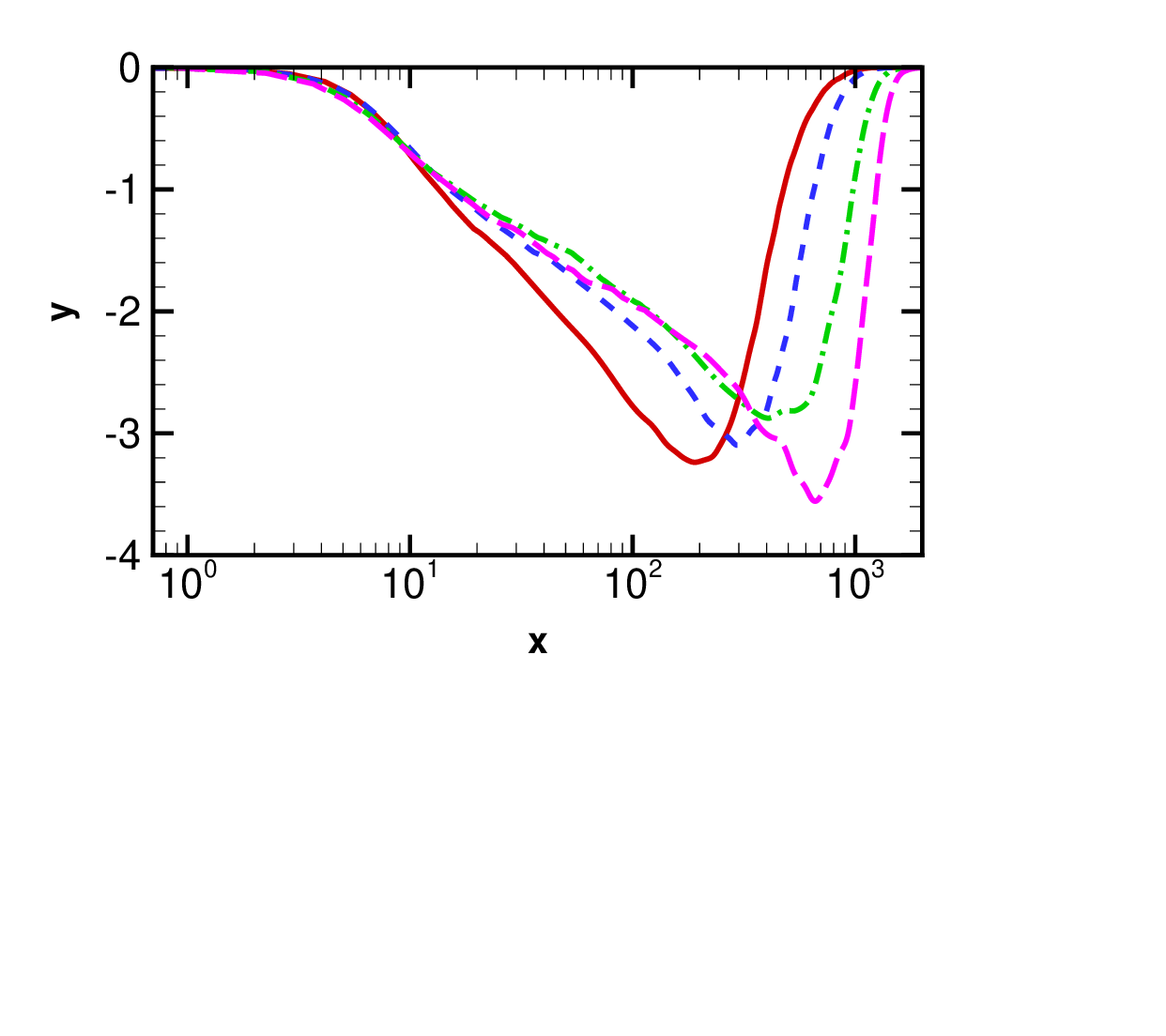}\label{stats_inner_d}}}

\caption{Profiles of (a-c) r.m.s. values of velocity fluctuations and (d) Reynolds shear stress with inner scaling at: {\redsolid}, $x/D=0.21$; {\bluedashed}, $x/D=0.43$; {\greendashdotted}, $x/D=0.71$; {\magentadashed}, $x/D=0.99$.}
\label{stats_inner}
\end{figure}

Figure~\ref{stats_outer} presents the r.m.s. values of velocity fluctuations $u'$, $v'$, $u'_\theta$ and the Reynolds shear stress $\overbar{u'v'}$ with outer scaling at the same four $x$-locations. A near-wall peak, henceforth referred to as the inner peak, is observed in the profiles of both the streamwise and azimuthal velocity components. This peak becomes weaker in the downstream direction and, in the case of the azimuthal velocity, vanishes at the last two stations plotted. Note that the inner peak was not captured in the measurement of \citet{balantrapu2021structure} due to its close proximity to the wall. On the other hand, a broader peak can be found in all r.m.s. profiles in the wake region of the TBL. Similar to the inner peak, the strength of the outer peak also decreases, albeit more slowly, in the downstream direction. However, this is accompanied by an outward shift and broadening of the peak, resulting in a growth of turbulence intensity in the outer region of $0.4\lesssim y/\delta \lesssim0.9$. In figure~\ref{stats_inner}, the same velocity r.m.s. and Reynolds shear-stress profiles are plotted with inner scaling. They exhibit a degree of self-similarity within the very near-wall region ($y^+<10$), a characteristic also observed in the inner-scaled mean streamwise velocity profiles. The inner peaks in the streamwise and azimuthal velocity fluctuations are located at $y^+\approx12$ and $y^+\approx25$, respectively. While their strengths decrease in the streamwise direction, the locations of the inner peaks remain unchanged in wall units. Conversely, the outer peaks move further outward with increasing strength and width as $x/D$ increases. The appearance of outer peaks in turbulence intensity profiles is a general feature of TBLs with APG, which energizes large-scale motions \citep{monty2011parametric}. More discussions about energy distributions in the boundary layer are provided in Section~\ref{pre_spect_TBL}.

\subsection{Self-similarity with embedded-shear-layer scaling}
\label{self_TBL}

The results in the previous section show a lack of self-similarity in velocity statistics profiles with either inner scaling or outer scaling with $\delta$ as the length scale, which is expected given the highly nonequilibrium nature of the boundary layer. \citet{balantrapu2021structure} showed improved outer-scaling using the displacement thickness as the length scale, but the degree of profile collapse is still limited. They subsequently demonstrated good self-similarity by using the embedded-shear-layer scaling proposed by \citet{schatzman2017experimental}. In this section, the validity of embedded-shear-layer scaling for the present axisymmetric APG TBL is further evaluated using the LES data, which include all three velocity components.  

\begin{figure}
\centering
{\psfrag{x}[][]{{$\eta$}}
\psfrag{y}[][]{{$U^*$}}\includegraphics[width=.56\textwidth,trim={0cm 0.5cm 0.0cm 2cm},clip]{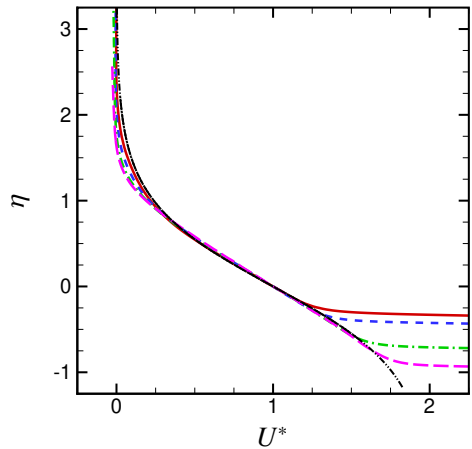}}
\caption{Mean streamwise velocity profiles with embedded-shear-layer scaling \citep{schatzman2017experimental} at four streamwise positions: {\redsolid}, $x/D=0.21$; {\bluedashed}, $x/D=0.43$; {\greendashdotted}, $x/D=0.71$; {\magentadashed}, $x/D=0.99$. The line {\blackdashdotdot} represents the function $U^*=1-\tanh{(\eta)}$.}
\label{u_scaling}
\end{figure}

Based on the embedded-shear-layer scaling \citep{schatzman2017experimental}, the similarity variables are $\eta=(y-y_{\text{IP}})/\delta_{\omega}$ and $U^*=(U_e-U)/U_d$, where the subscript $\text{IP}$ denotes the outer inflection point of the mean streamwise velocity profile, $\delta_{\omega}=(U_e-U)_{\text{IP}}/(\dif{U}/\dif{y})_{\text{IP}}$ is the local vorticity thickness of the embedded shear layer, and $U_d=(U_e-U)_{\text{IP}}$ is the local mean streamwise velocity defect at the inflection point. In figure~\ref{u_scaling}, the mean streamwise velocity profiles in figure~\ref{mean_u_profile} are presented in terms of the embedded-shear-layer scaling. The location $\eta=0$ corresponds to the outer inflection point in the mean streamwise velocity profile. All four profiles show good similarity over a wide range of the wall-normal distance except in the vicinity of the wall due to the wall effect. This confirms the scaling results of \citet{balantrapu2021structure}, whose experimental data does not include the near-wall region, and is consistent with the data of \cite{schatzman2017experimental} for the APG TBLs on a 2-D hump. As the location moves downstream, the self-similar behavior extends further toward the wall in the $\eta$-coordinate. In addition, the self-similar profiles are seen to collapse well onto the function $U^*=1-\tanh{(\eta)}$, and the $\eta$-range of collapse also expands along the streamwise direction.

\begin{figure}
\centering

\sidesubfloat[]{
{\psfrag{x}[][]{{$u'_{\text{rms}}/U_d$}}
\psfrag{y}[][]{{$\eta$}}\includegraphics[width=0.46\textwidth,trim={1cm 0.2cm 2cm 0.8cm},clip]{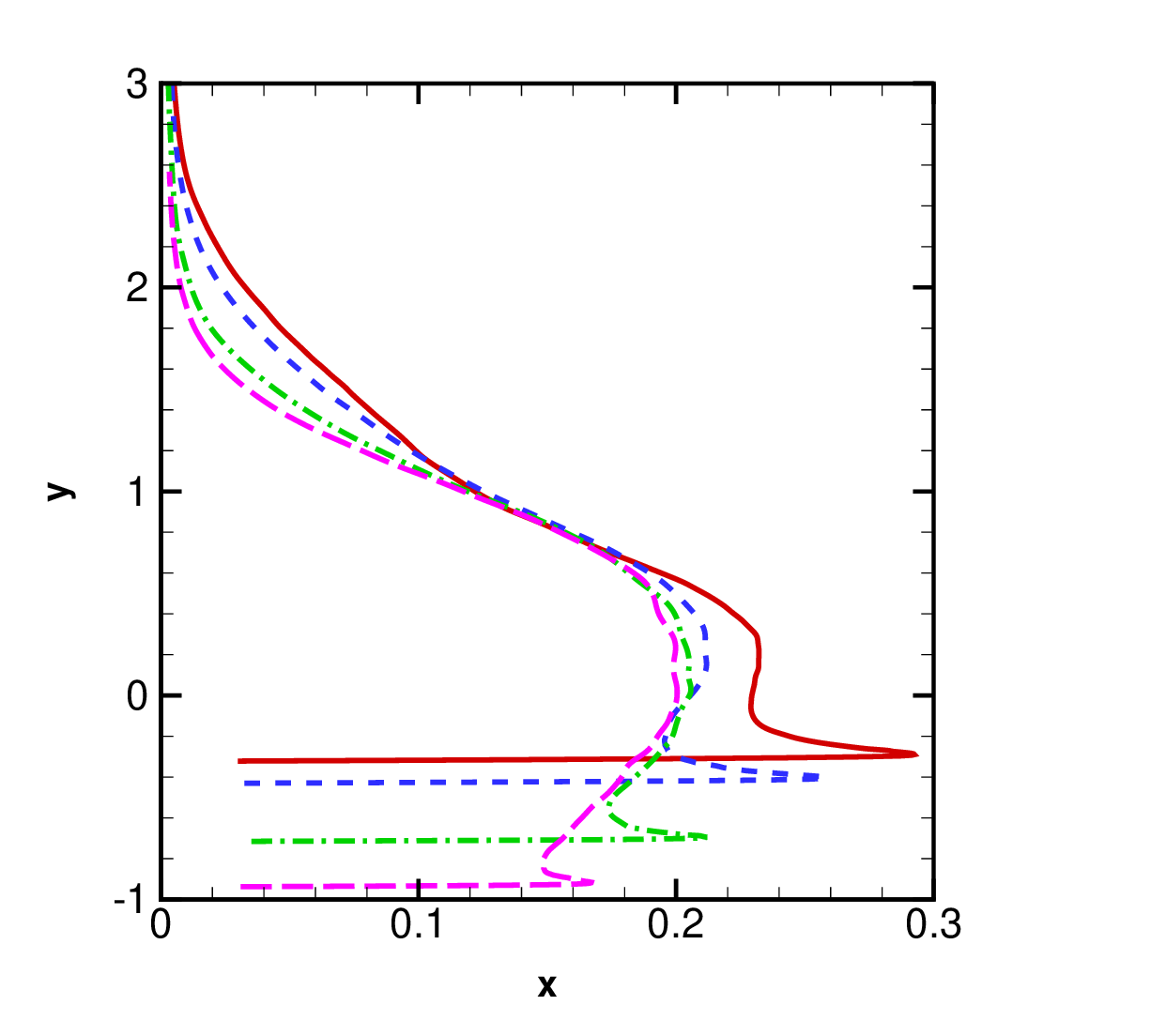}\label{stats_scaling_a}}}
\sidesubfloat[]{
{\psfrag{x}[][]{{$v'_{\text{rms}}/U_d$}}
\psfrag{y}[][]{{$\eta$}}\includegraphics[width=0.46\textwidth,trim={1cm 0.2cm 2cm 0.8cm},clip]{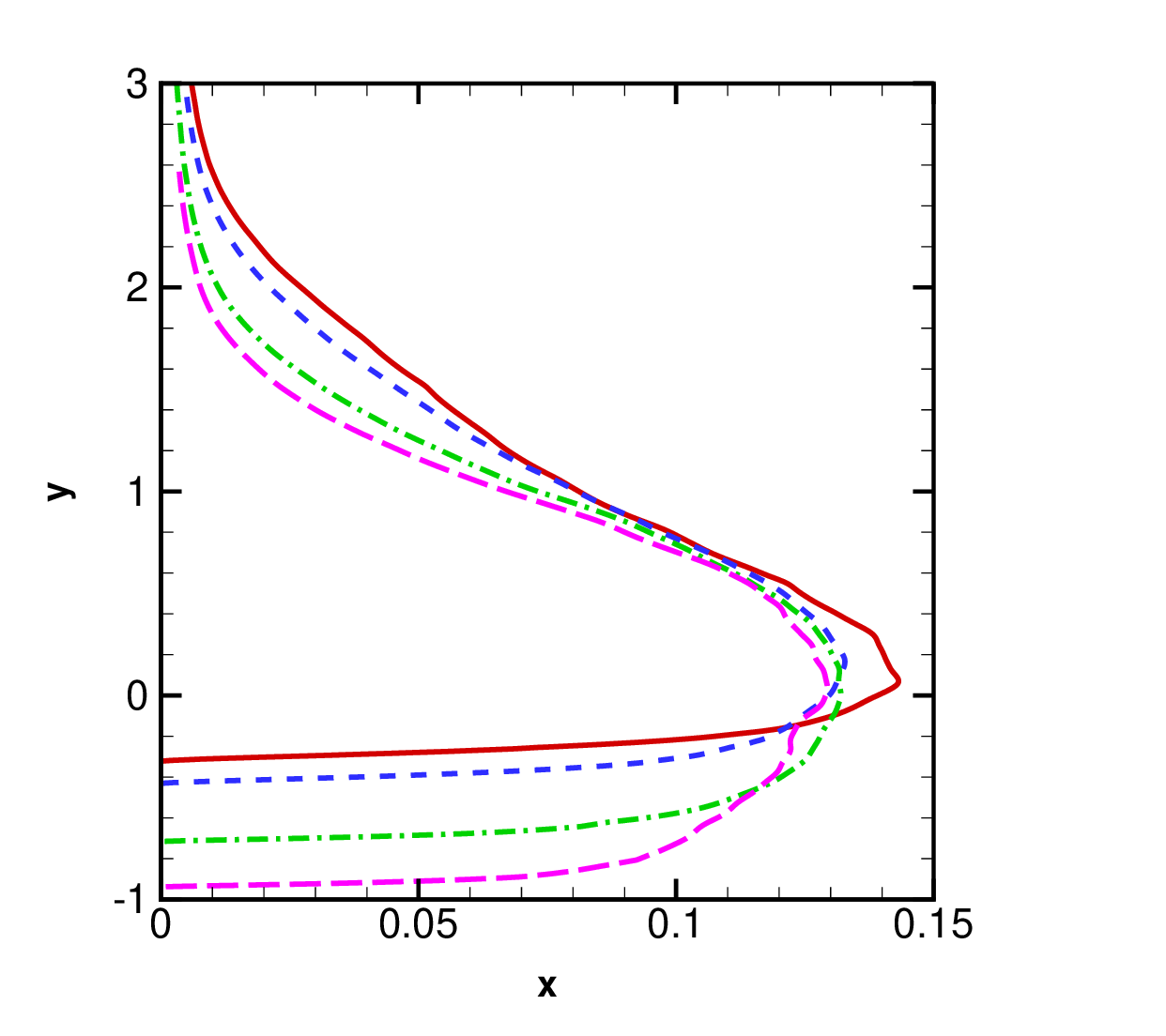}\label{stats_scaling_b}}}

\sidesubfloat[]{
{\psfrag{x}[][]{{$u'_{\theta, \text{rms}}/U_d$}}
\psfrag{y}[][]{{$\eta$}}\includegraphics[width=0.46\textwidth,trim={1cm 0.6cm 2cm 0.8cm},clip]{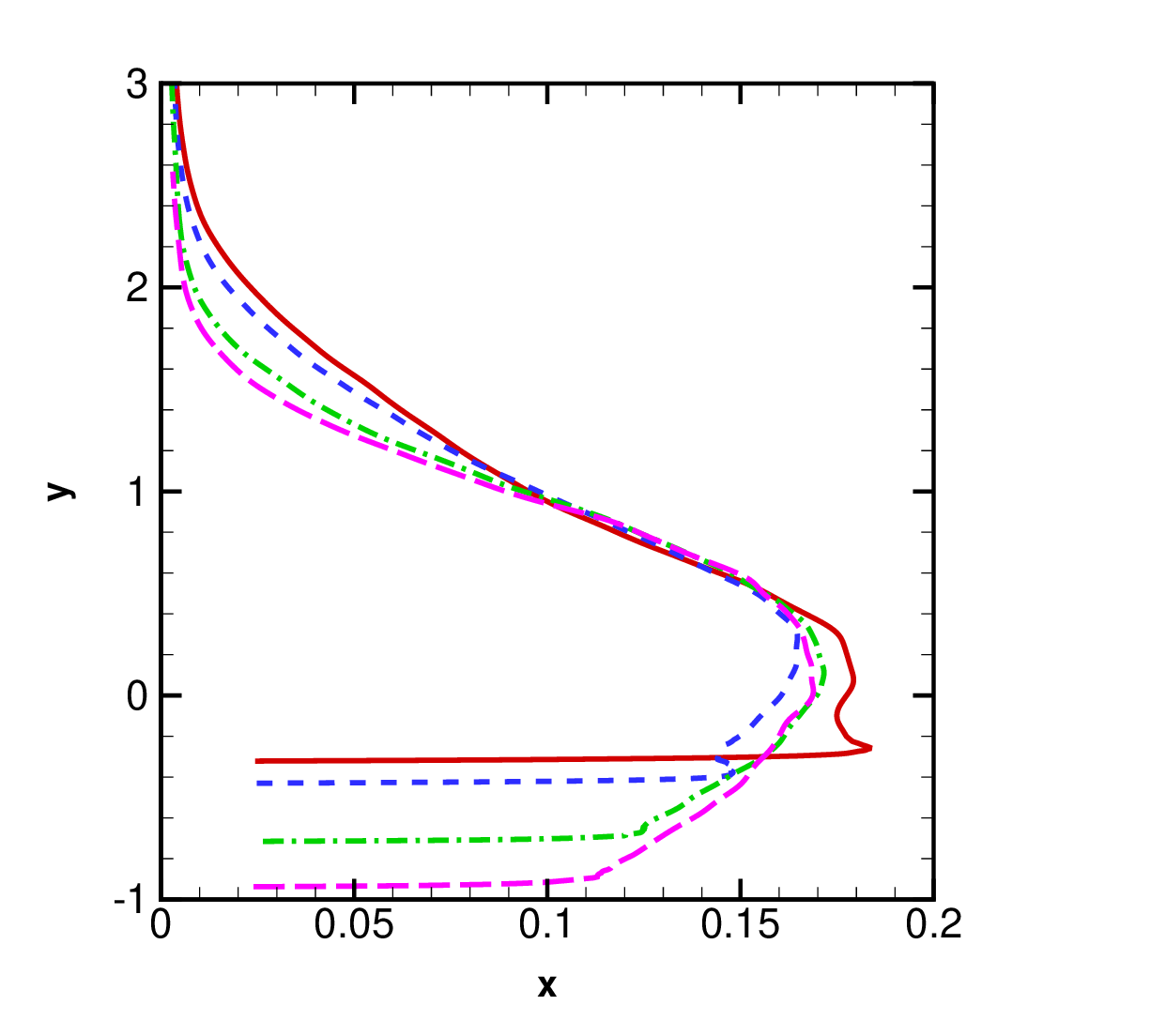}\label{stats_scaling_c}}}
\sidesubfloat[]{
{\psfrag{x}[][]{{$\overbar{u'v'}/U_d^2$}}
\psfrag{y}[][]{{$\eta$}}\includegraphics[width=0.46\textwidth,trim={1cm 0.6cm 2cm 0.8cm},clip]{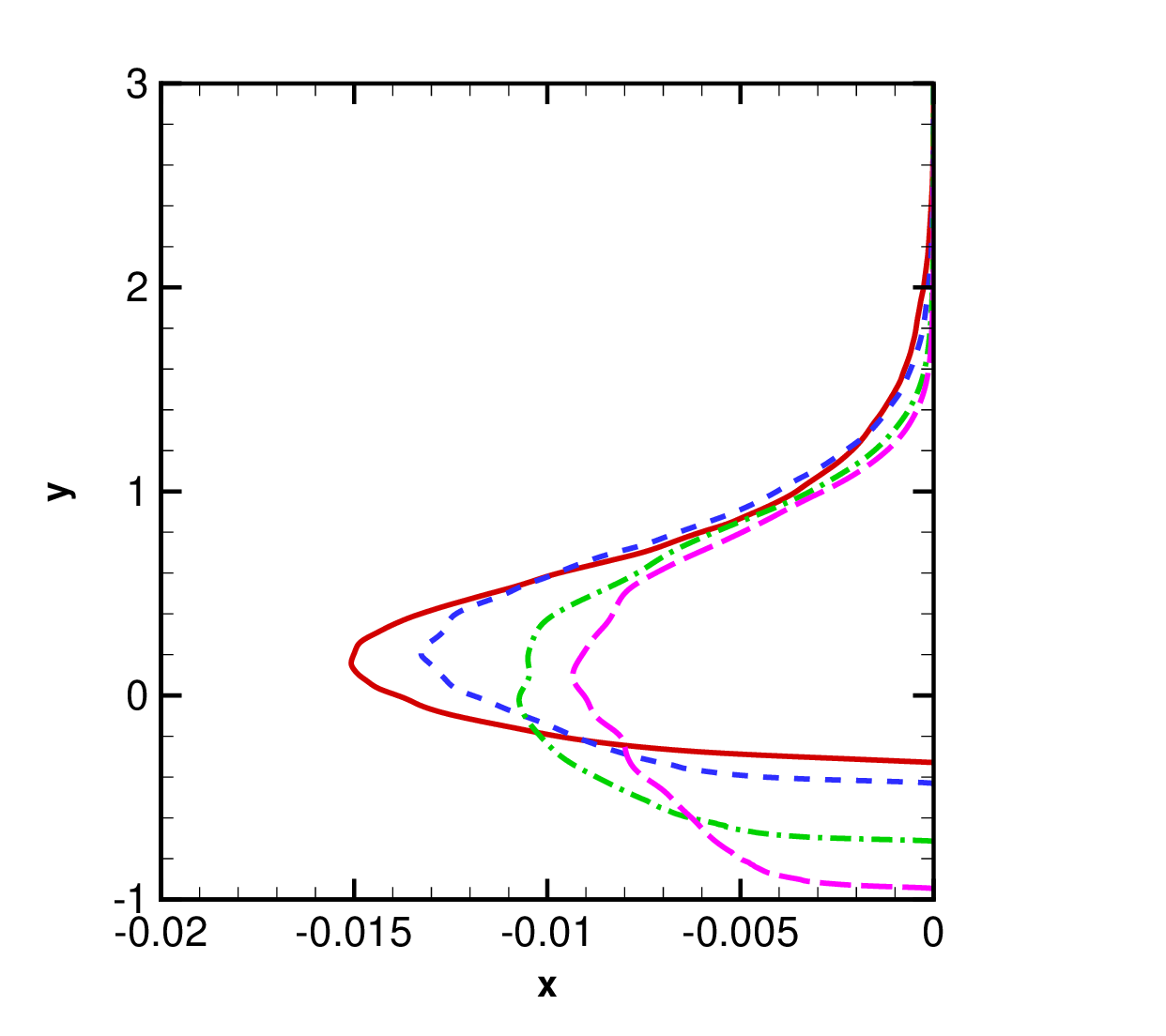}\label{stats_scaling_d}}}
\caption{Profiles of (a-c) r.m.s. values of the three components of velocity fluctuations and (d) the Reynolds shear stress $\overbar{u'v'}$ with embedded-shear-layer scaling at: {\redsolid}, $x/D=0.21$; {\bluedashed}, $x/D=0.43$; {\greendashdotted}, $x/D=0.71$; {\magentadashed}, $x/D=0.99$.}
\label{stats_scaling}
\end{figure}

Figure~\ref{stats_scaling} displays the r.m.s. values of the three components of velocity fluctuations and the Reynolds shear stress $\overbar{u'v'}$ with the embedded-shear-layer scaling. The profiles of all three r.m.s. velocity components exhibit some degree of similarity in the region above the inflection point ($\eta \geq 0$), similar to that shown by \citet{balantrapu2021structure} with single-hotwire velocity data. The failure of the shear-layer scaling in the near-wall region is again expected due to the wall effect, and the larger deviations of the profiles at $x/D=0.21$ near the inflection point result from the influence of the upstream corner flow at the beginning of the cone, where the pressure gradient is extremely large and varies rapidly (cf. figure~\ref{Cp}). Similarity in velocity fluctuations is improved at the downstream stations as the APG variation becomes more mild. Furthermore, it is worth noting that the outer peaks in the r.m.s. velocity profiles are all located at $\eta \approx 0.2$, which is slightly above the inflection point, and the peak values relative to the mean streamwise velocity defect remain relatively constant. The peaks in the streamwise-component profiles are $u'_{rms}/U_d \approx 0.21$, which is consistent with \citet{balantrapu2021structure} and the values for the APG TBLs on a 2-D ramp investigated by \citet{schatzman2017experimental}. Large deviations are observed among the different Reynolds shear-stress profiles around their peaks, suggesting a more restricted region of validity of embedded-shear-layer scaling.

\subsection{Azimuthal wavenumber spectra of streamwise velocity fluctuations}
\label{pre_spect_TBL}

\begin{figure}
\centering

\sidesubfloat[]{
{\psfrag{b}[][]{{$y^+$}}
\psfrag{d}[][]{{$\lambda^+_{\theta}$}}\includegraphics[width=0.46\textwidth,trim={0.1cm 0.3cm 0.1cm 0.3cm},clip]{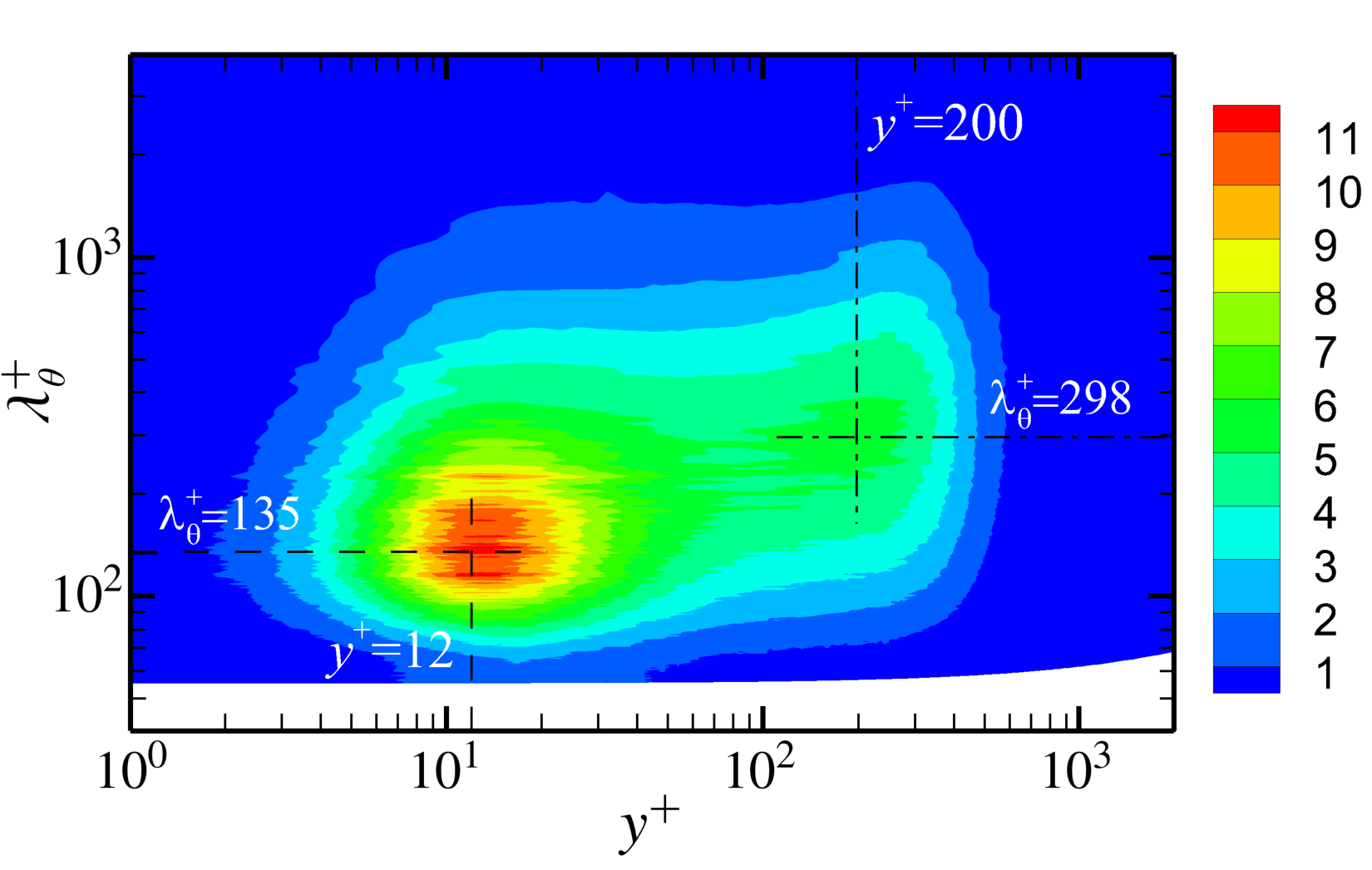}\label{pre_spect_a}}}
\sidesubfloat[]{
{\psfrag{b}[][]{{$y^+$}}
\psfrag{d}[][]{{$\lambda^+_{\theta}$}}\includegraphics[width=0.46\textwidth,trim={0.1cm 0.0cm 0.1cm 0.1cm},clip]{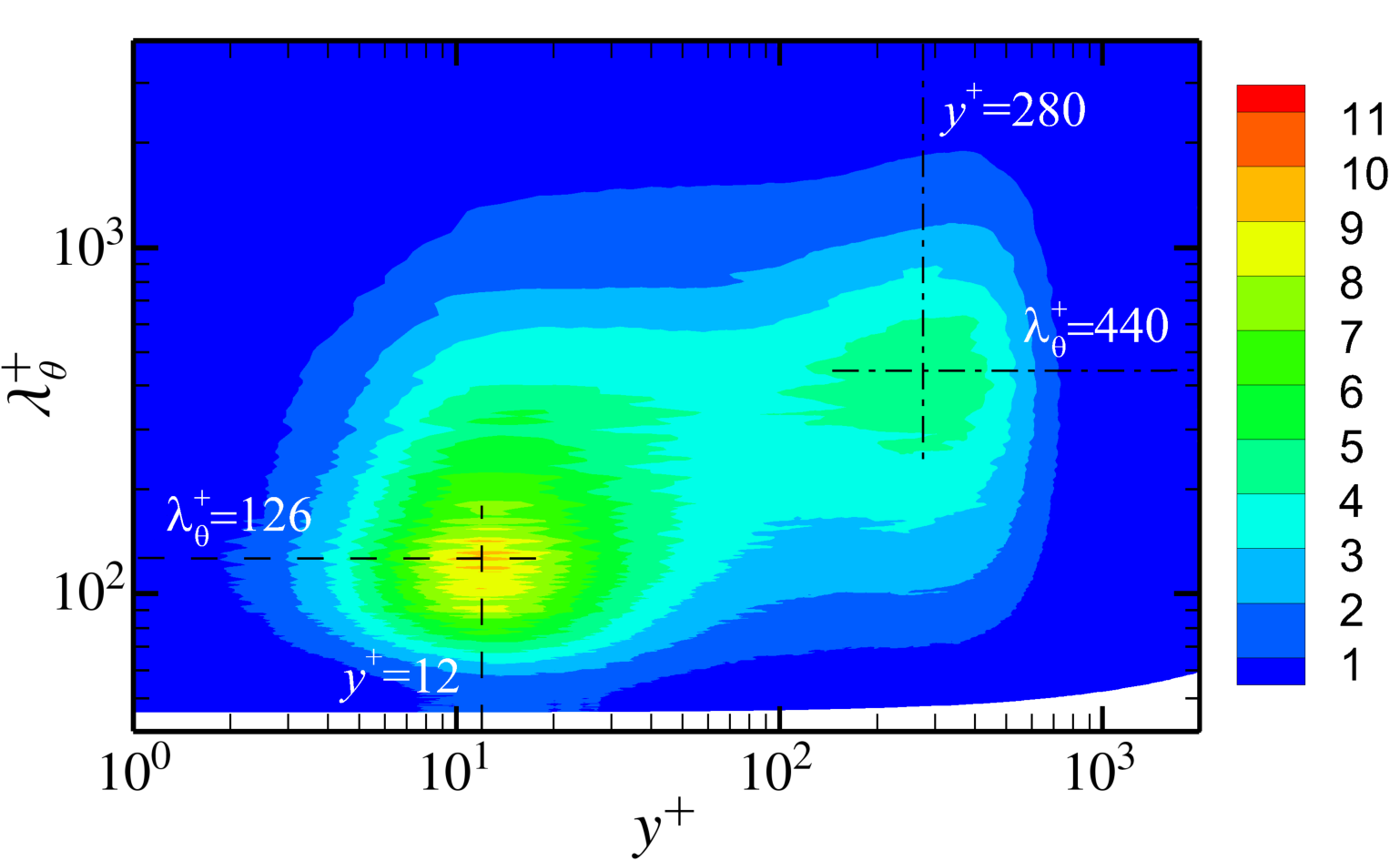}\label{pre_spect_b}}}

\sidesubfloat[]{
{\psfrag{b}[][]{{$y^+$}}
\psfrag{d}[][]{{$\lambda^+_{\theta}$}}\includegraphics[width=0.46\textwidth,trim={0.1cm 0.0cm 0.1cm 0.2cm},clip]{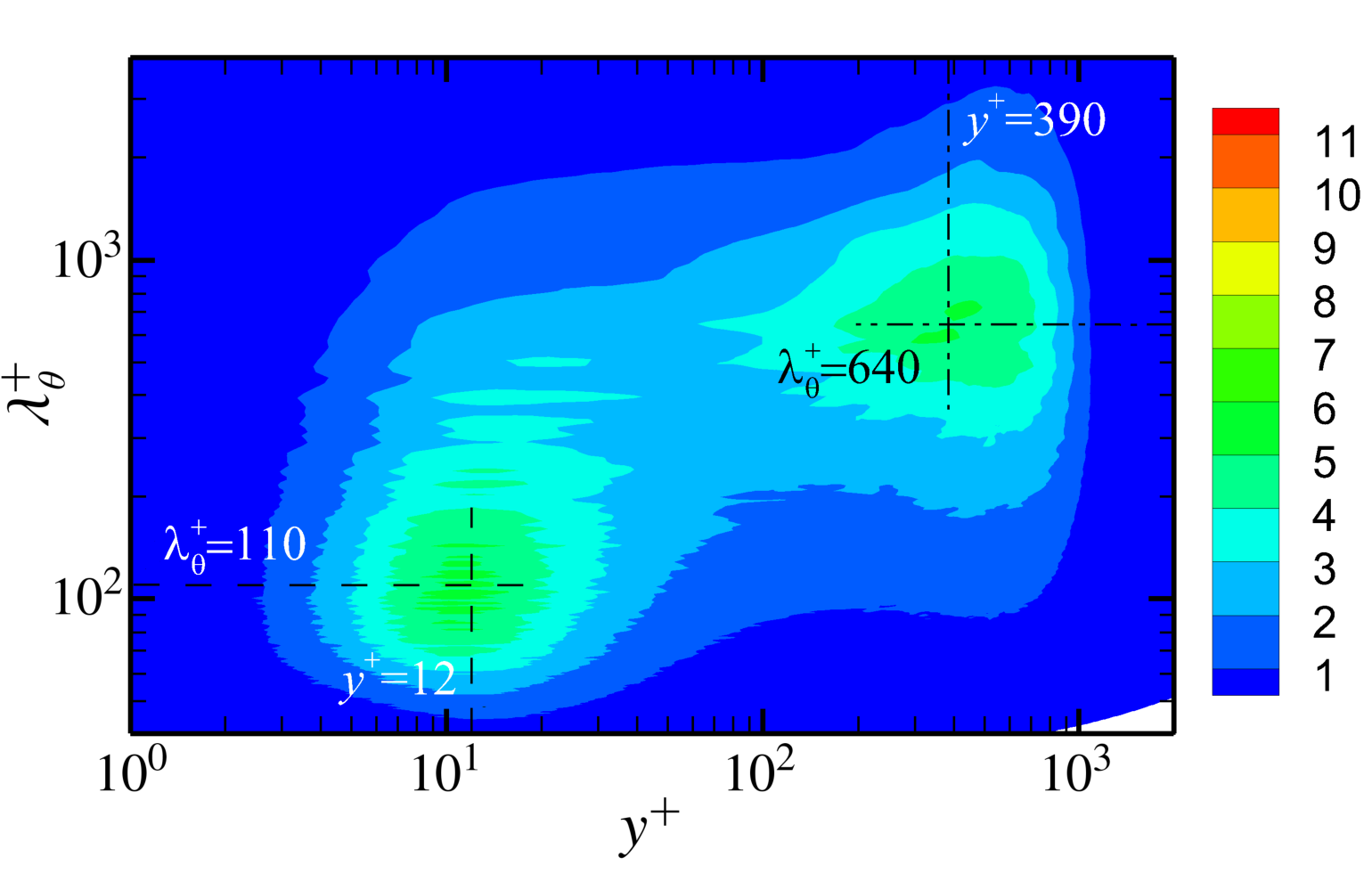}\label{pre_spect_c}}}
\sidesubfloat[]{
{\psfrag{b}[][]{{$y^+$}}
\psfrag{d}[][]{{$\lambda^+_{\theta}$}}\includegraphics[width=0.46\textwidth,trim={0.1cm 0.4cm 0.1cm -0.1cm},clip]{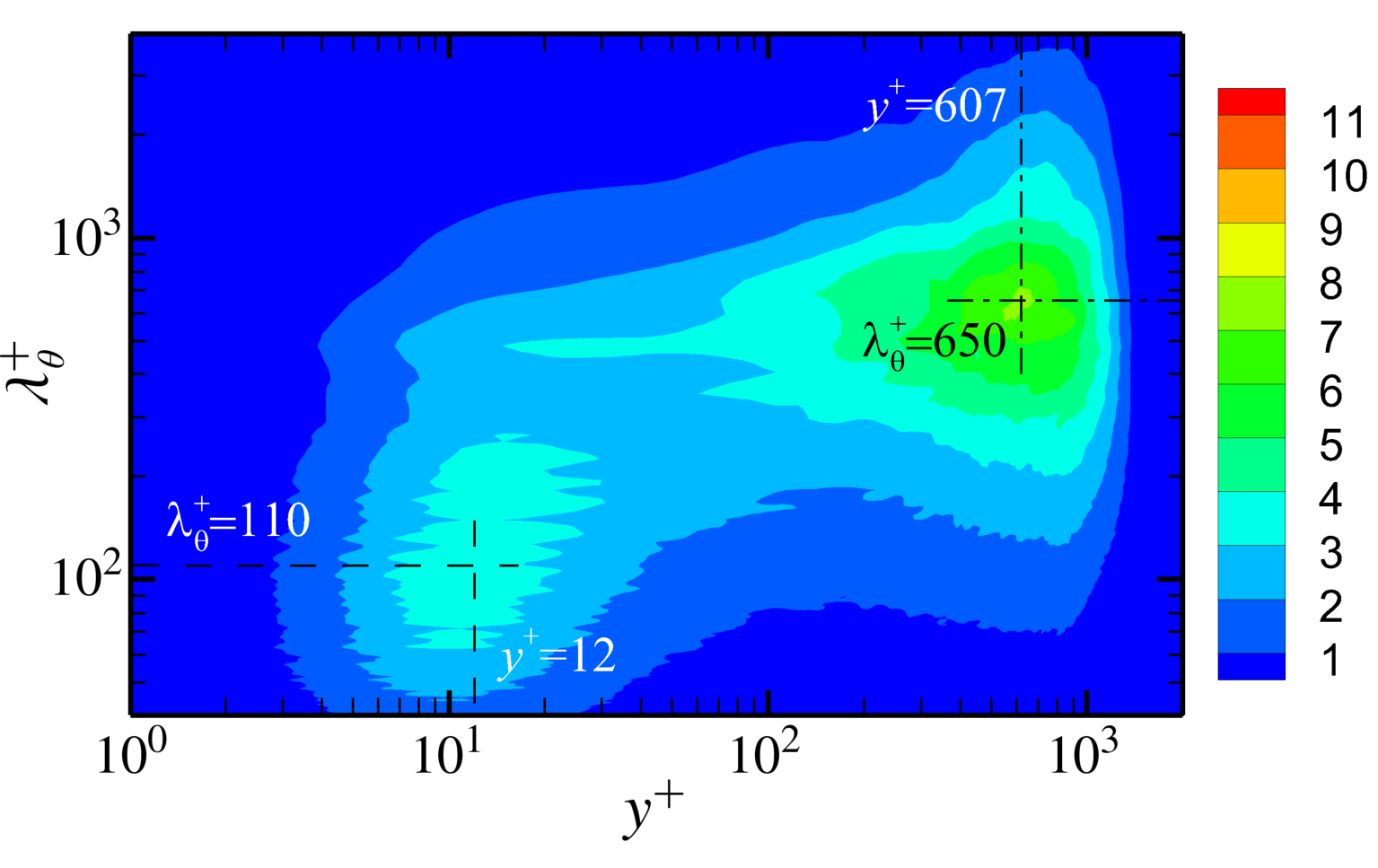}\label{pre_spect_d}}}

\caption{Pre-multiplied energy spectra $k_{\theta}\phi_{uu}/u_\tau^2$ as a function of the wall-normal coordinate and azimuthal wavelength in wall units at: (a) $x/D=0.21$, (b) $x/D=0.43$, (c) $x/D=0.71$ and (d) $x/D=0.99$. The intersections of the dashed lines and dash-dotted lines indicate the approximate locations of the inner and outer peaks, respectively.}
\label{pre_spect}
\end{figure}

The profiles of velocity fluctuations in figure~\ref{stats_inner} provide an overall measure of the energy distributions within the TBL. To explore the evolution of energy distribution among different length scales, the pre-multiplied azimuthal wavenumber spectra of the streamwise velocity fluctuations, $k_{\theta}\phi_{uu}(k_{\theta})/u_\tau^2$, are examined. Figure~\ref{pre_spect} shows the pre-multiplied spectra as a function of $y^+$ and the azimuthal wavelength in wall units, $\lambda^+_\theta=(2\pi/k_\theta)(u_\tau/\nu)$, at the four streamwise locations. It can be observed that there are two energy peaks at different wall-normal locations, which is consistent with the r.m.s. results shown in figure~\ref{stats_inner_a}. The azimuthal wavelengths and the wall-normal locations of these two peaks are marked in the figure. The first peak, located at $y^+\approx 12$ with azimuthal wavelengths of 110 to 135 wall units, is called the inner peak and is associated with elongated near-wall streaks. While the azimuthal wavelength of the inner peak, and hence the spacing between the streaks in wall units are slightly decreased in the downstream direction, they are not significantly different from those in ZPG flat-plate TBLs. This is consistent with the experimental results of \citet{harun2013pressure} but contrary to the findings of \citet{lee2009structures} from DNS of an APG flat-plate TBL at a lower Reynolds number. The latter showed a significant increase in the spanwise spacing between streaks under the influence of a strong APG. It can also be noted from figure~\ref{pre_spect} that although the positions of the inner peaks are similar at different streamwise locations, the strengths of the peaks decrease rapidly in the downstream direction.

The secondary peak in the azimuthal wavenumber spectra, referred to as the outer peak, is located in the wake region of the TBL, which is different from that for a ZPG TBL \citep{mathis2009large, wang2018spanwise}. As the TBL evolves toward downstream, the azimuthal wavelength of the outer peak and its distance to the wall both grow, indicating that increasingly larger-scale structures further away from the wall are energized. However, the growth of the azimuthal length scale becomes slower in the downstream. The azimuthal wavelength normalized by the outer length scale, $\lambda_\theta/\delta$, is around 0.4, which is smaller than the spanwise length scale in ZPG flat-plate TBLs (e.g., $\approx0.9$ in \citet{wang2018spanwise}). Furthermore, the strength of the outer peak increases gradually toward downstream, in agreement with the trend for the turbulence intensity and Reynolds shear stress seen in figure~\ref{stats_inner}.

\subsection{Two-point correlations of streamwise velocity fluctuations}
\label{sec_corr}

\begin{figure}
\centering
{\includegraphics[width=.8\textwidth,trim={0 0.2cm 0 0cm},clip]{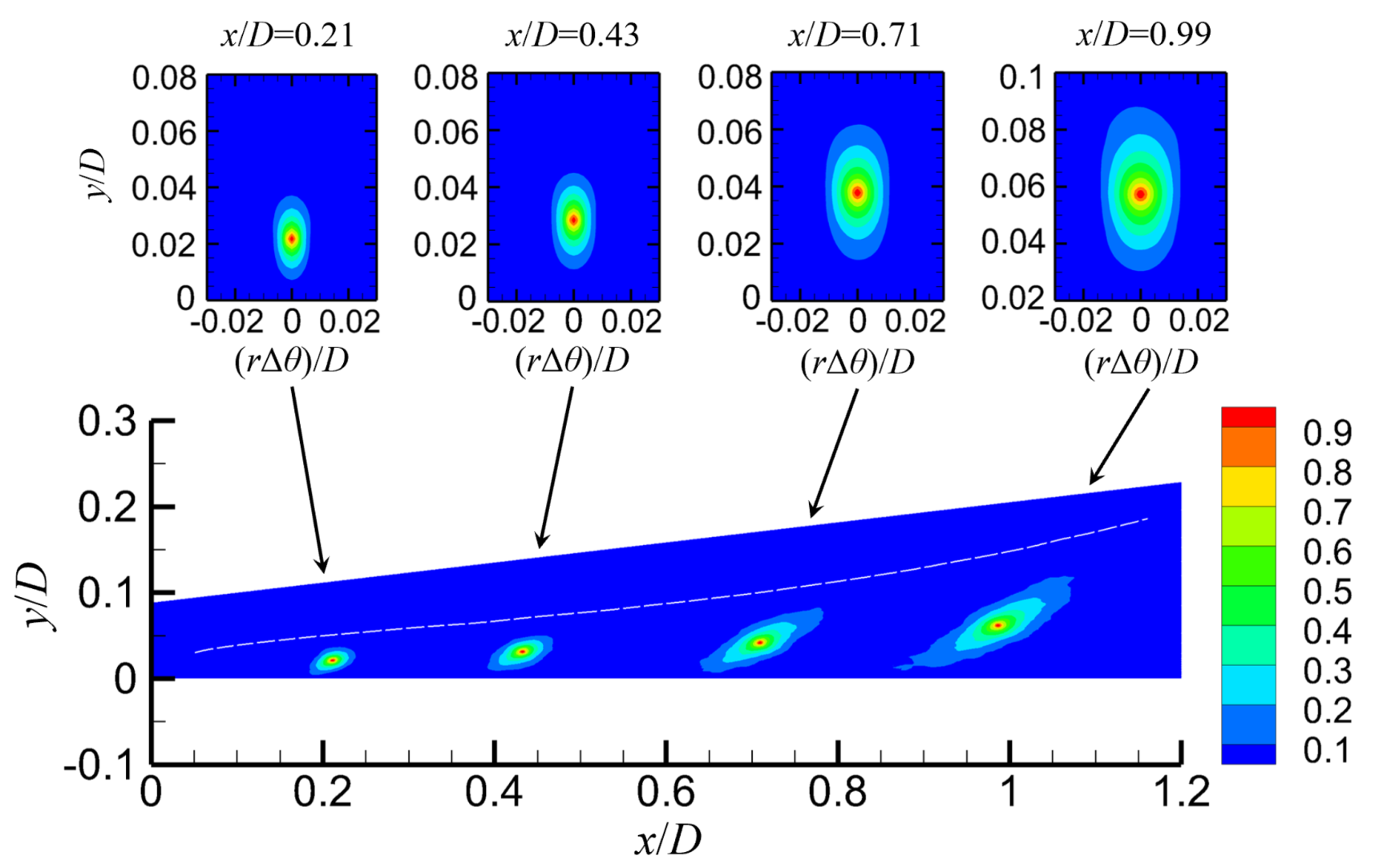}
}
\caption{Coefficients of two-point correlations of streamwise velocity fluctuations in the streamwise-wall-normal ($x$-$y$) and cross-flow ($y$-$\theta$) planes with origins at four streamwise stations and $y/\delta=0.4$. The white dashed line represents the edge of the boundary layer.}
\label{up_04delta}
\end{figure}

In this section, the structure of the tail-cone TBL is discussed further in terms of two-point spatial correlations. Figure~\ref{up_04delta} shows the two-point correlations of the streamwise velocity fluctuations in the streamwise-wall-normal ($x$-$y$) and cross-flow ($y$-$\theta$) planes anchored at the previous four streamwise stations with $y/\delta=0.4$, which is in the vicinity of the outer peaks of the energy spectra. The two-point spatial correlation coefficient $C_{uu}$ is defined as:
\begin{equation}
    C_{uu}(x, y, \Delta x, \Delta y, \Delta \theta) = \frac{\overbar{\langle u'(x, y, \theta,t)u'(x+\Delta x, y+\Delta y, \theta+\Delta \theta,t)\rangle}}{\sqrt{\overbar{\langle u'^2(x,y,\theta,t)\rangle}}\sqrt{\overbar{\langle u'^2(x+\Delta x, y+\Delta y, \theta+\Delta \theta,t)\rangle}}}.
\end{equation}
It is independent of the azimuthal angle $\theta$ because of spatial homogeneity. The correlation contours illustrate significant growth of turbulence length scales with growing boundary-layer thickness, indicated by the dashed line, in the downstream direction. Moreover, the evolving contour shapes in the $x$-$y$ plane indicate increasing anisotropy of large-scale structures toward the downstream. 

\begin{figure}
\centering

\sidesubfloat[]{
{\includegraphics[width=0.46\textwidth,trim={0.1cm 0.1cm 0.2cm 0.3cm},clip]{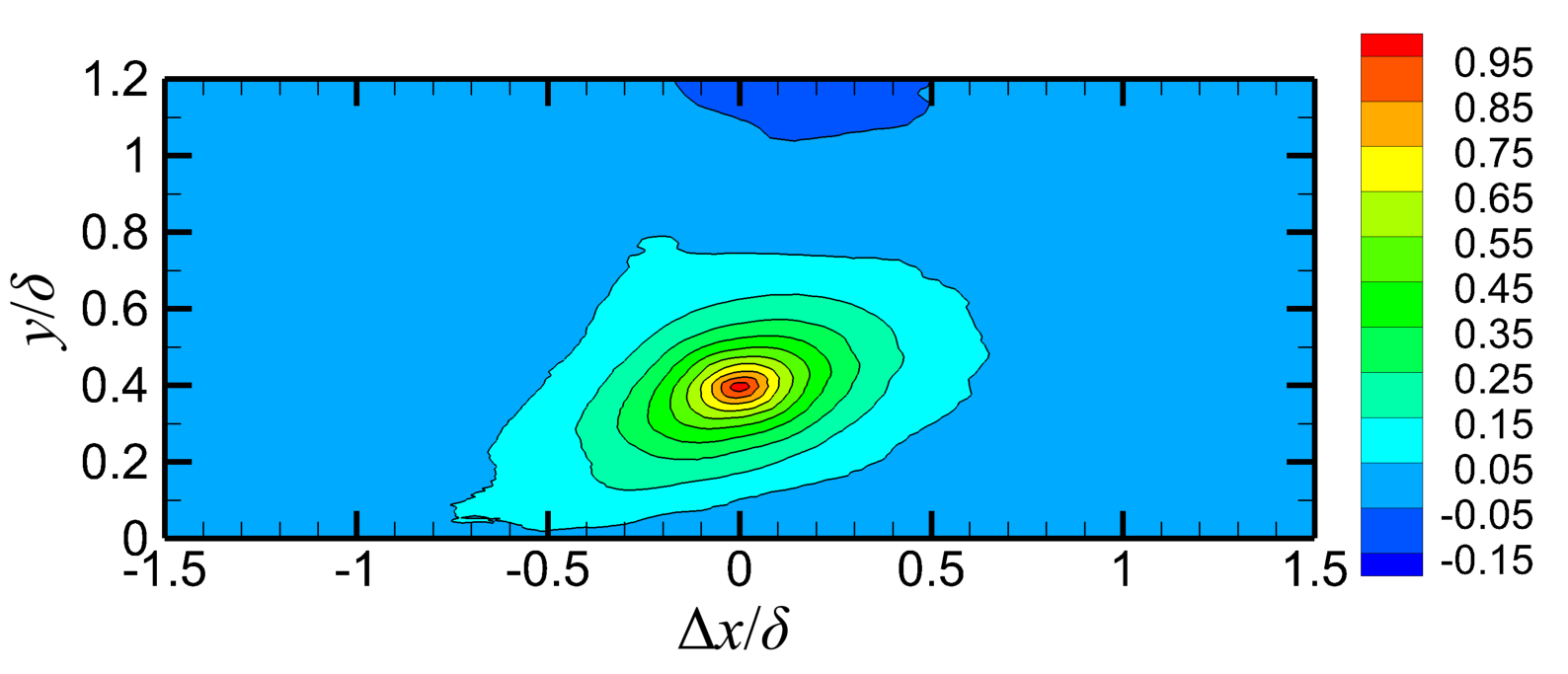}\label{up_corr_xy_a}}}
\sidesubfloat[]{
{\includegraphics[width=0.46\textwidth,trim={0.2cm 0.3cm 0.2cm 0.5cm},clip]{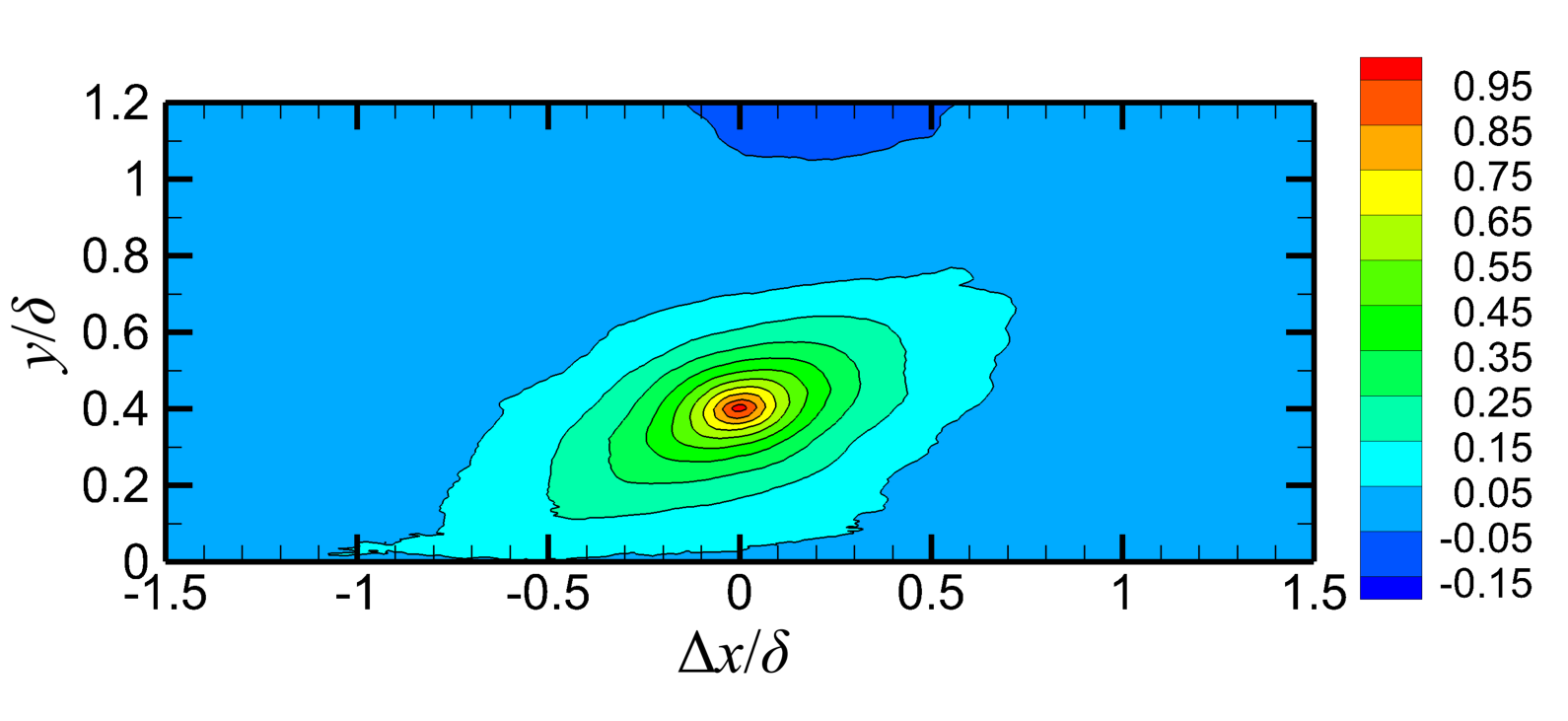}\label{up_corr_xy_b}}}

\sidesubfloat[]{
{\includegraphics[width=0.46\textwidth,trim={0.2cm 0.4cm 0.2cm 0.5cm},clip]{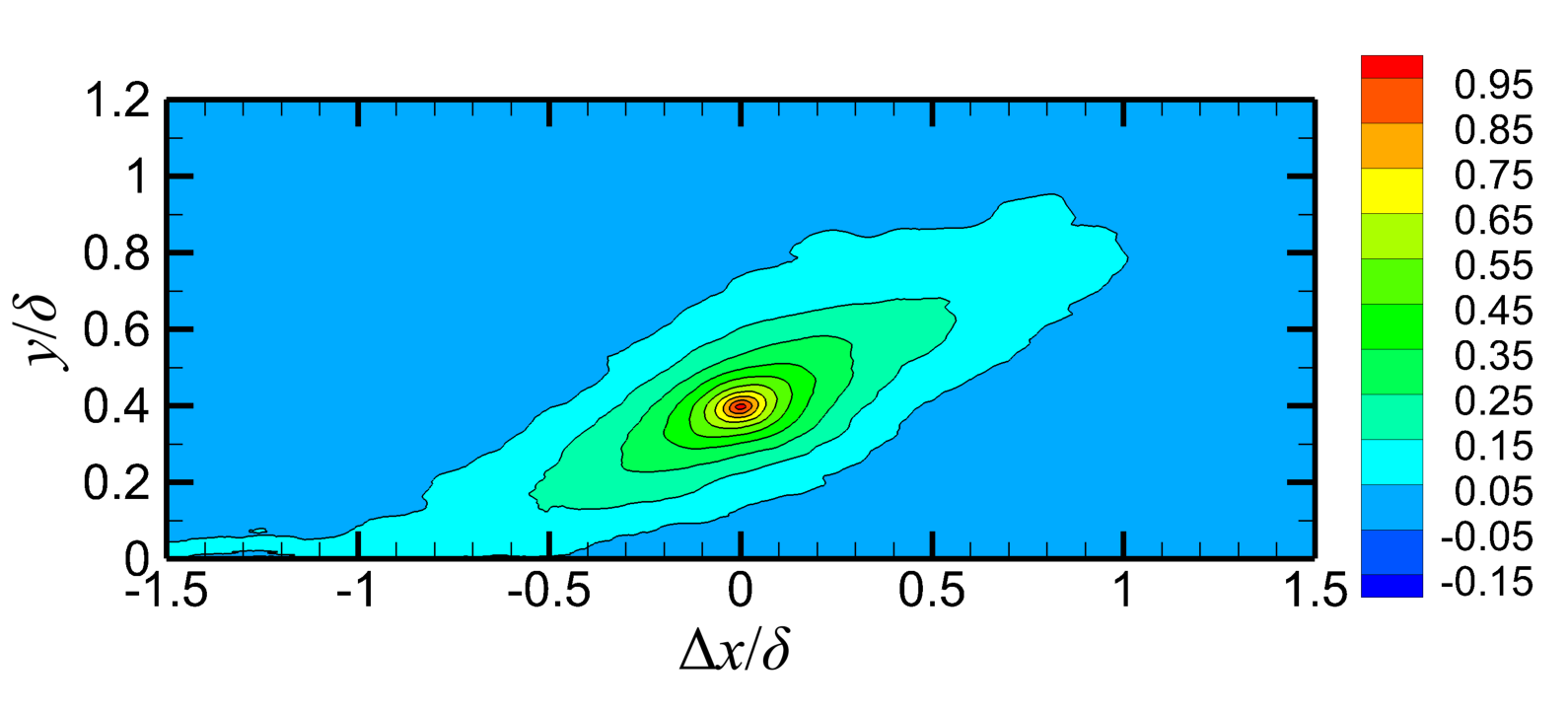}\label{up_corr_xy_c}}}
\sidesubfloat[]{
{\includegraphics[width=0.46\textwidth,trim={0.2cm 0.2cm 0.1cm -0.1cm},clip]{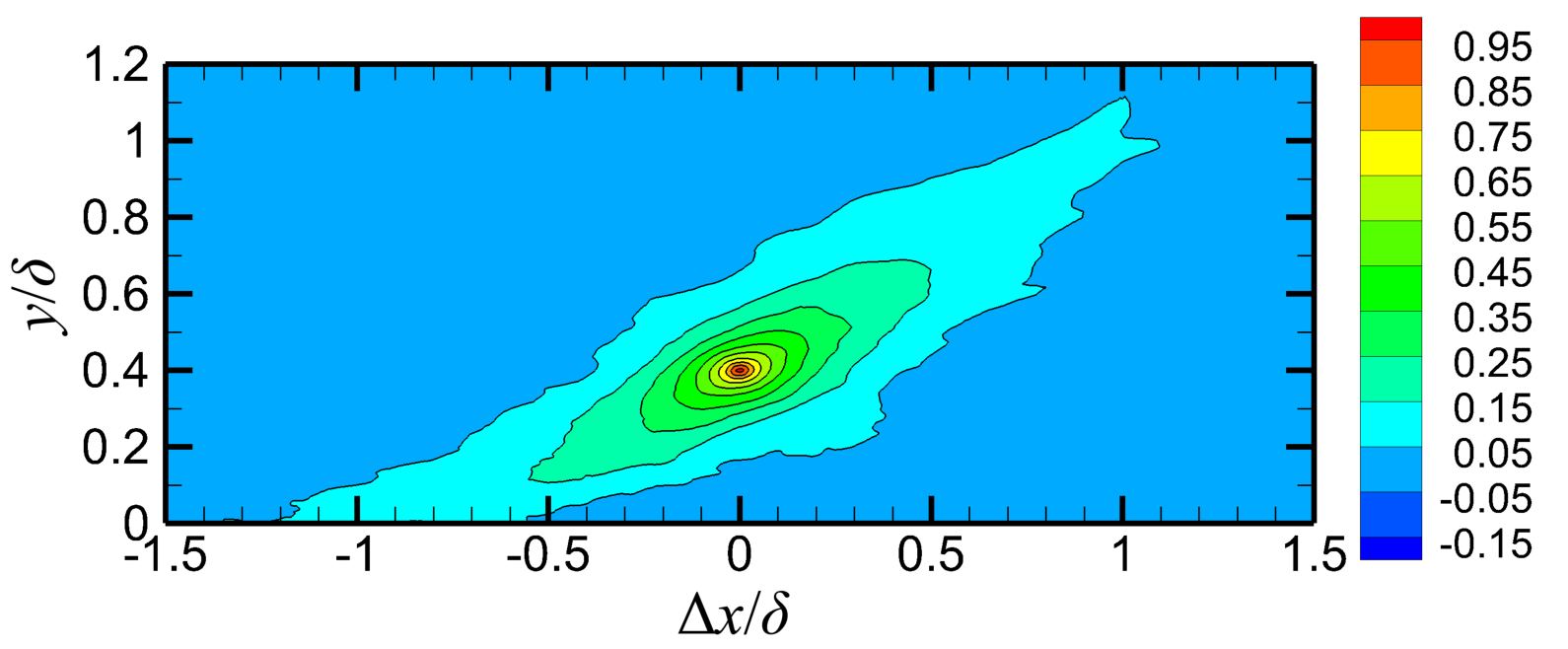}}\label{up_corr_xy_d}}

\caption{Coefficients of two-point correlations of streamwise velocity fluctuations in the $x$-$y$ plane with anchor points located at $y/\delta=0.4$ and (a) $x/D=0.21$; (b) $x/D=0.43$; (c) $x/D=0.71$; (d) $x/D=0.99$.}
\label{up_corr_xy_04}
\end{figure}

\begin{figure}
\centering

\sidesubfloat[]{
{\includegraphics[width=0.46\textwidth,trim={0.1cm 0.1cm 0.2cm 0.3cm},clip]{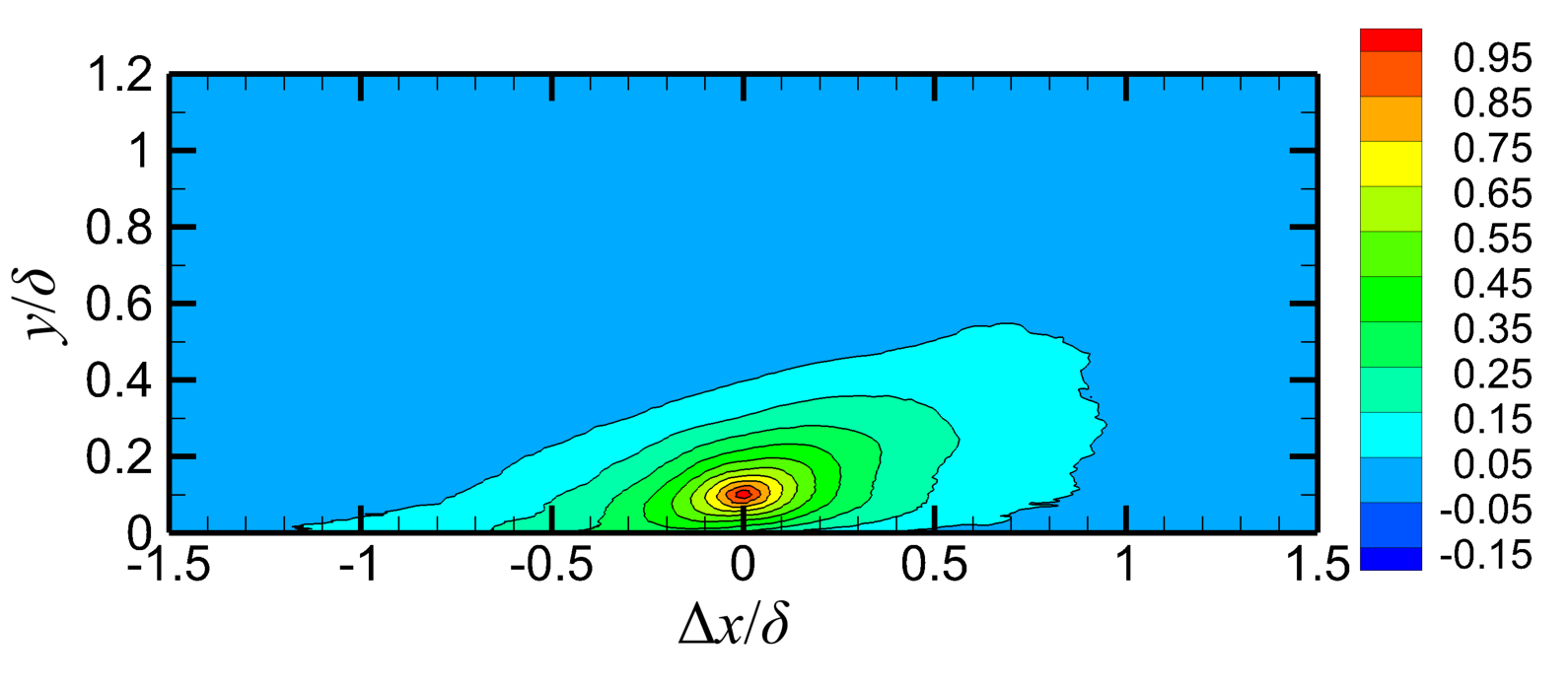}\label{up_corr_xy_a2}}}
\sidesubfloat[]{
{\includegraphics[width=0.46\textwidth,trim={0.2cm 0.2cm 0.2cm 0.3cm},clip]{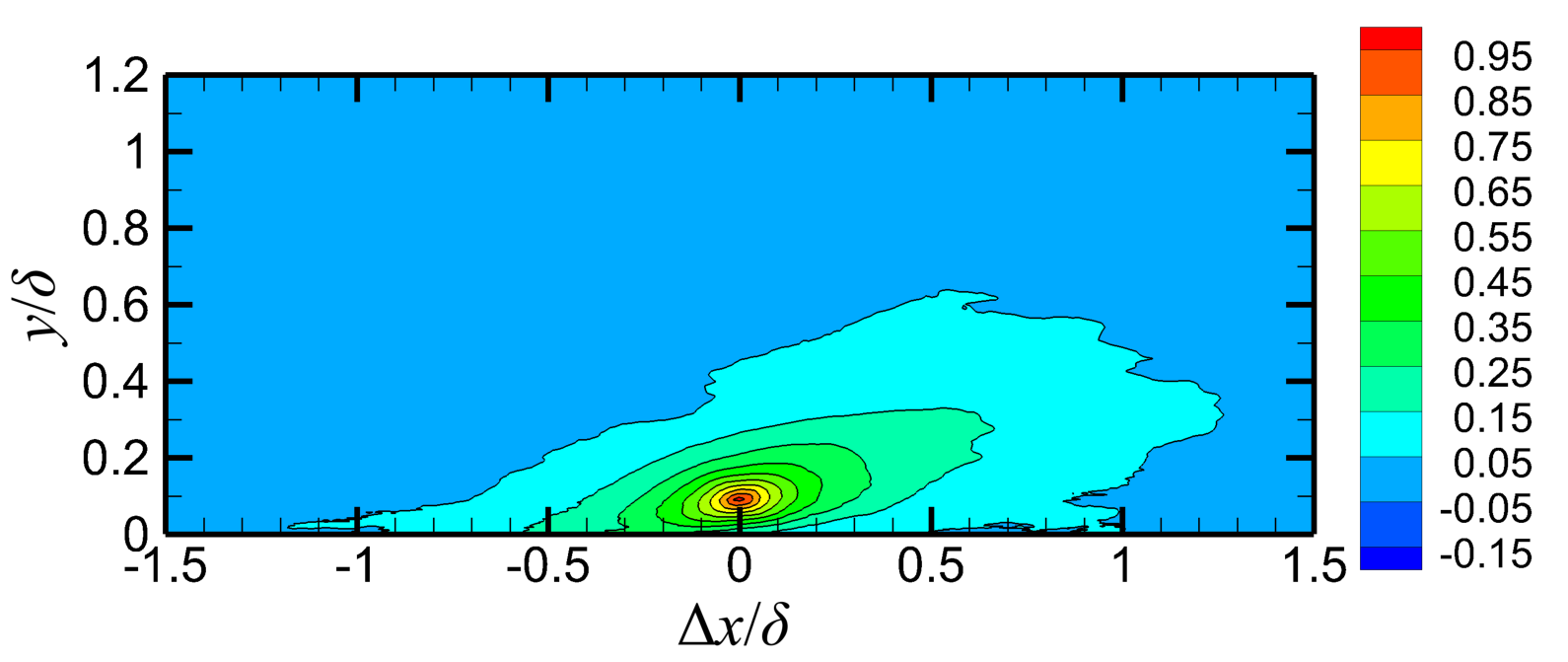}\label{up_corr_xy_b2}}}

\sidesubfloat[]{
{\includegraphics[width=0.46\textwidth,trim={0.2cm 0.1cm 0.2cm 0.5cm},clip]{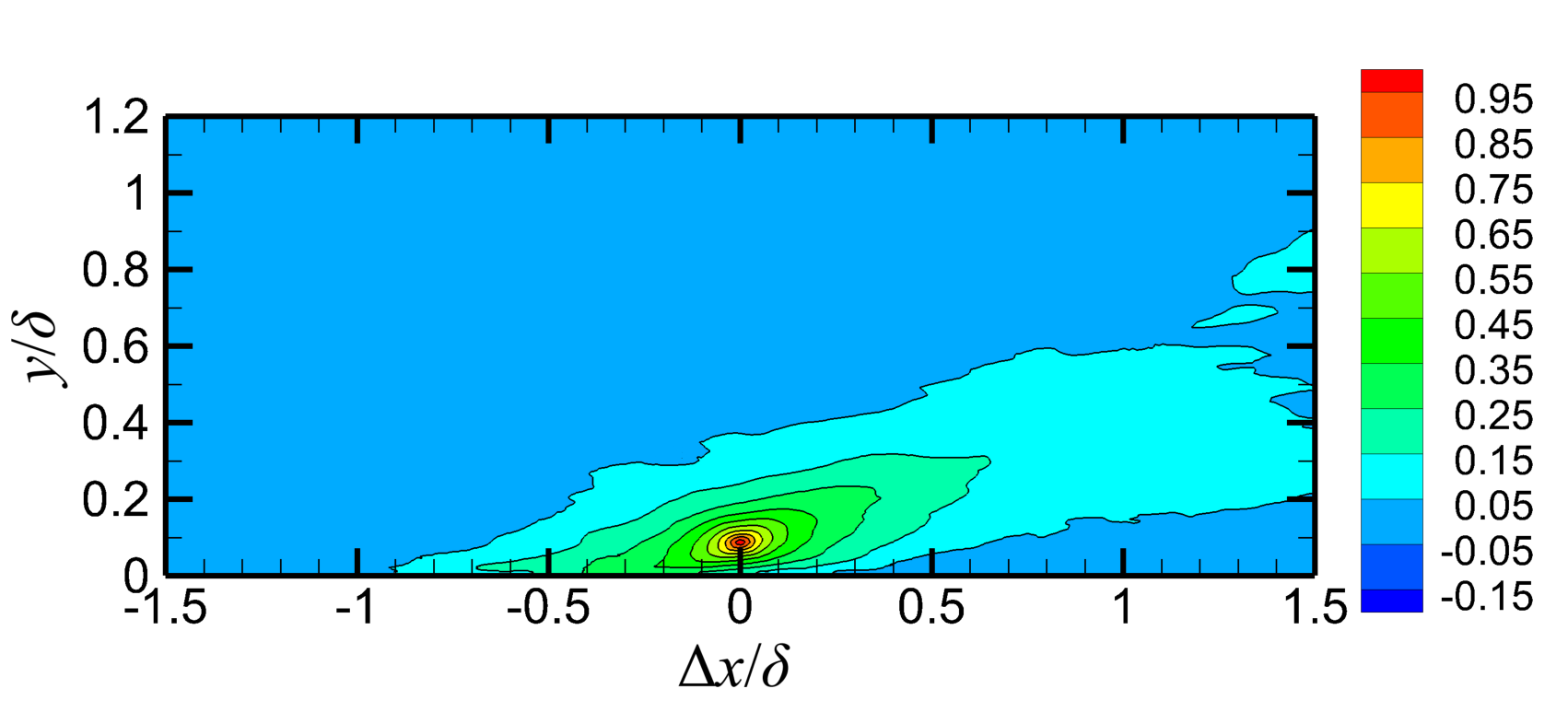}\label{up_corr_xy_c2}}}
\sidesubfloat[]{
{\includegraphics[width=0.46\textwidth,trim={0.2cm 0.2cm 0.1cm 0.3cm},clip]{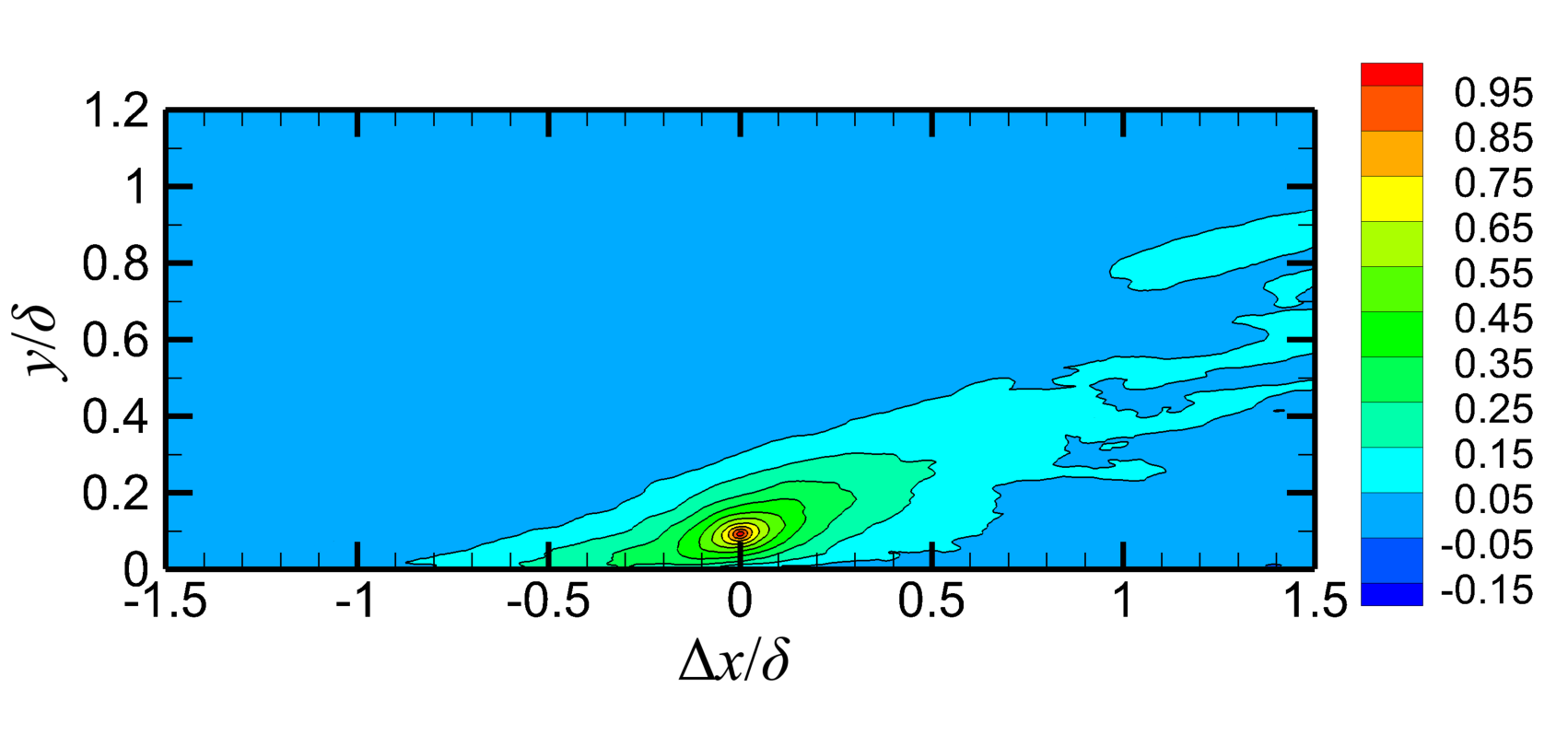}}\label{up_corr_xy_d2}}

\caption{Coefficients of two-point correlations of streamwise velocity fluctuations in the $x$-$y$ plane with anchor points located at $y/\delta=0.1$ and (a) $x/D=0.21$; (b) $x/D=0.43$; (c) $x/D=0.71$; (d) $x/D=0.99$.}
\label{up_corr_xy_01}
\end{figure}

To examine the size, shape and orientation of turbulence structures relative to the boundary-layer thickness, the two-point correlations in the $x$-$y$ plane in figure~\ref{up_04delta} are replotted in figure~\ref{up_corr_xy_04} with the coordinates normalized by the local $\delta$. Around all four anchor locations, the high-level contours have similar shapes and orientations while their enclosed areas decrease relative to $\delta$ in the downstream direction, indicating that small-scale turbulence is not strongly affected by the boundary-layer deceleration (see also figure~\ref{up_04delta}). On the other hand, the shapes of the lower level contours at larger separations vary greatly with $x$, becoming more elongated along the major axis of the oval-like contours. The contours span a large portion of the boundary layer at each streamwise station, and the correlation lengths relative to $\delta$, estimated by the decay of $C_{uu}$ (e.g., to $e^{-1}$), are only modestly decreased toward the downstream despite large changes in the low-level contour shapes. Figure~\ref{up_corr_xy_01} shows the two-point velocity correlations closer to the wall at $y/\delta=0.1$. Compared to their outer-region counterparts in figure~\ref{up_corr_xy_04}, the correlation contours are more elongated and lean more toward the wall with smaller inclination angles, which is consistent with the known hairpin vortex structure in TBLs. It should be mentioned that the $x$-$y$ correlations of the wall-normal and azimuthal velocity components, not shown here for brevity, exhibit the same trend of variations with the streamwise position as the streamwise velocity correlations.

\begin{figure}
\centering
{\psfrag{a}[][]{\large{$x/D$}}
\psfrag{b}[][]{\large{$\gamma$ (deg)}}
\includegraphics[width=.6\textwidth,trim={0.6cm 5.3cm 0.3cm 1.5cm},clip]{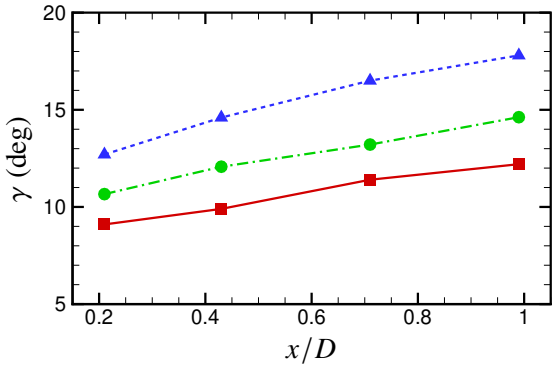}
}
\caption{Overall inclination angle $\gamma$ of $C_{uu}$ contours anchored at four streamwise stations and three wall-normal locations: \textcolor{red}{\rule{0.5em}{0.5em}}, $y/\delta=0.1$; {\sqcirgreen}, $y/\delta=0.2$; {\sqtri}, $y/\delta=0.4$.}
\label{inclination}
\end{figure}

The overall inclination angles of the correlations at $y/\delta=0.1$, $0.2$ and $0.4$ are quantified in figure~\ref{inclination} for the four streamwise stations. The angles are determined using the approach proposed by \citet{christensen2005characteristics} based on a least-squares fit of a line through the points farthest upstream and downstream of each of the $C_{uu}=0.35$, 0.45, 0.55, 0.65 and 0.75 isocontours. The angle between this fitted line and the positive $x$-axis, $\gamma$, is representative of the inclination of the hairpin vortex organization in the TBL \citep{christensen2005characteristics}. Consistent with the measurements of \citet{volino2020non} for APG TBLs on a 2-D ramp, the results indicate that the inclination angle increases in the downstream direction. At each streamwise station, the inclination angle increases with the wall-normal distance, and its value at $y/\delta=0.4$ exceeds the typical value of approximately $11^\circ$ in plane-wall turbulent channel flows and ZPG TBLs at the same position \citep{christensen2005characteristics, sillero2014two,volino2020non}. This increase is attributed to the effect of APG \citep{lee2017large, volino2020non}.

\begin{figure}
\centering

\sidesubfloat[]{
{\includegraphics[width=0.46\textwidth,trim={0.1cm 0.2cm 0.2cm 0.0cm},clip]{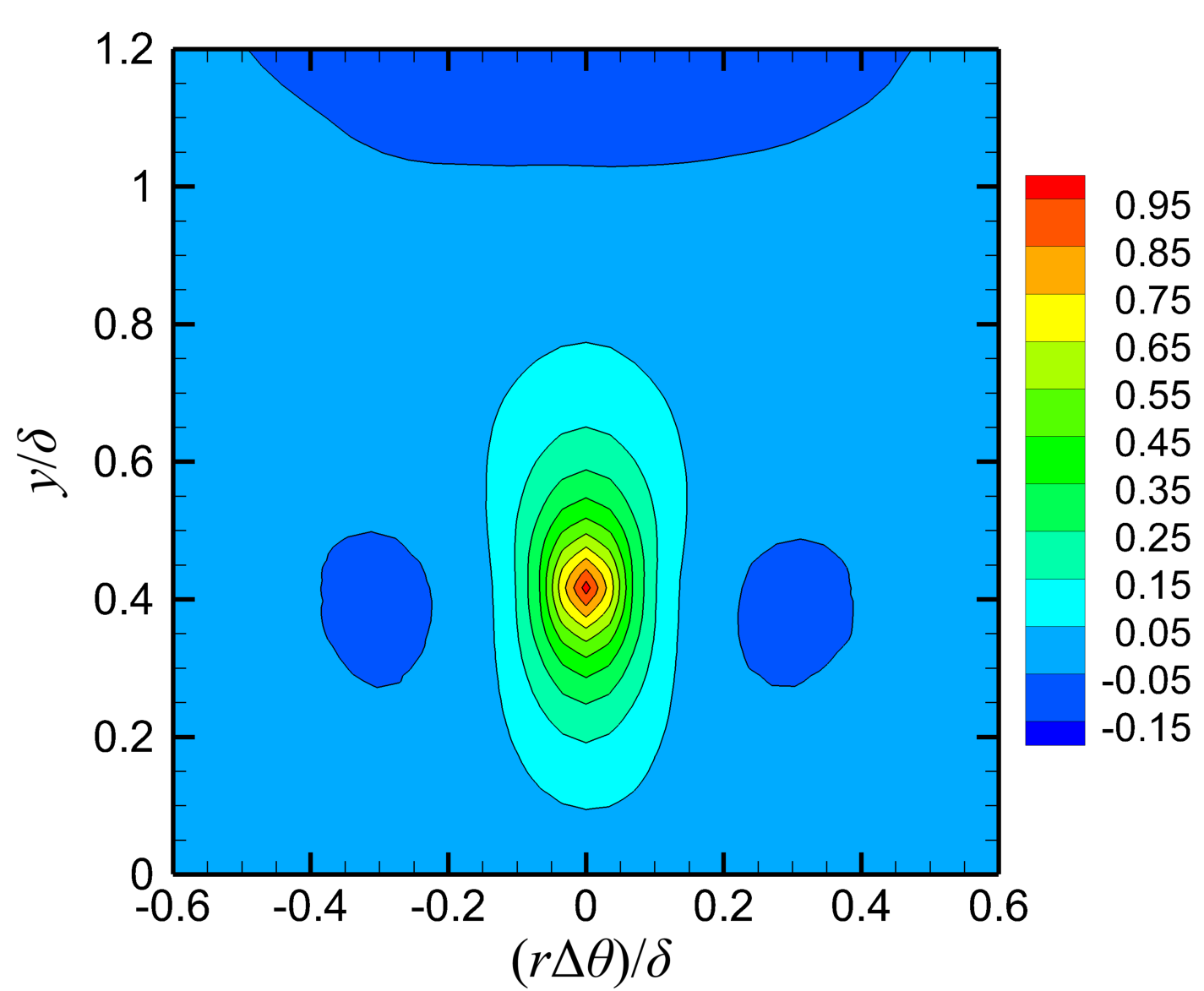}\label{up_corr_yt_a}}}
\sidesubfloat[]{
{\includegraphics[width=0.46\textwidth,trim={0.4cm 0.1cm 0.4cm 0.1cm},clip]{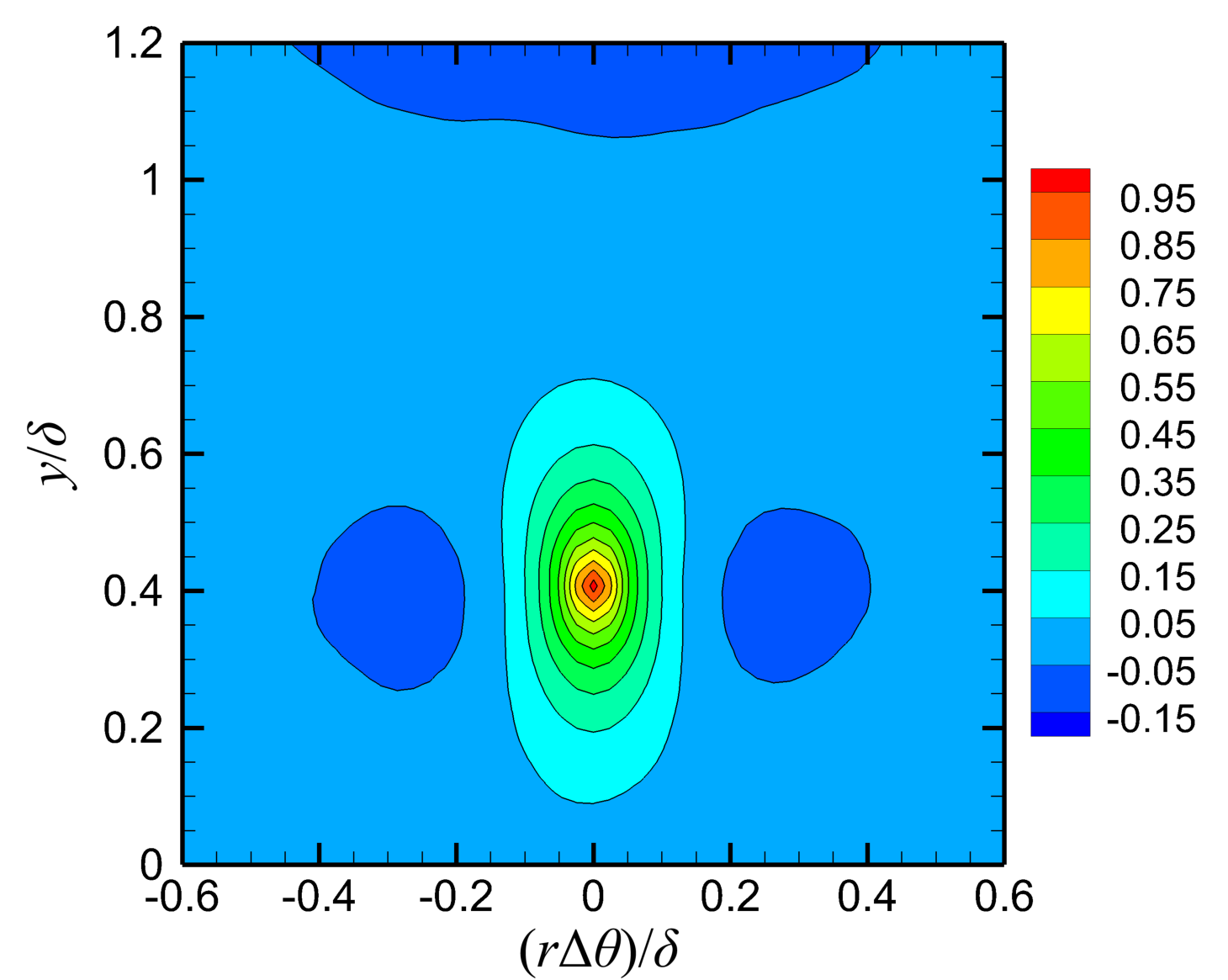}\label{up_corr_yt_b}}}

\sidesubfloat[]{
{\includegraphics[width=0.46\textwidth,trim={0.2cm 0.2cm 0.2cm 0.0cm},clip]{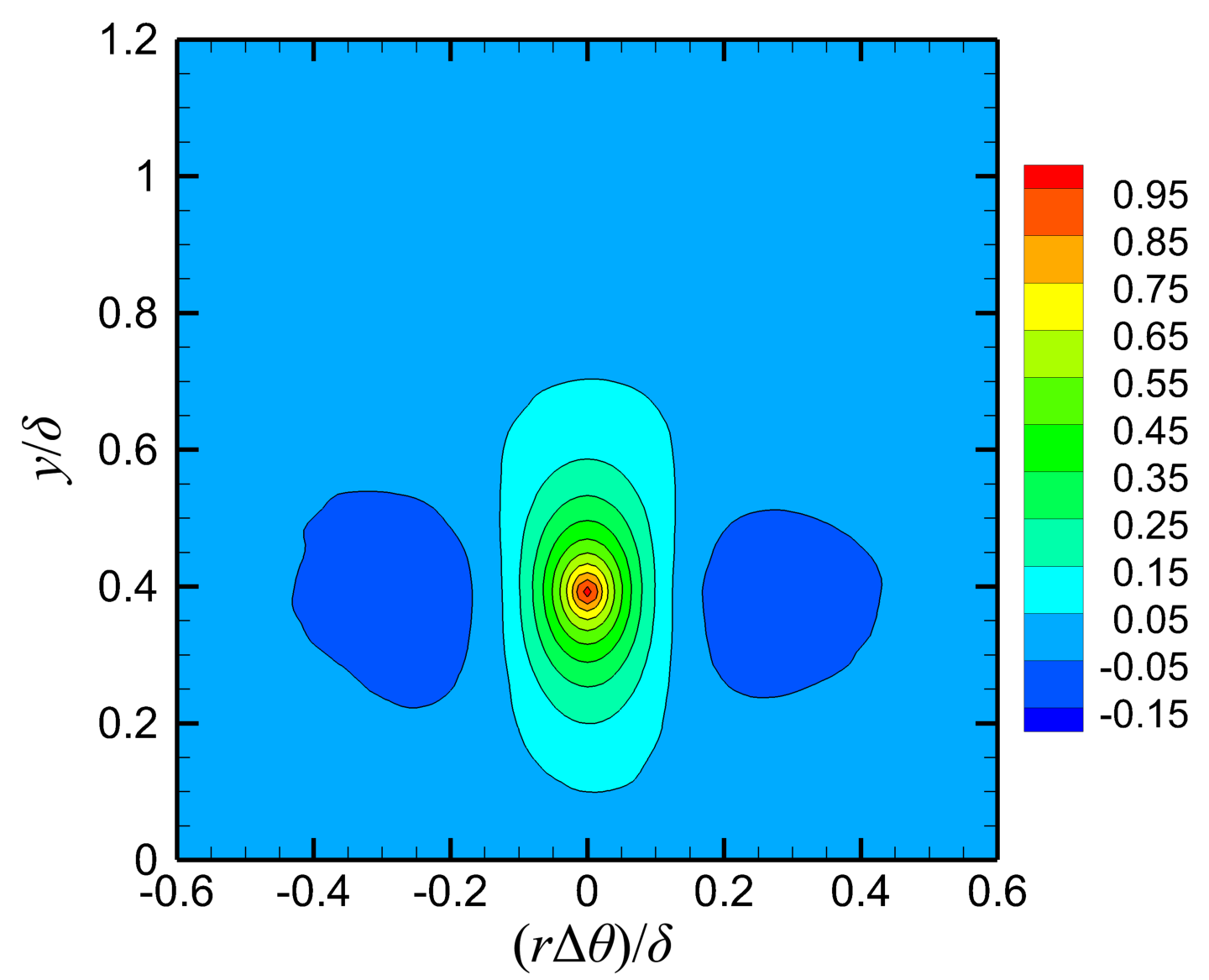}\label{up_corr_yt_c}}}
\sidesubfloat[]{
{\includegraphics[width=0.46\textwidth,trim={0.2cm 0.1cm 0.2cm 0.1cm},clip]{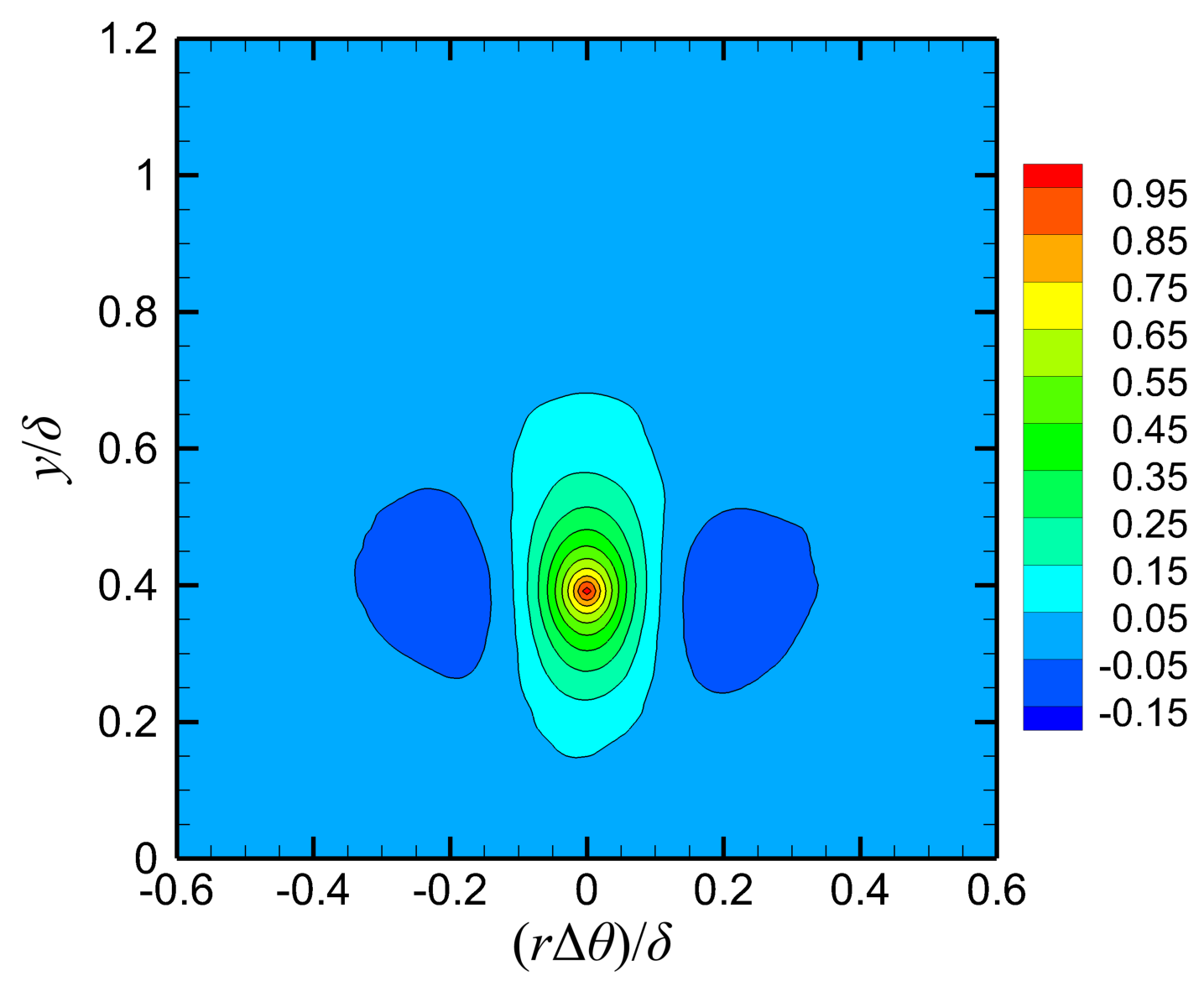}\label{up_corr_yt_d}}}

\caption{Coefficients of two-point correlations of streamwise velocity fluctuations in $y$-$\theta$ planes with anchor points located at $y/\delta=0.4$ and (a) $x/D=0.21$; (b) $x/D=0.43$; (c) $x/D=0.71$; (d) $x/D=0.99$.}
\label{up_corr_yt}
\end{figure}

Figure~\ref{up_corr_yt} shows $\delta$-scaled streamwise-velocity correlations in the cross-flow plane. As in the $x$-$y$ plane, the correlation contours exhibit modest decreases in length scales relative to the local boundary-layer thickness in the downstream direction. However, the shapes of the correlation contours, elongated in the wall-normal direction, are very similar at the four locations, in contrast to their counterparts in the $x$-$y$ plane. Negative lobes are observed on both sides of the positive contours. The distance between alternating positive and negative correlation regions decreases to approximately $0.4\delta$ at the last downstream station, which is much smaller than that in a ZPG flat-plate TBL \citep{sillero2014two}.

\begin{figure}
\centering

\sidesubfloat[]{
{\psfrag{x}[][]{{$\Delta x/\delta$}}
\psfrag{y}[][]{{$C_{uu}$}}\includegraphics[width=0.3\textwidth,trim={6.5cm 5.6cm 2.0cm 1.8cm},clip]{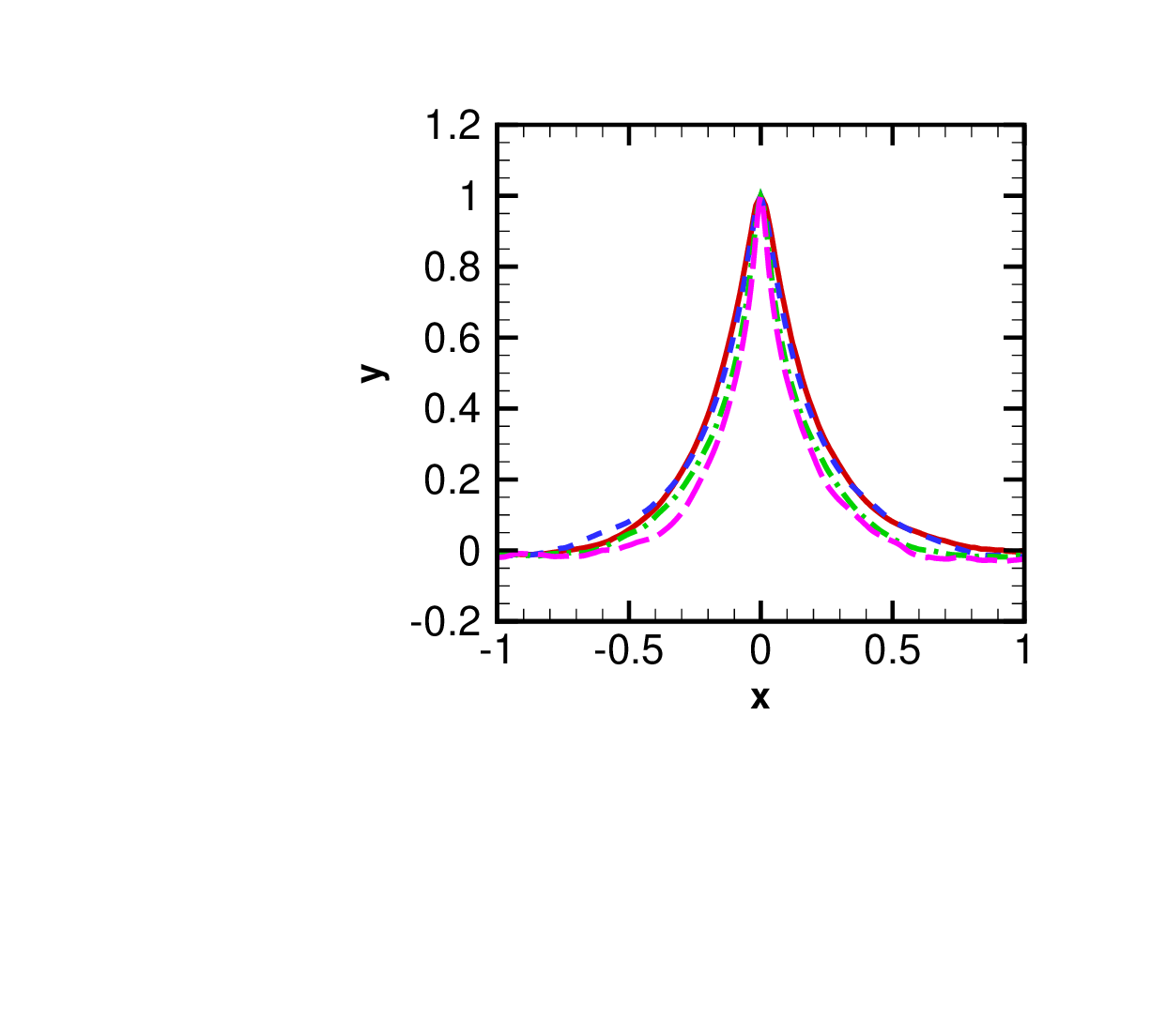}\label{up_corr_1d_a}}}
\sidesubfloat[]{
{\psfrag{x}[][]{{$\Delta y/\delta$}}
\psfrag{y}[][]{{$C_{uu}$}}\includegraphics[width=0.3\textwidth,trim={6.5cm 5.6cm 2.0cm 1.8cm},clip]{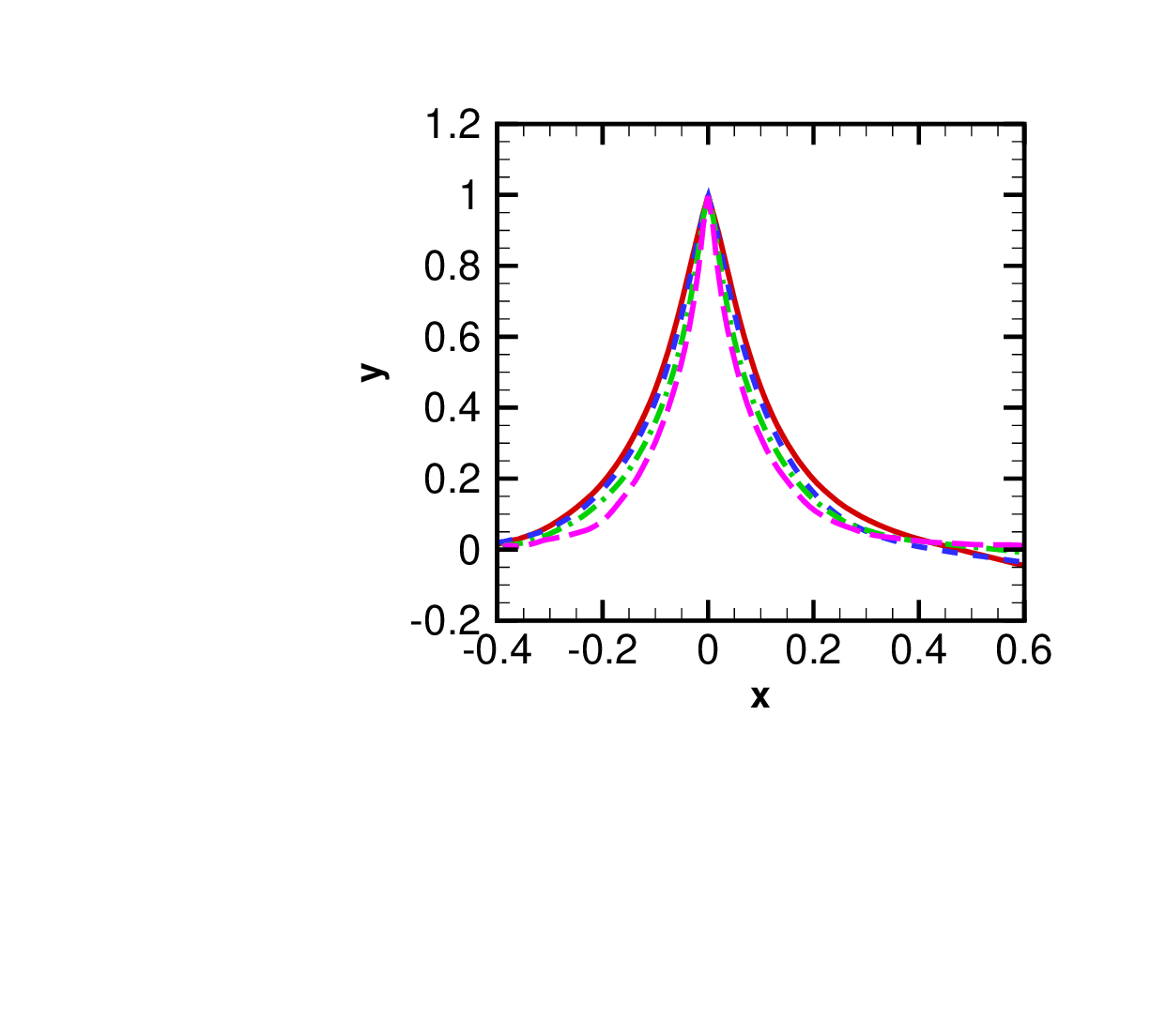}\label{up_corr_1d_b}}}
\sidesubfloat[]{
{\psfrag{x}[][]{{$(r\Delta \theta)/\delta$}}
\psfrag{y}[][]{{$C_{uu}$}}\includegraphics[width=0.3\textwidth,trim={6.5cm 5.6cm 2.0cm 1.8cm},clip]{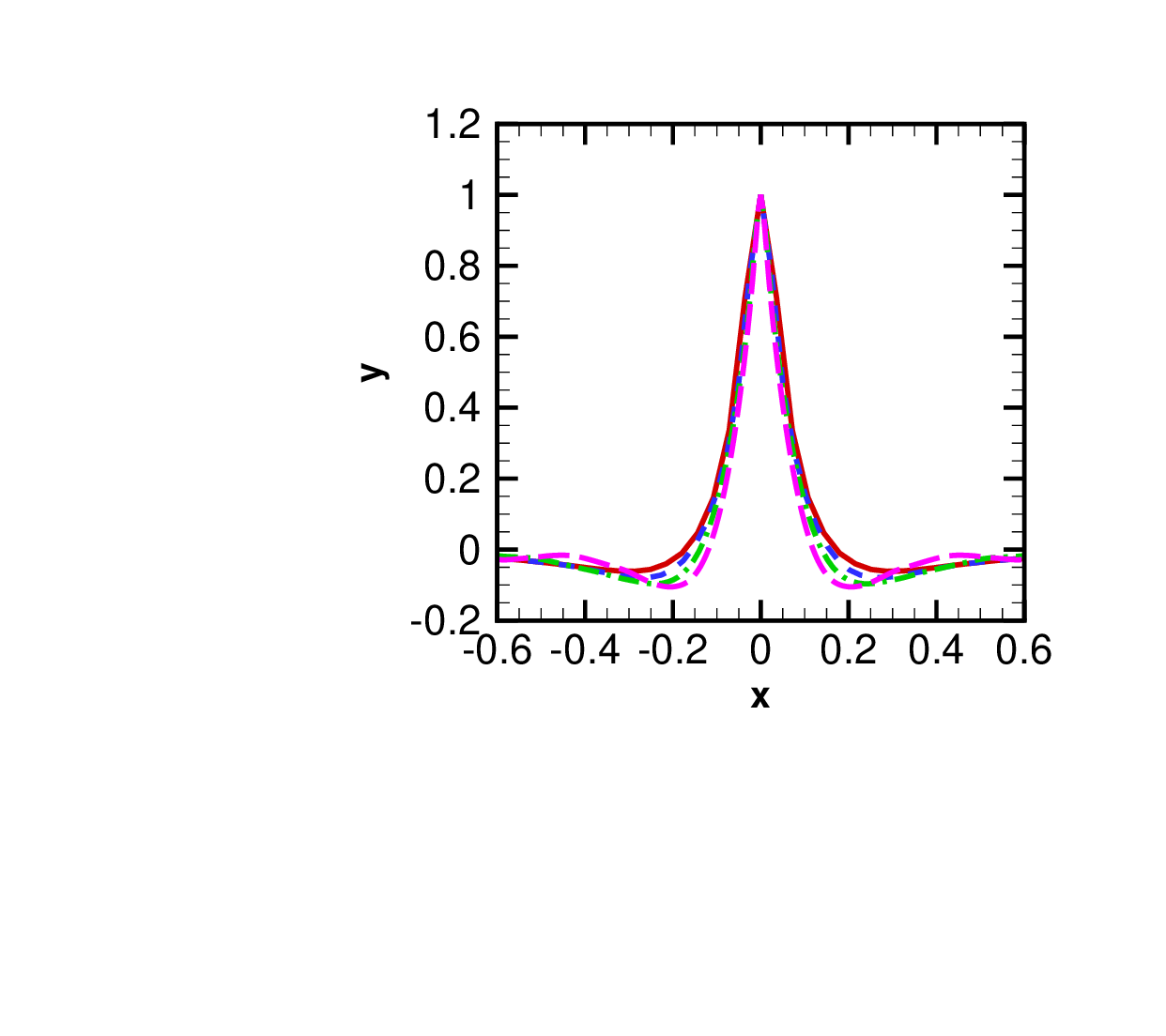}\label{up_corr_1d_c}}}

\caption{Correlation coefficients of streamwise velocity fluctuations as a function of (a) streamwise separation, (b) wall-normal separation, and (c) azimuthal separation. The lines in each plot represent different anchor locations at $y/\delta=0.4$ and: {\redsolid}, $x/D=0.21$; {\bluedashed}, $x/D=0.43$; {\greendashdotted}, $x/D=0.71$; {\magentadashed}, $x/D=0.99$.}
\label{up_corr_1d}
\end{figure}

\begin{figure}
\centering

\sidesubfloat[]{
{\psfrag{x}[][]{{$\Delta x/\delta$}}
\psfrag{y}[][]{{$C_{uu}$}}\includegraphics[width=0.3\textwidth,trim={6.5cm 5.6cm 2.0cm 1.8cm},clip]{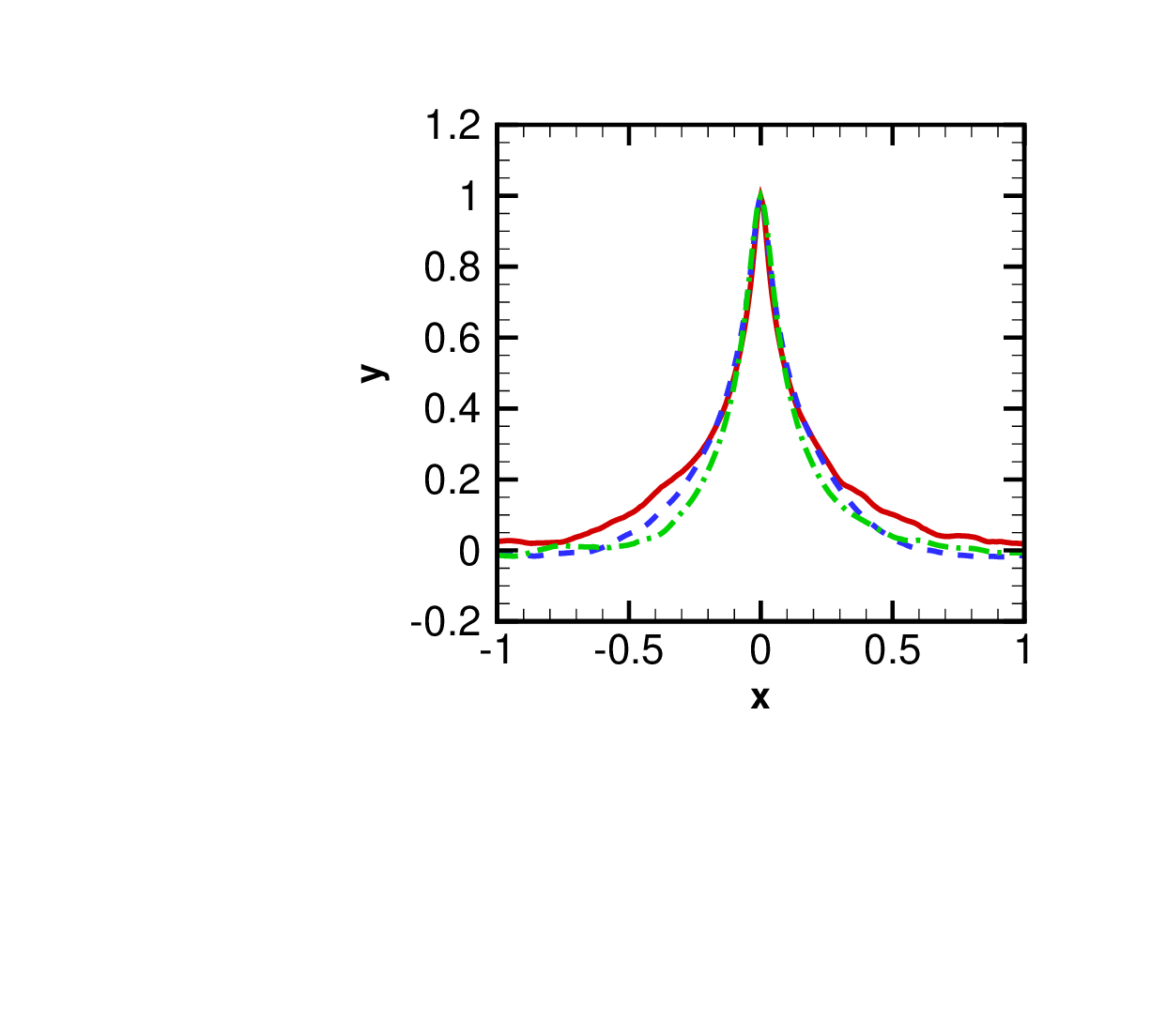}\label{up_corr_7th_a}}}
\sidesubfloat[]{
{\psfrag{x}[][]{{$\Delta y/\delta$}}
\psfrag{y}[][]{{$C_{uu}$}}\includegraphics[width=0.3\textwidth,trim={6.5cm 5.6cm 2.0cm 1.8cm},clip]{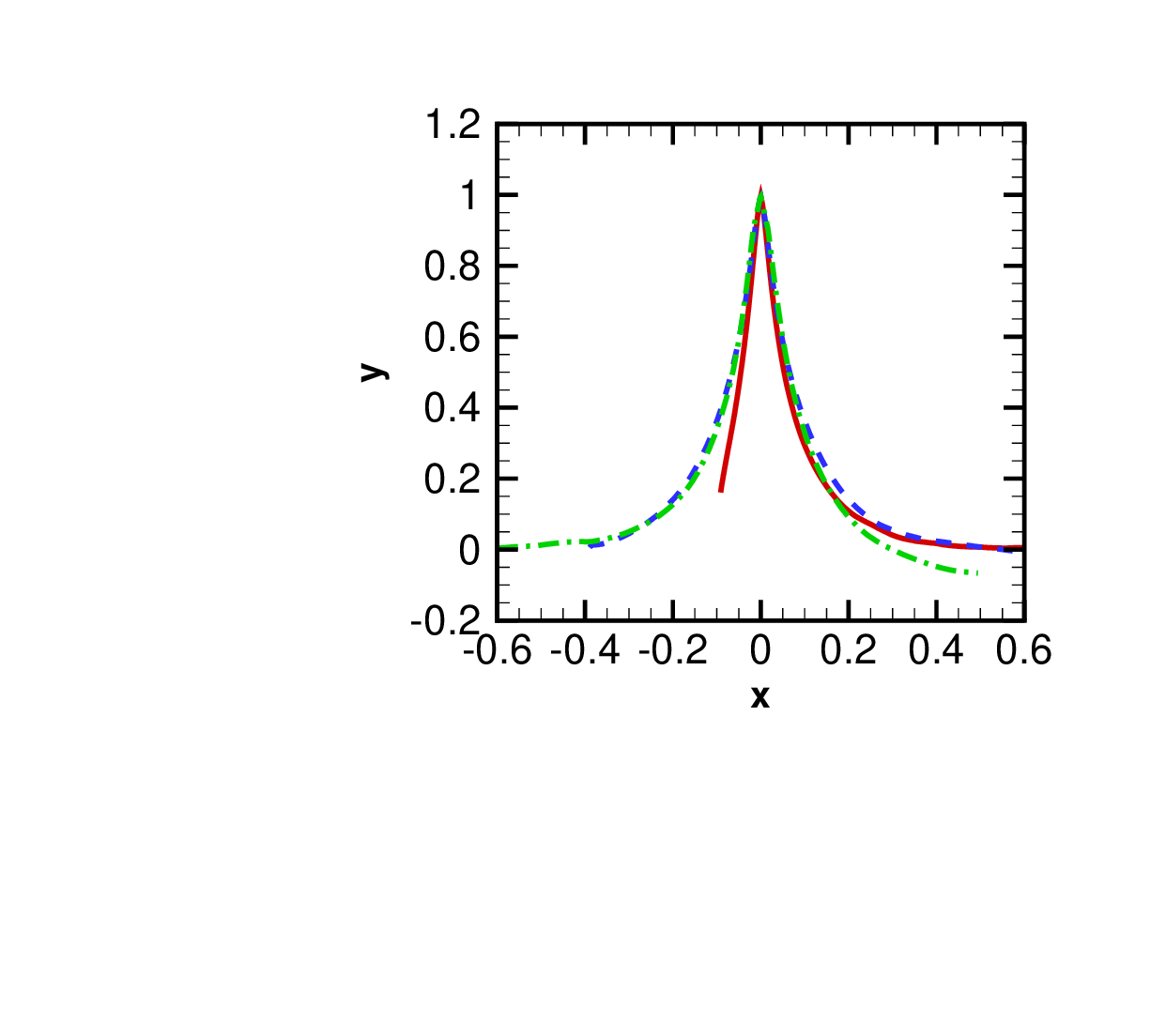}\label{up_corr_7th_b}}}
\sidesubfloat[]{
{\psfrag{x}[][]{{$(r\Delta \theta)/\delta$}}
\psfrag{y}[][]{{$C_{uu}$}}\includegraphics[width=0.3\textwidth,trim={6.5cm 5.6cm 2.0cm 1.8cm},clip]{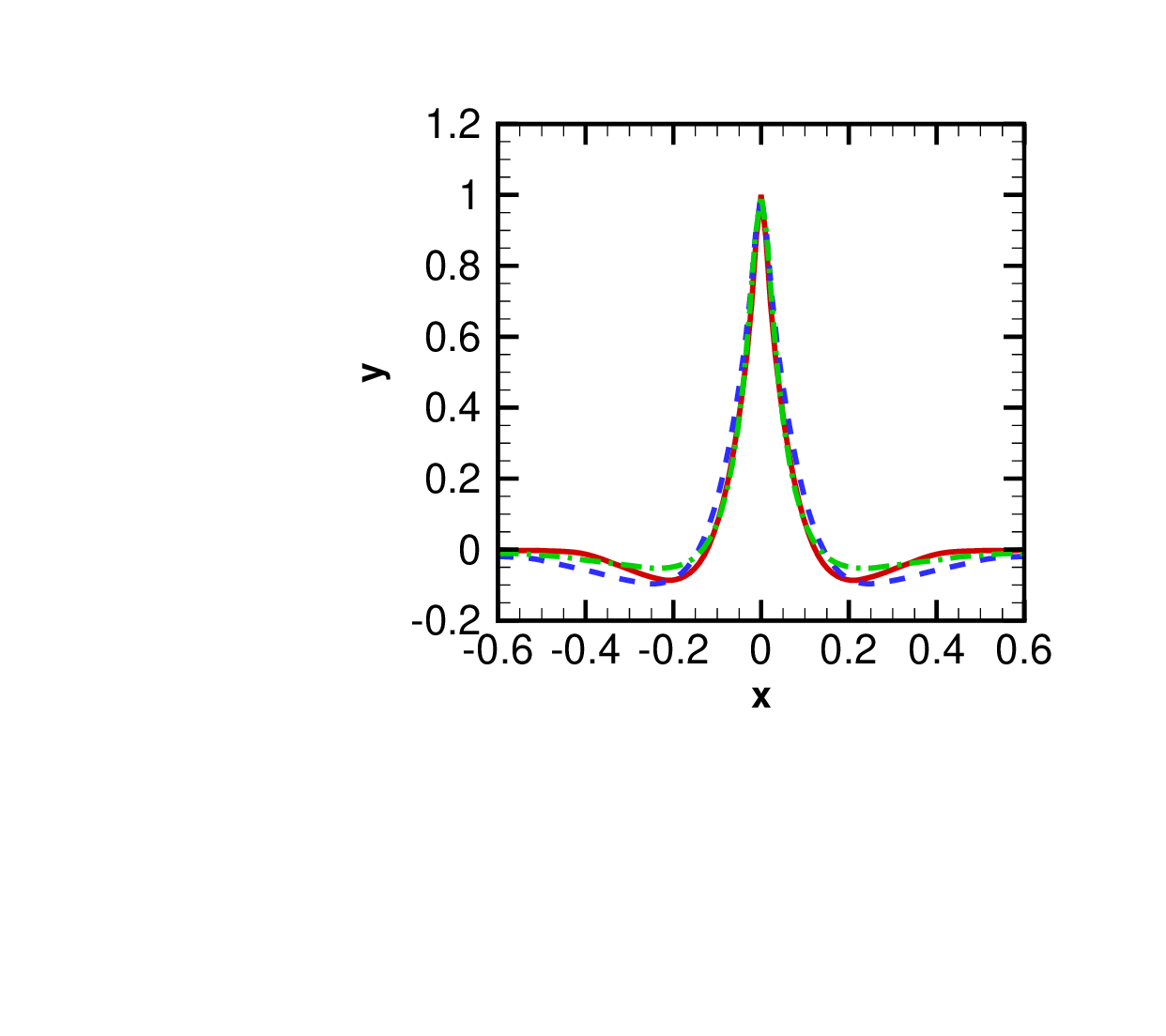}\label{up_corr_7th_c}}}

\caption{Correlation coefficients of streamwise velocity fluctuations as a function of (a) streamwise separation, (b) wall-normal separation, and (c) azimuthal separation. The lines in each plot represent different anchor locations at $x/D=0.71$ and: {\redsolid}, $y/\delta=0.1$; {\bluedashed}, $y/\delta=0.4$; {\greendashdotted}, $y/\delta=0.7$.}
\label{up_corr_7th}
\end{figure}

The spatial correlations at different positions in the boundary layer are compared more quantitatively in figures~\ref{up_corr_1d} and \ref{up_corr_7th}. Figure~\ref{up_corr_1d} illustrates the correlations anchored at the previous four streamwise locations at $y/\delta=0.4$ as one-dimensional functions of spatial separations. Each plot corresponds to separation in one direction, and the four curves in each plot represent the four anchor locations. It can be seen that for all anchor locations, the correlation length in the streamwise direction is larger than that in the wall-normal direction (note the different abscissa ranges used in the plots for clarity), and the azimuthal correlation length is the smallest. As the anchor point moves downstream, a reduction in correlation lengths relative to $\delta$ is evident in all three directions. In relative terms, the azimuthal correlation is less changed than the correlations in the other two directions. Additionally, as shown in figure~\ref{up_corr_1d_c} the negative lobes in the azimuthal correlations become more pronounced and the separation distance is reduced toward downstream. In figure~\ref{up_corr_7th}, the one-dimensional correlations anchored at three different wall-normal locations at $x/D = 0.71$ are compared. The correlations are remarkably similar at the three vastly different positions: $y/\delta=0.1$, 0.4 and 0.7. For the streamwise correlation, a small reduction with increasing distance to the wall can be noticed at large separations, leading to a small decrease of the correlation length. The correlations in the other two directions barely vary with the wall-normal distance. These findings are consistent with the results of the flat-plate TBL with strong APG reported by \citet{lee2017large}.

\subsection{Integral length scales}

\begin{figure}
\centering

\sidesubfloat[]{
{\psfrag{x}[][]{{$y/\delta$}}
\psfrag{y}[][]{{$\Lambda_x/\delta$}}\includegraphics[width=0.3\textwidth,trim={3.3cm 1.0cm 5.5cm 0.8cm},clip]{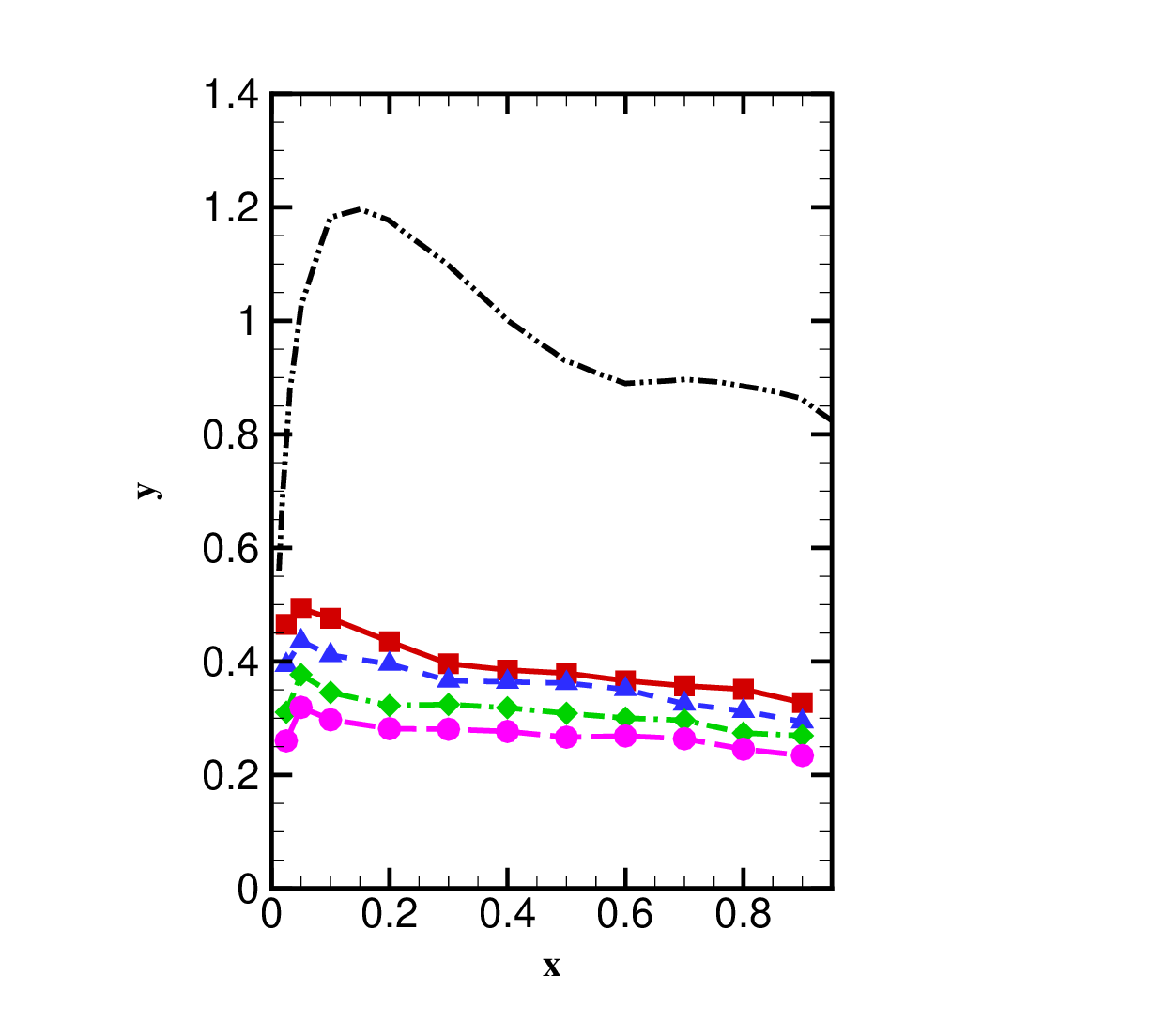}\label{up_corr_xloc_outer_a}}}
\sidesubfloat[]{
{\psfrag{x}[][]{{$y/\delta$}}
\psfrag{y}[][]{{$\Lambda_y/\delta$}}\includegraphics[width=0.3\textwidth,trim={3.3cm 1.0cm 5.5cm 0.8cm},clip]{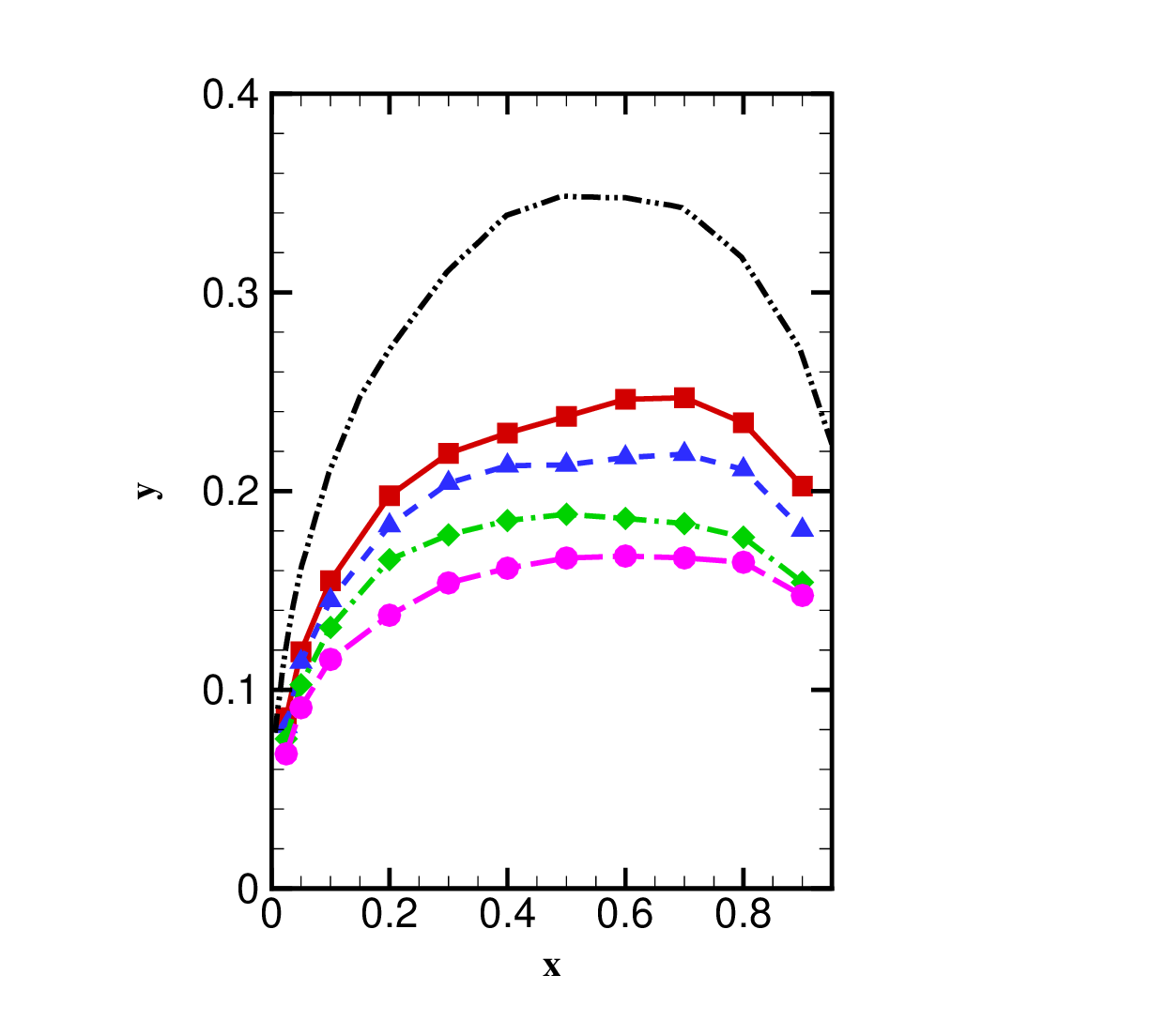}\label{up_corr_xloc_outer_b}}}
\sidesubfloat[]{
{\psfrag{x}[][]{{$y/\delta$}}
\psfrag{y}[][]{{$\Lambda_{\theta}/\delta$}}\includegraphics[width=0.3\textwidth,trim={3.3cm 1.0cm 5.5cm 0.8cm},clip]{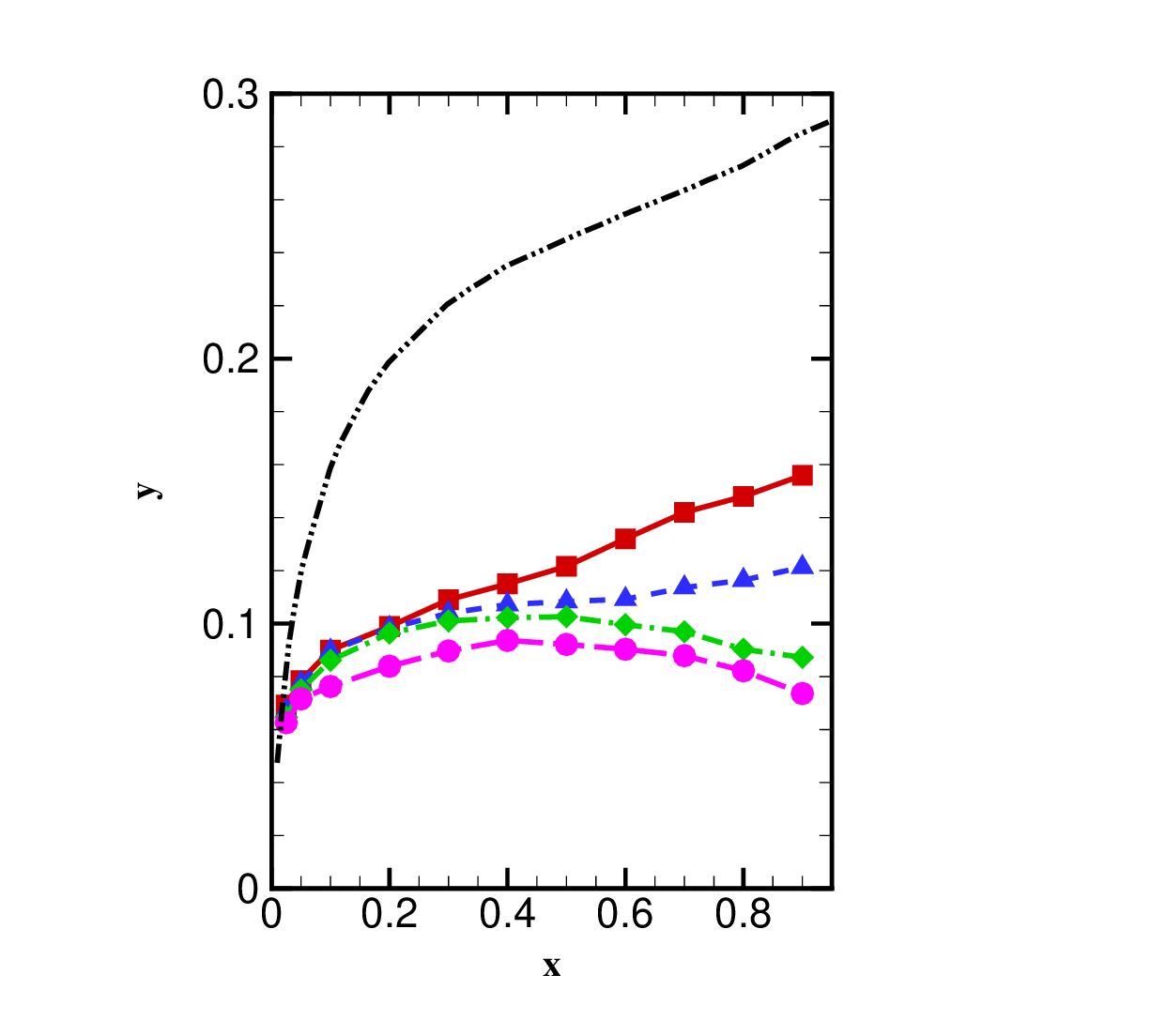}\label{up_corr_xloc_outer_c}}}

\caption{Outer-scaled integral lengths in (a) streamwise, (b) wall-normal, and (c) azimuthal directions as a function of $y$ at four streamwise locations: {\redsolid} with \textcolor{red}{\rule{0.5em}{0.5em}}, $x/D=0.21$; {\bluedashed} with {\sqtri}, $x/D=0.43$; {\greendashdotted} with {\sqdiamond}, $x/D=0.71$; {\magentadashed} with {\sqcirmag}, $x/D=0.99$. The chain-dot-dot lines are from the DNS of a ZPG flat-plate TBL at $Re_\tau \approx 2000$ \citep{sillero2014two}, with the azimuthal correlation replaced by spanwise correlation.}
\label{up_corr_xloc_outer}
\end{figure}

To further elucidate the three-dimensional structure of the boundary layer, the integral lengths (or correlation lengths) along the streamwise, wall-normal and azimuthal directions are evaluated based on the spatail correlations as 
\begin{equation}
    \Lambda_x=\scaleobj{.8}{\int} C_{uu}(x, y, \Delta x) \mathrm{d}(\Delta x),
\end{equation}
\begin{equation}
    \Lambda_y=\scaleobj{.8}{\int} C_{uu}(x, y, \Delta y) \mathrm{d}(\Delta y),
\label{length_y}    
\end{equation}
\begin{equation}
    \Lambda_\theta=\scaleobj{.8}{\int} C_{uu}(x, y, \Delta \theta) r \mathrm{d}(\Delta \theta),
\end{equation}
respectively.  The integrations are taken between the intersections with $C_{uu}=0.05$, and in the wall-normal direction it is bounded by $y/\delta=0$ and 1 at the two ends. This definition is consistent with that used by \citet{sillero2014two} to facilitate a comparison with their flat-plate TBL results. Figure~\ref{up_corr_xloc_outer} shows the outer-scaled integral lengths of the streamwise velocity as functions of the wall-normal distance at four streamwise locations, along with the DNS data of \citet{sillero2014two} for a ZPG flat-plate TBL at $Re_\tau=2000$. The integral lengths in the APG axisymmetric TBL are all markedly shorter than those in the ZPG TBL. Consistent with the observation from figure~\ref{up_corr_1d}, all the integral lengths decrease in the downstream direction. The wall-normal variations of streamwise and wall-normal integral lengths are similar among the different streamwise locations, and exhibit largely the same trend as those in the ZPG TBL. In the near-wall region, however, the streamwise integral lengths in the APG axisymmetric TBL show a much weaker growth in $y$ and reach peak values closer to the wall than that of the ZPG TBL. Note that the decline of the wall-normal correlation lengths near the boundary-layer edge is caused by capping the upper bound of the integral (\ref{length_y}) at $y/\delta=1$, rather than a true decrease in the physical length scale. For the azimuthal integral length, similar values across different streamwise locations are observed within the near-wall region. However, their wall-normal variations in the outer region show diverging trends for different streamwise locations. At $x/D=0.21$, the integral length increases almost linearly away from the wall at a rate similar to that of the reference ZPG TBL. This rate of increase is much reduced at $x/D=0.43$. Further downstream, at $x/D=0.71$ and $0.99$, the integral lengths begin to decrease above a wall-normal distance of $y/\delta\approx 0.4$. This qualitatively different behavior is likely due to the increasing effect of transverse curvature towards the end of the tail cone.

\begin{figure}
\centering

\sidesubfloat[]{
{\psfrag{x}[][]{{$y/\delta$}}
\psfrag{y}[][]{{$\Lambda_x^+$}}\includegraphics[width=0.3\textwidth,trim={3.3cm 1.0cm 5.5cm 0.8cm},clip]{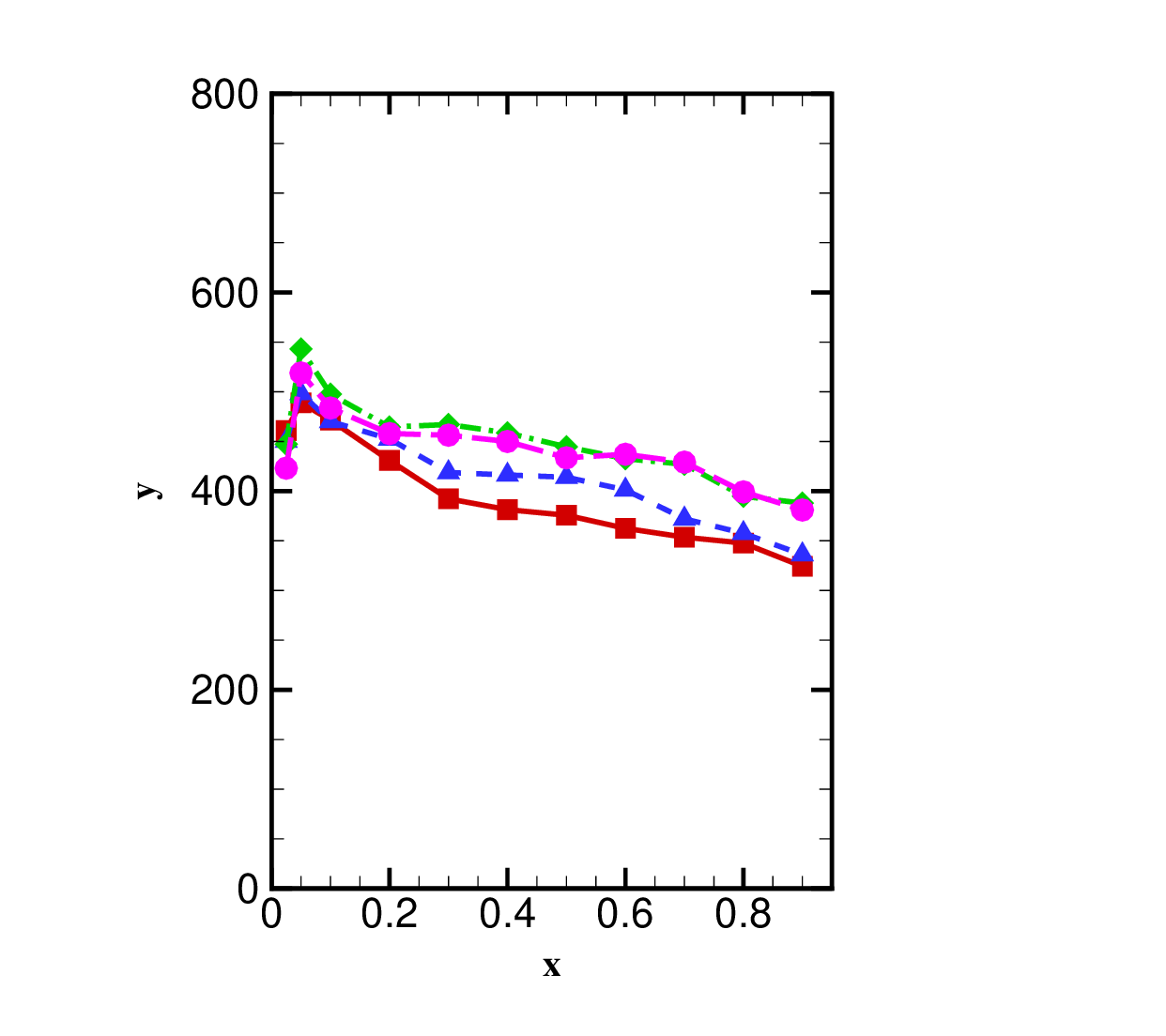}\label{up_corr_xloc_inner_a}}}
\sidesubfloat[]{
{\psfrag{x}[][]{{$y/\delta$}}
\psfrag{y}[][]{{$\Lambda_y^+$}}\includegraphics[width=0.3\textwidth,trim={3.3cm 1.0cm 5.5cm 0.8cm},clip]{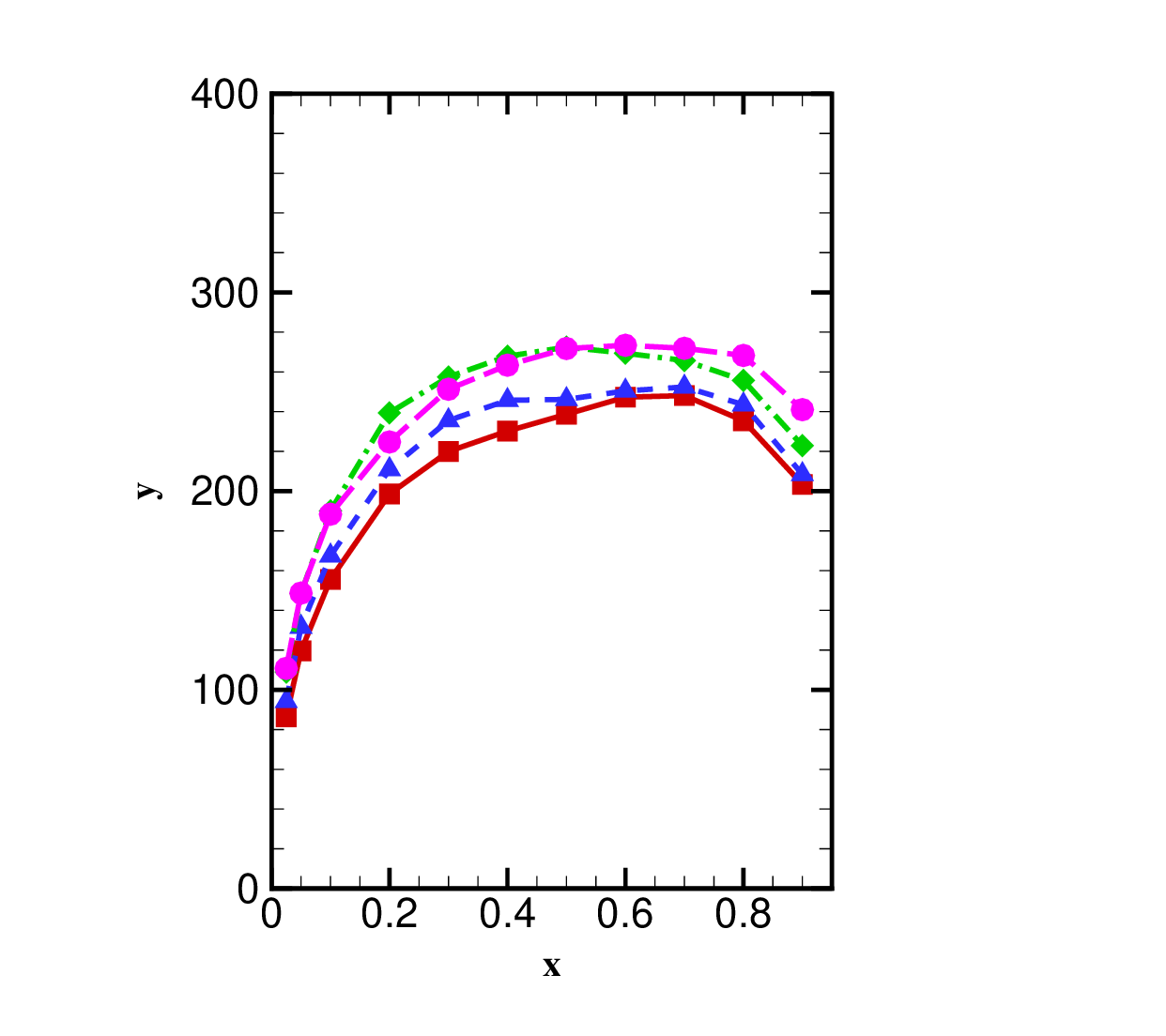}\label{up_corr_xloc_inner_b}}}
\sidesubfloat[]{
{\psfrag{x}[][]{{$y/\delta$}}
\psfrag{y}[][]{{$\Lambda_{\theta}^+$}}\includegraphics[width=0.3\textwidth,trim={3.3cm 1.0cm 5.5cm 0.8cm},clip]{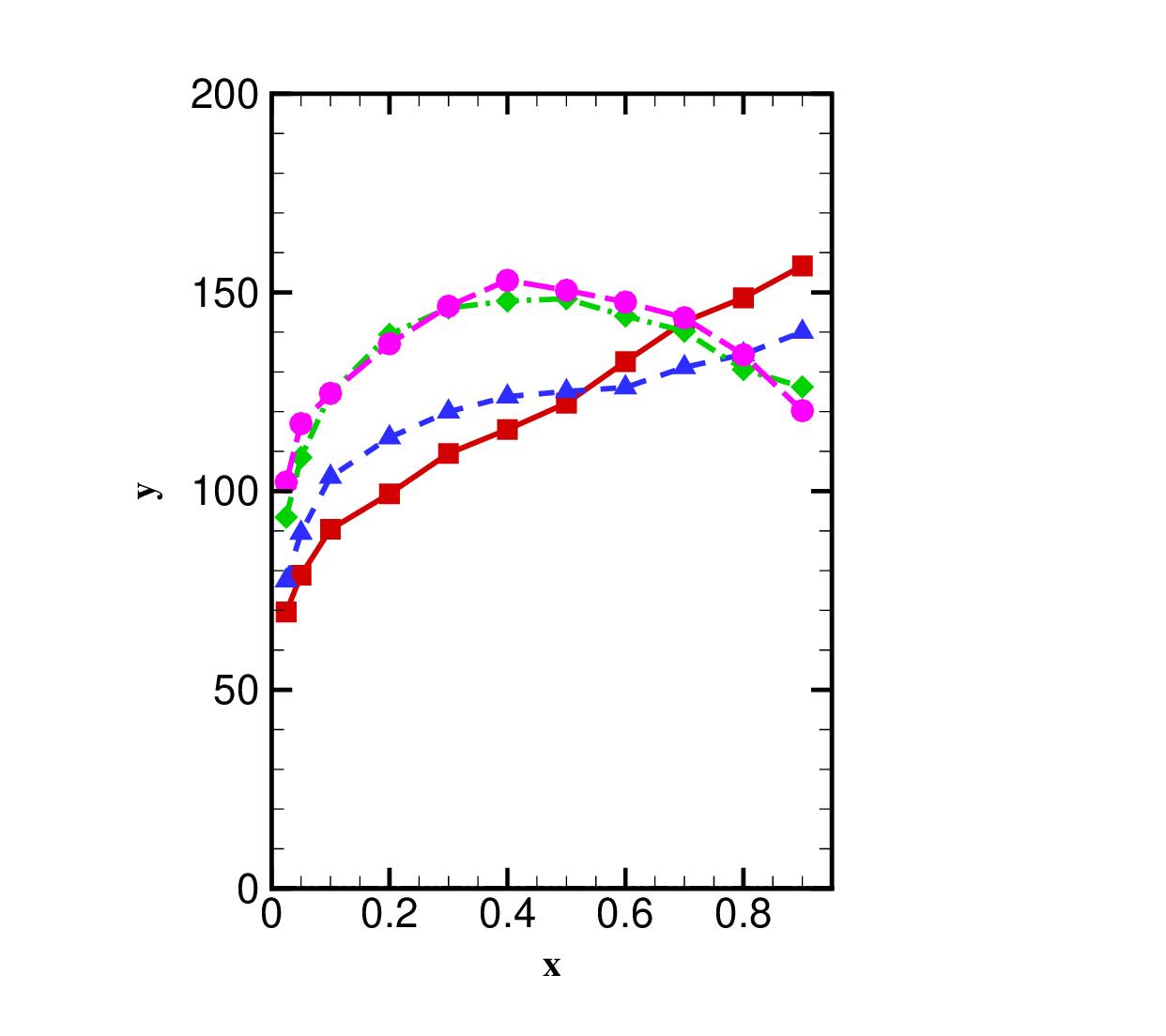}\label{up_corr_xloc_inner_c}}}

\caption{Inner-scaled integral lengths in (a) streamwise, (b) wall-normal, and (c) azimuthal directions as a function of $y$ at four streamwise locations: {\redsolid} with \textcolor{red}{\rule{0.5em}{0.5em}}, $x/D=0.21$; {\bluedashed} with {\sqtri}, $x/D=0.43$; {\greendashdotted} with {\sqdiamond}, $x/D=0.71$; {\magentadashed} with {\sqcirmag}, $x/D=0.99$.}
\label{up_corr_xloc_inner}
\end{figure}

In figure~\ref{up_corr_xloc_inner}, the distributions of the integral lengths in wall units are depicted. A notable observation is that their streamwise variations exhibit a trend opposite to that of their outer-scaled counterparts seen in figure~\ref{up_corr_xloc_outer}. The inner-scaled integral lengths show a mild increase in the downstream direction except for the azimuthal integral length in the outer region of the boundary layer, where the opposite is noted.

\section{Conclusion}\label{sec:conclusion}

In this study, large-eddy simulation has been employed to investigate the characteristics and structure of an axisymmetric TBL under strong APG. The TBL is on the tail cone of a BOR at zero angle of attack, as in the experimental setup described by \citet{balantrapu2021structure, balantrapu2023wall}. The BOR length is 3.17 times the diameter $D$, of which the last 1.17$D$ is the tail cone that has a $20^\circ$ half apex angle. The Reynolds number based on the free-stream velocity and the BOR length is $1.9\times 10^6$. To reduce the computational expense, an equilibrium wall model is employed in the nose and midsection of the BOR, while the tail-cone section is computed with wall-resolved LES. Detailed comparisons of velocity and wall-pressure statistics between the numerical and experimental results have verified the accuracy of the simulation, and demonstrated the efficacy of the zonal wall-modeled LES approach in accurately capturing the spatio-temporal characteristics of the nonequilibrium, axisymmetric TBL under the influences of APG and transverse curvature. The simulation data significantly enhances the experimental measurements documented by \citet{balantrapu2021structure, balantrapu2023wall} for this fundamentally interesting flow, and provides new insights into the turbulence physics.

Utilizing flow-field data from the LES, the statistics of the nonequilibrium axisymmetric TBL on the tail cone and the evolution of turbulence structures are investigated in a region where the momentum-thickness Reynolds number varies from 3,420 to 12,600 and the friction Reynolds number varies from 955 to 1,650. The Clauser pressure-gradient parameter $\beta$ based on the wall pressure increases progressively downstream from 4.61 to 11.2, whereas its value based on the pressure at the boundary-layer edge is within a much narrower range of 4.15 to 6.03 due to the increasing pressure variation across the decelerating boundary layer.The transverse curvature parameters $\delta/r_s$ and $r_s^+$ are within 0.12 to 0.94 and 1,760 to 8,260, respectively, indicating that the curvature effect is relatively small.  The major findings of this investigation are:

\begin{itemize}
    \item Similar to planar TBLs under APG \citep{harun2011boudnary}, the inner-scaled mean streamwise velocity profiles of the axisymmetric TBL exhibit a distinguishable logarithmic region. However, this region is characterized by a shorter range and a slightly steeper slope than the classical logarithmic law. As the APG grows downstream, the profiles fall below the standard logarithmic law, leading to a more pronounced wake region.

    \item With the embedded-shear-layer scaling \citep{schatzman2017experimental}, the mean streamwise velocity profiles demonstrate very good self-similarity over a wide range of wall-normal distance, confirming the scaling results of \citet{balantrapu2021structure}. The three components of turbulence intensity also exhibit a reasonable degree of self-similarity. However, the application of embedded-shear-layer scaling to Reynolds shear stress is more restrictive and only valid in a smaller portion of the outer region.

    \item The pre-multiplied azimuthal-wavenumber spectra of streamwise velocity fluctuations have two distinct peaks along the wall-normal coordinate.  The inner peak, located at $y^+\approx 12$, is characterized by a mean azimuthal wavelength of approximately 100 wall units. This wavelength, indicative of the average spacing between near-wall elongated streaks, is similar to that in ZPG planar TBLs but shows a slight decrease in the downstream direction as the intensity of the peak decays significantly. The outer peak, positioned in the wake region, increases in strength, wavelength and distance from the wall in the downstream direction as increasingly larger-scale motions are energized. The azimuthal wavelength of the peak is less than one half of the corresponding spanwise wavelength in a ZPG TBL. 

    \item Large-scale turbulence structures based on the two-point correlations of streamwise velocity fluctuations show rapid growth and elongation in the thickening axisymmetric TBL. However, relative to the local boundary-layer thickness, the structures decrease in size toward downstream, accompanied by increasing inclination angles that are significantly larger than typical values in plane channel flows and ZPG TBLs.

    \item The integral lengths in all three directions relative to $\delta$ are qualitatively similar to, but substantially smaller than those in ZPG planar TBLs at comparable Reynolds numbers. While the streamwise integral length decreases with the wall-normal distance in the outer layer, the wall-normal and azimuthal integral lengths increase with the wall-normal distance except at downstream stations, where the trend is reversed in the outer region for the azimuthal length.
    
\end{itemize}

\section*{Acknowledgments}

This research was supported by the Office of Naval Research under grants N00014-17-1-2493 and N00014-20-1-2688, with Drs. Ki-Han Kim and Yin Lu Young as Program Officers. Computer time was provided by the U.S. Department of Defense High Performance Computing Modernization Program (HPCMP) and Center for Research Computing (CRC) at the University of Notre Dame. The authors gratefully acknowledge Drs.~William Devenport, Nathan Alexander and N. Agastya Balantrapu for providing experimental data and helpful discussions. D.Z. is grateful to Dr.~Xinyue E. Zhao for her help with data post-processing. A Portion of this work in a preliminary form was presented in AIAA Paper 2020-2989, AIAA Aviation 2020 forum (online).

\section*{Declaration of Interests}
The authors report no conflict of interest.


\bibliographystyle{jfm}
\bibliography{reference}

\end{document}